\documentclass[11pt,a4paper]{article}
\usepackage{jheppub}
\usepackage{amsmath}
\usepackage{amssymb}
\usepackage{amsthm}
\usepackage{graphicx}
\usepackage{slashed}
\usepackage{setspace}
\usepackage{floatflt}
\usepackage{hyperref}
\usepackage{multirow}

\newcommand{\cN}{\mathcal{N}}

\newcommand{\cO}{\mathcal{O}}

\newcommand{\ov}{\overline}

\newcommand{\drawsquare}[2]{\hbox{%
\rule{#2pt}{#1pt}\hskip-#2pt%  left vertical
\rule{#1pt}{#2pt}\hskip-#1pt%  lower horizontal
\rule[#1pt]{#1pt}{#2pt}}\rule[#1pt]{#2pt}{#2pt}\hskip-#2pt%  upper horizontal
\rule{#2pt}{#1pt}}% right vertical
\newcommand{\fund}{\raisebox{-.5pt}{\drawsquare{6.5}{0.4}}}%  fund
\newcommand{\Ysymm}{\raisebox{-.5pt}{\drawsquare{6.5}{0.4}}\hskip-0.4pt%
\raisebox{-.5pt}{\drawsquare{6.5}{0.4}}}%  symmetric second rank
\newcommand{\Yasymm}{\raisebox{-3.5pt}{\drawsquare{6.5}{0.4}}\hskip-6.9pt%
\raisebox{3pt}{\drawsquare{6.5}{0.4}}}%  antisymmetric second rank
\newcommand{\antifund}{\overline{\fund}}

\usepackage{color}

\newcommand{\ur}{u^c_L}
\newcommand{\dr}{d^c_L}
\newcommand{\Er}{e^c_L}
\newcommand{\Nr}{\nu^c_L}

\title{Implications of String Constraints for Exotic Matter and $Z'$\,s Beyond the Standard Model}

\author{Mirjam Cveti\v c$^{1,2}$,}
\author{James Halverson$^1$ and}
\author{Paul Langacker$^{3,4}$}

 \affiliation{$^1$Department of Physics and Astronomy, \\University of Pennsylvania,
  Philadelphia, PA 19104-6396, USA \vspace{.25cm} } 
 \affiliation{$^2$Center for Applied Mathematics and Theoretical Physics,\\
University of Maribor, Maribor, Slovenia \vspace{.25cm} }
 \affiliation{$^3$School of Natural Science, Institute for Advanced Study, \\ Einstein Drive, Princeton, NJ 08540, USA \vspace{.25cm} }
 \affiliation{$^4$Department of Physics, Princeton University, Princeton, NJ 08544, USA}

\emailAdd{cvetic@cvetic.hep.upenn.edu}
\emailAdd{jhal@physics.upenn.edu}
\emailAdd{pgl@ias.edu}

\abstract{Global consistency of string compactifications places constraints on the chiral matter spectrum of a gauge
theory which include those necessary for the absence of cubic nonabelian anomalies, but also contain some additional
conditions. In the class of theories we study, some of these
are present in a field theory augmented by anomalous $U(1)$'s
and Chern-Simons terms, but  some  are genuinely not present in field theory. Their violation  has  phenomenological implications, rendering inconsistent many quiver gauge theories with the chiral 
matter spectrum of the MSSM. The inconsistent MSSM quivers often violate the constraints in a particular way that
is suggestive of what matter must be added for  consistency. The preferred matter additions are
MSSM singlets with anomalous $U(1)$ charge, hyperchargeless $SU(2)$ triplets, quasichiral Higgs or lepton isodoublet pairs,  quasichiral quark isosinglet pairs, and nonabelian singlets with charge $\pm1$. Smaller numbers of quark isodoublet
pairs, lepton pairs with charges ($\pm 1,\pm 2$), and chiral fourth families are also found. We
present the results of systematic analyses including multiplicity counts of matter beyond the standard model and also study the 
possibility of using the  singlets for a dynamical perturbative $\mu$-term or for neutrino mass. We also systematically study the appearance of additional non-anomalous $U(1)'$ symmetries in the low energy theory  and find that family non-universality is very common. These new physics effects may be observable at the LHC even for a large string scale $M_s$ close to the Planck scale.}

\begin{document}
\begin{flushright}
{\small \tt
  \tt UPR-1231-T
}
\end{flushright}
\maketitle

\section{Introduction}

String theory is a consistent UV-complete theory that naturally gives rise to gauge theories. This means, in particular, that gauge theories arising
from some string theories automatically satisfy
field theoretic anomaly cancellation conditions. For example, in type IIa orientifold 
compactifications\footnote{For original work on intersecting branes, see \cite{Blumenhagen:2000wh, Aldazabal:2000dg, Aldazabal:2000cn, Blumenhagen:2001te,Cvetic:2001nr,Cvetic:2001tj}. For extensive reviews, see \cite{Blumenhagen:2005mu,Blumenhagen:2006ci}. For a shorter introduction to intersecting branes and
their non-perturbative D-instanton effects, see \cite{Cvetic:2011vz}.},  gauge theories live on D6-branes which wrap three-cycles in 
the internal dimensions, while chiral matter appears at the intersection of two D6-branes.
Consistency of the compactification imposes constraints on the chiral matter (known as tadpole cancellation conditions) which ensure 
the cancellation of non-abelian gauge anomalies (i.e., anomalies involving the cube of a single
simple group factor). The absence of abelian, mixed, and mixed gauge-gravitational
anomalies is also ensured, sometimes utilizing the   generalized Green-Schwarz mechanism,
in which the triangle anomalies are cancelled by Chern-Simons couplings.

However, global consistency conditions often impose extra constraints
 which are not present in standard field theoretic gauge theories\footnote{There is also the well-known distinction that perturbative intersecting-brane constructions only lead to matter fields transforming as bifundamentals, symmetric or antisymmetric products, and
adjoints. Gauge singlets can arise as bifundamentals of $U(1)$ or antisymmetrics of $SU(2)$, but in both
cases they have anomalous $U(1)$ charge that frequently plays an important phenomenological role.}.  
This occurs even in the ``local" or ``bottom-up" approach \cite{Antoniadis:2000ena,Aldazabal:2000sa}, which allows for the study of issues
related to particle physics while ignoring global aspects of the compactification, such as the 
specification of a particular compactification manifold. Despite the fact that it ignores global
aspects of a compactification, the bottom-up approach still contains string theoretic input
and constraints.
Many of the
new constraints are captured in ``augmented'' gauge field theory, i.e., field theory
augmented by ``anomalous'' $U(1)$ symmetries and Chern-Simons couplings\footnote{``Anomalous'' symmetries are those in which the apparent anomalies are cancelled by the Chern-Simons terms. Henceforth, we usually  omit the parentheses.}. It is precisely this
type of field theory that occurs in many string compactifications. 

In this paper we use the language of quiver gauge theories to specify
the gauge symmetry and matter content of such a field theory. A quiver is a graph of nodes and directed edges, where the nodes represent gauge factors and directed edges represent chiral matter\footnote{For an example of a quiver with the exact MSSM spectrum (i.e.,  without  right-handed neutrinos), see \cite{Cvetic:2011vz}. For a discussion of the advantages of this approach to the string landscape, see \cite{Cvetic:2010dz}.}.
In the type II intersecting
brane context,
the quiver nodes are the gauge symmetries on D-branes, the edges are matter fields living at the intersection
of branes, and the anomalous $U(1)$'s are the (trace) $U(1)$ of a $U(N)$ gauge symmetry on a brane. The
Chern-Simons 
couplings are a generic feature of the D-brane effective  action, and  so augmented
field theories of the type described are  strongly motivated by string theory.

Though many aspects of stringy quiver gauge theories can be captured by considering an augmented
field theory,  there are additional ``stringy'' consistency constraints that are not required in field theory (given our current knowledge). The constraints necessary for tadpole cancellation, which
ensures the net cancellation of D-brane charge on the compact internal dimensions via Gauss'
law, provide two such examples. The simplest of these requires that (in the absence of symmetric
and antisymmetric products)
there are an equal number of $\textbf{2}$ and $\ov{\textbf{2}}$ representations, where the
bar denotes the charge under the trace $U(1)$ of $U(2)$. Though the analogous constraint for $\textbf{3}$ and $\ov{\textbf{3}}$ ensures the absence of $SU(3)^3$ triangle anomalies, the 
constraint on doublets is not present in field theory, as there is never an $SU(2)^3$ anomaly.
To make this point more concrete, we will present a quiver gauge theory with the exact MSSM spectrum and additional $U(1)$'s, the anomalies of which
can be cancelled by the introduction of Chern-Simons terms. This quiver does not suffer from any
gauge anomalies and appears to have no field theoretic pathology, but does in fact violate the constraint that requires than same number of $\textbf{2}$ and $\ov{\textbf{2}}$. The violation 
is due to the precise structure of the MSSM spectrum with respect to anomalous $U(1)$'s,  so the constraints have implications for the structure of matter both in and beyond
the standard model.

We discuss the constraints in a broad context because, though they have been extensively
studied in the context of type IIa orientifold compactifications with intersecting branes, 
they also appear in other patches of the landscape.
This includes type IIb orientifold compactifications, as one might expect via T-duality, but
in addition the same constraints appear in the non-geometric rational conformal field theory phase of type II \cite{Bianchi:2000de,Dijkstra:2004ym,Anastasopoulos:2010hu}, complete with
a ``mod 3" feature that is surprising on first sight. Since the stringy constraints are closely
related to the presence of anomalous $U(1)$ factors, it would be interesting
to see if they can be derived in smooth heterotic compactifications in the case where
the holomorphic vector bundle $V$ has structure group $U(N)$. Such compactifications are S-dual to type I string compactifications with D9-branes and D5-branes, which in turn are mirror symmetric to the type IIa intersecting braneworlds we typically have in mind \cite{Blumenhagen:2005zg}. 

Since the stringy constraints have implications for the structure of matter (with respect to anomalous $U(1)$'s),
one might choose to explore the phenomenology of those MSSM quivers which
satisfy the constraints necessary for string consistency. 
At the level of the spectrum, such issues were investigated in
\cite{Anastasopoulos:2006da}, which included an analysis of possible MSSM hypercharge
embeddings\footnote{\cite{Anastasopoulos:2006da} also allows for exotics whose
mass is not protected by the electroweak scale. In addition to other differences from their work, our work allows for the most general exotics afforded by the brane spectrum, rather than specific subsets.}. These identify the hypercharge with a linear combination of
anomalous $U(1)$'s which itself is anomaly free. The anomalous
$U(1)$ gauge bosons generically obtain a large Stuckelberg mass, but the $U(1)$ symmetries  survive as global selection rules in the effective
action, forbidding certain superpotential terms. In type II string compactifications, those couplings are forbidden in string perturbation theory, though D-instanton effects can generate non-perturbative corrections \cite{Blumenhagen:2006xt,Ibanez:2006da,Florea:2006si} to otherwise forbidden couplings, such as those involving the $\mu$ parameter or a large Majorana seesaw mass
for right-handed neutrinos. Quiver analyses at the level of couplings including
the implications of D-instantons were carried out in a number of works\footnote{Previous work on type II motivated MSSM-containing quivers often denoted the left-chiral fields $u_L^c$, $d_L^c$,  $e_L^c$, and $\nu_L^c$ by  $u_R$, $d_R$,  $E_R$, and $N_R$, respectively.}, beginning with \cite{Ibanez:2008my}. For MSSM-like quivers, semi-realistic
Yukawa couplings and mass hierarchies were studied in \cite{Anastasopoulos:2009mr,Cvetic:2009yh,Cvetic:2009ez,Cvetic:2009ng}. Issues of proton decay \cite{Cvetic:2009ez, Cvetic:2009ng}, a non-perturbatively generated Dirac neutrino mass operator
$LH_u\,\Nr$ \cite{Cvetic:2008hi},
a non-perturbatively generated Weinberg operator $LLH_uH_u$ \cite{Kiritsis:2009sf, Cvetic:2010mm}, and
singlet-extended standard models \cite{Cvetic:2010dz} have also been investigated in
this context. For an analysis of $SU(5)$ GUT quivers at the level of couplings, including
some globally consistent models, see \cite{Anastasopoulos:2010hu}.

In this work we take a related but different approach, studying what matter beyond the exact MSSM
is preferred by the stringy constraints. We do this by systematically 
studying what matter could be added to otherwise inconsistent MSSM quivers  to render them
consistent. Phenomenologically, such an analysis is important because, though the MSSM is minimal, it is by no means the only option: sometimes simple extensions solve phenomenological problems,
and in any case additional matter or interactions may survive as apparently accidental remnants of the underlying construction. Matter beyond the standard model is one of the major possibilities for new
physics to be observed at the LHC, and it is interesting that the constraints we study prefer some
possibilities over others. We emphasize that we do not add matter arbitrarily, but
instead in the most general way afforded by the stringy constraints. This work can be viewed as a small step in exploring the interesting subset of the string landscape that is consistent with the SM or MSSM but allows additional TeV-scale physics.

For the simplest
class of MSSM quivers, which have three nodes,
we will explicitly show that the violation of stringy constraints takes a form which  strongly suggests what matter should be added to the quiver for the sake of string consistency. We continue with a systematic four-node analysis, demonstrating
the ``preferences" of the stringy constraints for matter beyond the standard model. The
most likely matter additions in the analysis are MSSM singlets charged under the anomalous
$U(1)$'s, triplets of $SU(2)$ with weak hypercharge $Y=0$,  quasichiral Higgs or lepton isodoublet pairs,
quasichiral quark isosinglet pairs, and nonabelian singlets with charge $\pm1$. 
(By quasichiral pairs we mean fields that transform as vector pairs under the SM gauge group, but which are chiral under additional anomalous or non-anomalous $U(1)$ factors.)
Quasichiral $\dr$ $\ov{\dr}$ pairs are far more 
likely than their up-type quark counterparts.
Even less common than the latter are quasichiral quark isodoublet
pairs and lepton isodoublet pairs with charges ($\pm 1,\pm 2$). 
Often the SM singlets
couple perturbatively to the quasichiral  pairs in such as way as to generate
masses for the pair when the singlet acquires a vacuum expectation value (VEV), the most familiar example being a dynamical $\mu$-term for a Higgs pair.

There are also a number of
additions that correspond to an ordinary chiral fourth family, or to a  chiral family
in which the electric charges are shifted by $\pm 1$. The
SM chiral extensions
lead to large corrections to precision electroweak observables~\cite{Kribs:2007nz,Erler:2010sk,Baak:2011ze} except for
small parameter regions with delicate cancellations. More seriously, experimental limits
on their masses require large Yukawa couplings to the Higgs doublets, leading to
Landau poles at one (two) loops for the Yukawa (gauge) couplings at low (e.g., 10-1000 TeV) scales~\cite{Godbole:2009sy},
effectively eliminating the possibility of any perturbative connection to the underlying string construction
unless the string scale is very low. We will therefore focus on the quasichiral pairs\footnote{We also exclude from consideration the subset of the four-node
quivers which lead to fractionally-charged color-singlet particles.}, which do not affect precision
observables unless there are large mass splittings or mixings, and which can acquire large masses by the couplings to MSSM singlet fields\footnote{For LHC implications of such quasichiral pairs, see, e.g.,~\cite{Kang:2007ib,Nath:2010zj,Atre:2011ae,Athron:2011wu,Alves:2011wf,Gopalakrishna:2011ef}.}, avoiding the problem of Landau poles.

 We present results
for the likelihood of all matter fields allowed in the analysis and also for other phenomenological considerations, such as distinguishing between various types of singlets.

Many extensions of the SM involve additional $U(1)'$ gauge symmetries in the low energy theory. These include not
only string constructions, but also many grand unified theories, alternative theories of
electroweak breaking, etc. The associated $Z'$ gauge bosons are
excellent candidates for new physics to be observed at the LHC, and the extra gauge symmetry often has implications for  extended Higgs, neutralino, and fermion sectors; the $\mu$ problem; supersymmetry breaking and mediation; for neutrino mass; and for cosmology. (For  recent reviews, see~\cite{Langacker:2008yv,Nath:2010zj}.)
In the theories we study in this paper, Chern-Simons coupling typically give a string scale Stuckelberg mass to $U(1)$ gauge bosons, though in some cases a non-anomalous $U(1)$
is left massless. We require that one such\footnote{Massless non-anomalous $U(1)'$s may also emerge as subgroups of an underlying non-abelian symmetry. In some, but not all, cases the weak hypercharge and/or additional $U(1)'s$
considered here may emerge in this way, e.g., by the splitting between the $SU(5)$ branes in a $U(5)$
stack of D6-branes.} non-anomalous $U(1)$ is present and identifiable
as weak hypercharge. In some cases there exists another massless $U(1)$
factor, which should be interpreted as a $U(1)'$ gauge symmetry, with a gauge boson that
can obtain a low scale mass via the ordinary Higgs mechanism.
We systematically study the appearance of  $U(1)'$ factors in three-node and four-node quivers,
which are typically accompanied by
quasichiral exotics to ensure anomaly cancellation. In many cases the $U(1)'$ couplings are not family universal.
This implies flavor changing $Z'$ couplings, such as those which have been suggested
as one explanation of possible anomalies in the neutral $B$ system~\cite{Langacker:2008yv,Barger:2009eq,Barger:2009qs,Everett:2009cn,Deshpande:2010hy,delAguila:2010mx}.

We emphasize that the additional $Z'$s that we are considering do not acquire Stuckelberg masses
and are therefore massless except for any mass obtained by the ordinary Higgs mechanism. In
this case, there can be a light $Z'$ even if the string scale in a given compactification is $\cO(M_{pl})$. This is in contrast to a different 
scenario~\cite{Antoniadis:2001np,Kiritsis:2002aj, Ghilencea:2002da,Berenstein:2006pk,Berenstein:2008xg,Armillis:2007tb,Kumar:2007zza,Dudas:2009uq,Anchordoqui:2011ag,Anchordoqui:2011eg} in which the light $Z'$
is associated with an anomalous $U(1)$ which obtains a Stuckelberg mass at the string scale, which is assumed to be  $\cO(\text{TeV})$. While extremely interesting, there is not a strong reason to
expect such a low string scale.

This paper is organized as follows.  In section \ref{sec:aft} we elaborate on the stringy constraints, and give an example of
a quiver which is consistent field theoretically, but does not satisfy the string tadpole conditions.
In section \ref{sec:three-node}, we study all three-node quivers with the exact MSSM spectrum, systematically adding up to five additional fields to those of the MSSM in order to satisfy the stringy conditions. We explicitly exclude vector pairs of fields (i.e., vector under both the gauge symmetries and anomalous $U(1)$'s), because such pairs always satisfy the stringy constraints
and typically acquire\footnote{Vector pairs have large masses at a generic point in moduli
space, though at special points they may become massless.} string scale masses. These three-node quivers are not realistic, because with our assumptions the  lepton and down-type Higgs doublets are indistinguishable, implying significant violation of lepton number and R-parity. Nevertheless, they illustrate other interesting issues,
such as the types of additional matter beyond the MSSM required by the stringy conditions, and the possibility of an additional $Z'$
with family non-universal couplings. In section \ref{sec:4-node} we  systematically study the
consistent four-node  quivers  for each realistic hypercharge embedding, again allowing up to 5 matter additions but no vector pairs.
Most of the matter additions are  MSSM singlets (with anomalous $U(1)$ charges), isotriplets with $Y=0$, or
quasichiral pairs, which are non-chiral under the MSSM gauge group. The largest number of the latter are
 lepton/Higgs doublets,  down-type isosinglet quarks,
and nonabelian singlets with charge $\pm 1$. There are smaller numbers of  pairs of up-type quark isosinglets, quark isodoublets, and
lepton/Higgs doublets with charges $(\pm 1,\pm 2)$. There are also examples of  chiral ordinary or charge-shifted fourth families,
and (as is typical in string theory) of 
fractionally charged color singlets (or particles which can form fractionally charged color-singlet bound states). We further examine the
numbers of quivers with additional properties, including a distinction between lepton and down-type Higgs doublets at the quiver level (necessary but not sufficient for lepton number conservation), how many MSSM-singlets $S_\mu$ can have perturbative
couplings $S_\mu H_u H_d$ that can lead to an NMSSM-type dynamical $\mu$ term, and how many can have the
types of Dirac couplings needed for right-handed neutrinos in a seesaw model. We also examine the subset of quivers involving
a non-anomalous $U(1)'$ gauge symmetry (in addition to hypercharge), with a  $Z'$ that does not acquire a Stuckelberg mass (i.e., which might be observable at the LHC even for a very large string scale comparable to the Planck scale), and find that the majority of these involve
family non-universal  couplings for at least one kind of fermion (because the families are quiver distinct), leading to FCNC effects that might be relevant to
the neutral $B$ system.
In appendix \ref{sec:stringy constraints} we sketch the origin and structure of the stringy tadpole and $U(1)$ masslessness
conditions, while appendix \ref{mattertables} contains tables of the possible fields that can occur for each type of quiver.

\section{Stringy Constraints and Augmented Field Theory}
\label{sec:aft}

The constraints on chiral matter necessary for consistency of a type II orientifold
(or related) string compactification are discussed in detail in appendix
\ref{sec:stringy constraints}, along with a brief sketch of their derivation.
The constraints include:
\begin{itemize}
	\item \emph{\textbf{Field Theory Anomaly Constraints}}: These include
 the cancellation of  non-abelian triangle anomalies, as well as the abelian, mixed, and mixed gauge-gravitational anomalies for non-anomalous $U(1)$'s, such as the hypercharge. 
 	\item \emph{\textbf{Augmented Field Theory Constraints}}: These are constraints arising
 	in some more specialized (augmented) field theory. One example would be the constraints that are required 
	for the cancellation of triangle and gravitational anomalies by Chern-Simons terms in 
	 the MSSM  augmented by anomalous $U(1)$'s. In practice, augmented field theory constraints can contain stringy inputs or motivations, such as the consideration of the anomalous $U(1)$'s and the choices of representations considered.
	\item \emph{\textbf{Stringy Constraints}}: These are constraints necessary for
		string consistency that are not required in field theory, given our current
		knowledge.
\end{itemize}

The gauge theories common in type II string compactifications can contain non-anomalous nonabelian and abelian factors, as in the standard model, but also generically include anomalous $U(1)$ factors
corresponding to the trace generator of a $U(N)$ factor. Fields of the same standard model representation can be charged differently with respect to the anomalous $U(1)$'s, corresponding to the appearance at the intersection of different pairs of D-branes.
We refer to such fields as ``quiver-distinct''. 
In addition, the Wess-Zumino component of the D-brane worldvolume action provides Chern-Simons
couplings which, after dimensional reduction, cancel the four-dimensional abelian and mixed anomalies associated with the anomalous $U(1)$'s. 

It is possible, of course, to consider a gauge theory with anomalous $U(1)$ factors without any reference to string theory. The anomalies can be cancelled by the introduction of four-dimensional Chern-Simons terms of the form  $\int \phi F \wedge F$, $\int \phi R \wedge R$ and $\int B \wedge F$, where $F$ is the field strength of an anomalous $U(1)$ gauge boson and the $0$-form $\phi$ and $2$-form $B$ both possess an axionic shift symmetry. This approach was studied in \cite{Anastasopoulos:2006cz,Mambrini:2009ad}, where \cite{Anastasopoulos:2006cz} gave the generic solution for the structure of Chern-Simons coefficients that cancel all abelian and mixed gauge anomalies. That study found no qualitative difference between
string theory and augmented field theory.  
It is also well known that terms of the form $\int \phi R \wedge R$ can cancel  mixed abelian-gravitational anomalies \cite{Cvetic:2001nr}. The constraints on Chern-Simons coefficients that ensure
the cancellation of anomalies are therefore augmented field theory constraints\footnote{Over the last few years, similar studies of augmented field theories and their relation to string theories (including constraints) were begun in \cite{Kumar:2009us,Kumar:2009ac,Kumar:2009ae}, which focus on the relationship between $6D$
supergravity theories and $6D$ F-theory compactifications. These theories do not
contain the $4D$ chiral matter which we are interested in.}.

There are two sets of constraints which we study in this paper, which are necessary for string consistency and a massless hypercharge boson. We state them here and discuss them in the
context of field theory, augmented field theory,  and stringy constraints, but refer the reader to appendix \ref{sec:stringy constraints} for a more in depth
discussion of their string theoretic origin. The well-known tadpole cancellation conditions
of type II string theory impose a constraint on the cycles wrapped by D-branes which ensures the net cancellation of D-brane charge in the compact internal space via Gauss' law. The condition
on D-brane cycles imposes constraints on chiral matter, given by
\begin{align}
	\label{eqn:chiral tadpole constraint}
	N_a \ge 2&: \qquad \# a - \# \ov a + (N_a+4)\,\, (\# \, \Ysymm_a - \#\, \ov \Ysymm_a) + (N_a-4) \,\, (\# \, \Yasymm_a - \# \, \ov \Yasymm_a) = 0 \notag \\ \notag \\
	N_a = 1&: \qquad \# a - \# \ov a + (N_a+4)\,\, (\# \, \Ysymm_a - \# \, \ov \Ysymm_a) = 0 \,\,\, \text{mod} \,\,\, 3,
\end{align}
where $N_a$ refers to a stack of $N_a$ coincident branes with $U(N_a)$ gauge symmetry, $a$ and
$\ov a$ are fundamentals and antifundamentals of $U(N_a)$, and Young Tableaux denotes
symmetric and antisymmetric representations as usual. 
For $N_a>2$ these are simply the conditions which ensure the absence of $SU(N_a)^3$ 
triangle anomalies, but the constraints for $N_a=2$ and $N_a=1$ are
stringy constraints.

The other constraints we study are related to Chern-Simons coupling of the form $\int B\wedge F$, which give a Stuckelberg mass to anomalous
$U(1)$ gauge bosons (and sometimes even non-anomalous bosons). If these couplings are absent for some non-anomalous linear
 combination $\sum q_x U(1)_x$, the corresponding boson does not obtain a Stuckelberg mass.
 The conditions on the chiral matter
\begin{equation}
\label{eqn:chiral masslessness constraint}
-q_aN_a\,\,(\#\Ysymm_a - \#\ov\Ysymm_a + \#\Yasymm_a - \#\ov\Yasymm_a) + \sum_{x\ne a} q_x N_x \,\, (\#(a,\ov x) - \#(a,x)) = 0
\end{equation}
for $N_a\ge 2$, and
\begin{equation}
\label{eqn:chiral masslessness constraint N1}
-q_a \,\,\frac{\#a - \#\ov a + 8(\#\Ysymm_a-\#\ov\Ysymm_a)}{3} + \sum_{x\ne a} q_x N_x \,\, (\#(a,\ov x) - \#(a,x)) = 0
\end{equation}
for $N_a=1$ and the constraints \eqref{eqn:chiral tadpole constraint} are necessary but
not sufficient for a massless gauge boson and for tadpole cancellation, respectively. We
require that the masslessness constraints are satisfied by a linear combination identifiable
as weak hypercharge. These constraints and the constraints
\eqref{eqn:chiral tadpole constraint} are necessary but not sufficient for a massless hypercharge
boson and tadpole cancellation, respectively.

Let us consider a particular quiver gauge theory with respect to these constraints. It is a three-node quiver, which means that it has three gauge factors, here taken
to be $U(3)_a\times U(2)_b \times U(1)_c$. For simplicity, we take all of the three families to have the structure
\begin{align}
\label{eqn:sick spectrum}
q_L: \,\, (a,\ov{b})_{1,-1,0} \qquad \ur: \,\, (\ov{a}, \ov c)_{-1,0,-1} \qquad \dr: \,\, (\ov{a}, c)_{-1,0,1} \notag \\
L: \,\, (\ov b, \ov c)_{0,-1,-1} \qquad \Er: \,\, (\Ysymm_c)_{0,0,2}, \qquad \qquad
\end{align}
and take $H_u: \,\, (b,c)$ and $H_d: \,\, (\ov b, \ov c)$ for the two Higgs doublets. We explicitly denote the charges of the fields under $U(1)_a$, $U(1)_b$ and $U(1)_c$ as subscripts.

In this quiver, a generic abelian symmetry $U(1)_G$ can be written as $U(1)_G = q_a\, U(1)_a + q_b\, U(1)_b + q_c\, U(1)_c$. Though the theory is free of non-abelian anomalies, we wish to study the abelian and mixed anomalies. Cancellation of these anomalies requires
\begin{align}
U(1)_G^3 \,\,\, \sim \,\,\, &6(q_a-q_b)^3 + 3(-q_a-q_c)^3 + 3(-q_a + q_c)^3 + 2(-q_b-q_c)^3 + (2q_c)^3 \notag \\
& = q_b(-18q_a^2 + 18q_a q_b - 8q_b^2 -6 q_b q_c-6 q_c^2) + q_c^2(-18q_a + 6 q_c) = 0 \notag \\
U(1)_G\, SU(2)^2 \,\,\, \sim \,\,\, & 3(q_a-q_b) + (-q_b-q_c) = 3q_a -q_c -4q_b = 0 \notag \\
U(1)_G\, SU(3)^2 \,\,\, \sim \,\,\, & 2(q_a-q_b) +(-q_a-q_c) + (-q_a + q_c) = -2q_b = 0 \notag \\
U(1) \,\, \text{grav} \,\,\, \sim \,\,\, & 6(q_a-q_b) + 3(-q_a-q_c) + 3(-q_a +q_c) + 2(-q_b-q_c) + 2q_c \notag \\
& = -8q_b = 0,
\end{align}
for which  the only solution has $q_b = 0$ and $3q_a-q_c = 0$. This non-anomalous $U(1)$ can be identified with hypercharge, and is (up to scaling) the Madrid embedding which will be discussed in section \ref{eqn:3-node madrid}. The theory also has two linearly independent
anomalous $U(1)$ symmetries.

By the appropriate introduction of Chern-Simons terms, either in augmented field theory or automatically in a string compactification, all of the anomalies associated with the anomalous $U(1)$'s can be cancelled.
After doing so, the theory is completely free of non-abelian, abelian, mixed abelian-gauge and mixed abelian-gravitational anomalies.
The spectrum is that of the exact MSSM, so there is no global $SU(2)$ anomaly \cite{Witten:1982fp} (i.e. there are not an odd number of $SU(2)$ doublets). From the point of view of augmented field theory,
this theory suffers no pathology.

However, this quiver does not satisfy all of the stringy constraints. From \eqref{eqn:sick spectrum}, all standard model fermions (save the vector-like 
Higgs pair) which are doublets of $SU(2)$ come as $\ov b$ rather than $b$, so that
\begin{equation}
\# b - \# \ov b + (6)(\# \Ysymm_b - \ov \Ysymm_b) + (-2) (\# \Yasymm_b - \ov \Yasymm_b) = -12.
\end{equation}
The constraint \eqref{eqn:chiral tadpole constraint} with $N_a = 2$, which is necessary for tadpole cancellation, requires that this quantity be zero. Therefore, this quiver does not satisfy the stringy constraints on chiral matter. Since $N_b=2$ in this quiver, we refer to the violation
of this constraint as having net ``$T_b$-charge". In analogy, the left hand side of \eqref{eqn:chiral masslessness constraint}
and \eqref{eqn:chiral masslessness constraint N1} are referred as ``M-charge" for convenience.
Via other three-node quivers, one can use similar arguments to argue that the $N_a=1$ constraint of \eqref{eqn:chiral tadpole constraint} is also
 stringy. Indeed, we will show in section \ref{sec:three-node} that these constraints are precisely the ones violated
by any inconsistent three-node MSSM quiver. We will also show that the structure of the violation strongly suggests
what type of matter should be added to the quiver to render it consistent. One would have no need
for such matter additions in augmented field theory.

Let us discuss two more examples which shed light on the constraints and their
phenomenological importance.
The four-node Madrid hypercharge embedding has $U(1)_Y = \frac{1}{6} U(1)_a + \frac{1}{2}U(1)_c + \frac{1}{2} U(1)_d$ 
and will be systematically studied in section \ref{sec:4-node}. Consider any quiver with the exact MSSM spectrum and
the four-node Madrid embedding which satisfies the constraints \eqref{eqn:chiral tadpole constraint},
\eqref{eqn:chiral masslessness constraint} and \eqref{eqn:chiral masslessness constraint N1}.
Now consider the addition of three fields that transform as $(c,\ov d)$. These are MSSM singlets, and therefore the quiver
is still free of $SU(3)^3$, $SU(3)^2 - Y$, $SU(2)^2-Y$, $Y^3$ and $Y-G-G$ anomalies. The singlets do not contribute to any
mixed anomalies involving hypercharge and anomalous $U(1)$'s, and any mixed abelian anomalies (which necessarily do not involve
hypercharge)
to which they contribute can be cancelled by Chern-Simons terms. The singlets are perfectly well behaved with respect to
field theoretic (anomaly) constraints. However, after the addition of the singlets, the quiver still satisfies
the constraints \eqref{eqn:chiral tadpole constraint} and \eqref{eqn:chiral masslessness constraint}, but now violates the 
constraints \eqref{eqn:chiral masslessness constraint N1}, since the M-charge is now $M_c=1$ and $M_d=-1$.
This is striking: these MSSM singlets, which are field theoretically mundane, necessarily force the hypercharge
boson to have a Stuckelberg mass. Similarly, the standard model vector pair
$(\ov{\textbf{3}},1)_{-\frac{2}{3}}$, $(\textbf{3},1)_{\frac{2}{3}}$ realized as $(\ov a, \ov c)$, $(a,d)$ forces
the hypercharge boson to obtain a Stuckelberg mass. This is necessary because the theory has $Y\,U(1)_c\,U(1)_c$ and
$Y\, U(1)_d \,U(1)_d$ triangle anomalies, which for anomaly cancellation requires the presence of the term $\int B\wedge F_Y$ which gives
the hypercharge boson a mass.

\section{Lessons from Three-Node Quivers} \label{sec:three-node}
An MSSM-containing quiver
must have at least three nodes, which we take to have $U(3)_a\times U(2)_b\times U(1)_c$ gauge symmetry, as motivated by type II compactifications. 
As mentioned in section \ref{sec:aft}, an anomalous $U(1)$ gauge boson always receives
a string scale Stuckelberg mass due to the presence of Chern-Simons couplings which
participate in anomaly cancellation, though
some times a non-anomalous $U(1)$ will be left massless. If there is precisely one such non-anomalous\footnote{In this context, a massless $U(1)$ is always non-anomalous, though a $U(1)$ gauge boson with a Stuckelberg mass could be either anomalous or anomaly
free~\cite{Ibanez:2001nd}.}  linear
combination
\begin{equation}
	U(1)_Y = q_a\, U(1)_a + q_b\, U(1)_b + q_c\, U(1)_c,
\end{equation}
that can be identified as hypercharge, then the
low energy gauge symmetry is $SU(3)\times SU(2) \times U(1)_Y$. Ensuring that a $U(1)$ is massless requires that the chiral matter in the theory satisfies the
constraints \eqref{eqn:chiral masslessness constraint} and 
\eqref{eqn:chiral masslessness constraint N1}.
For three-node
quivers there are only two such linear combinations (known as hypercharge embeddings~\cite{Anastasopoulos:2006da}) which allow for the spectrum of the MSSM. 
If a three-node quiver has the exact MSSM spectrum, the conditions for a massless hypercharge are always
satisfied, and the violation of the tadpole conditions suggests the addition of particular
matter fields to render the quiver consistent, i.e., to satisfy the stringy
constraints.

Let us first give
one illustrative example of the power of stringy constraints which applies to both embeddings. 
Consider a three-node quiver with the spectrum of the exact MSSM. If the quiver satisfies
the stringy constraints, then it is straightforward to show that it must have least one
particle $P=(\textbf{1},\textbf{2})_{-\frac{1}{2}}$ which forms a vector pair with $H_u$ under
all symmetries of the theory and is therefore present in the superpotential. In string
compactifications, such a term will generically have a string scale mass, giving rise to either a
very large $\mu$-term $H_d H_u$ or a  large lepton violating coupling $L H_u$ depending on whether
$P$ is identified as a  down-type Higgs or a lepton doublet. Both of these possibilities
have significant phenomenological drawbacks, but are the only possibilities allowed by the stringy
constraints for a three-node quiver with the spectrum of the exact MSSM. 
Thus, one requires either additional matter, or some  mechanism to suppress these couplings.

\subsection{The Madrid Embedding} \label{sec:3-node madrid} 
The first hypercharge embedding we examine is  the three-node Madrid embedding \cite{Ibanez:2001nd},
\begin{equation}
	\label{eqn:3-node madrid}
	U(1)_Y = \frac{1}{6}\, U(1)_a + \frac{1}{2}\, U(1)_c.
\end{equation}
It gives rise to the largest number of three-node MSSM quivers, including the quiver presented in
section \ref{sec:aft}; its natural extension  by adding a $U(1)_d$ node with $q_d=\frac{1}{2}$ gives rise to the largest
number of four-node quivers. A quiver with this hypercharge embedding and appropriate spectrum
might arise dynamically from brane splitting of a $U(5)\times U(1)_c$ GUT quiver, with $U(1)_a$ contained in $U(5)$.
This would be a flipped $SU(5)$ quiver (e.g.,~\cite{Jiang:2006hf}), 
rather than a Georgi-Glashow $SU(5)$  one~\cite{Georgi:1974sy}.
The $\textbf{10}$, realized as $\Yasymm_5$, would give rise to $\dr$, rather than $\ur$,
upon brane splitting.

All possible ways of realizing each MSSM field are listed in Table \ref{table:3-node madrid spectrum}.
Given this information, it is a straightforward exercise 
to enumerate all three-node MSSM quivers with the Madrid hypercharge embedding. 
Every such quiver 
satisfies the conditions necessary for a massless hypercharge. The constraints \eqref{eqn:chiral tadpole constraint} necessary for tadpole cancellation are not always satisfied, however. The $a$-node, $b$-node, and $c$-node  have $N_a=3,2,1$ in \eqref{eqn:chiral tadpole constraint}, respectively,
and we label the value of the left-hand side $T_a$, $T_b$, and $T_c$. A simple computation shows 
that any three-node MSSM quiver
in the Madrid embedding has $T_a$, $T_b$, and $T_c$ of the form
\begin{equation}
	\label{eqn:3-node madrid T}
	T_a = 0 \qquad \qquad T_b= \pm 2n \qquad \qquad T_c = 0 \, \text{mod}\, 3 \qquad \qquad \text{with} \,\,\, n\in \{0, \dots, 7\}.
\end{equation}
 However, a consistent  quiver must have $(T_a, T_b, T_c) = (0,0,0\,\text{mod}\,3)$  to satisfy \eqref{eqn:chiral tadpole constraint}. Thus, though $T_a$ and $T_c$ satisfy the constraint, $T_b$ only does so for quivers with $n=0$.
We henceforth refer to such overshooting as having net ``T-charge", and the analogous overshooting in
the conditions \eqref{eqn:chiral masslessness constraint} and \eqref{eqn:chiral masslessness constraint N1} ensuring a massless hypercharge as having net ``M-charge". The T-charges and M-charge of each possible realization of
MSSM fields are given in Table \ref{table:3-node madrid spectrum}.

\begin{table}[h]
	\centering
	\begin{tabular}{|c|c|c|c|c|c|c|c|} \hline
		Field & Transformation & $T_a$ & $T_b$ & $T_c$ & $M_a$ & $M_b$ & $M_c$ \\ \hline \hline
		$q_L$& $(a,\ov b)$ &$2$ &$-3$ &$0$ &$0$ &$-\frac{1}{2}$ &$0$\\ \hline
		$q_L$& $(a,b)$ &$2$ &$3$ &$0$ &$0$ &$-\frac{1}{2}$ &$0$\\ \hline
		$\ur$& $(\ov a,\ov c)$ &$-1$ &$0$ &$-3$ &$\frac{1}{2}$ &$0$ &$1$\\ \hline
		$\dr$& $\Yasymm_a$ &$-1$ &$0$ &$0$ &$-\frac{1}{2}$ &$0$ &$0$\\ \hline
		$\dr$& $(\ov a,c)$ &$-1$ &$0$ &$3$ &$-\frac{1}{2}$ &$0$ &$0$\\ \hline
		$H_u$& $(b,c)$ &$0$ &$1$ &$2$ &$0$ &$-\frac{1}{2}$ &$-\frac{1}{3}$\\ \hline
		$H_u$&  $(\ov b,c)$ &$0$ &$-1$ &$2$ &$0$ &$-\frac{1}{2}$ &$-\frac{1}{3}$ \\ \hline
		$L$& $(b,\ov c)$ &$0$ &$1$ &$-2$ &$0$ &$\frac{1}{2}$ &$\frac{1}{3}$\\ \hline
		$L$& $(\ov b,\ov c)$ &$0$ &$-1$ &$-2$ &$0$ &$\frac{1}{2}$ &$\frac{1}{3}$\\ \hline
		$\Er$& $\Ysymm_c$ &$0$ &$0$ &$5$ &$0$ &$0$ &$-\frac{4}{3}$\\ \hline
	\end{tabular}
	\caption{All possible MSSM field transformations for the hypercharge embedding 
		$U(1)_Y = \frac{1}{6}\, U(1)_a + \frac{1}{2}\, U(1)_c$, along with their contributions to the conditions
		necessary for tadpole cancellation and a massless $U(1)_Y$. Here and in Table \ref{table:3-node non-madrid spectrum} the rows labeled $L$ represent possible assignments
		for either the lepton doublets or for the $H_d$, which have the same MSSM quantum numbers. The two assignments for
		$H_u$ lead to equivalent quivers, so we will choose  $(b,c)$}
	\label{table:3-node madrid spectrum}
	\vspace{.3cm}
\end{table}

The form of tadpole overshooting in \eqref{eqn:3-node madrid T} strongly suggests what type of matter
should be added to the (inconsistent) quivers with $n\ne 0$. Interestingly, the fact that $T_b$ in \eqref{eqn:3-node non-madrid T} is even ensures that
any matter added to a three-node MSSM Madrid quiver to render it consistent will not introduce a global $SU(2)$ anomaly\footnote{See appendix \ref{sec:stringy constraints} for some brief, but
general, comments about global $SU(2)$ anomalies in this context.}.
The possible additional fields that can be
added to the quiver are listed in Table \ref{table:3-node arb} in appendix \ref{mattertables}.
One sees that the only allowed single-particle additions
are MSSM singlets with anomalous $U(1)$ charge or $SU(2)$ triplets with $Y=0$, since any of the other fields with $T_b\ne 0$
would induce T-charge on another node. Two-particle additions could include some combination of these fields,
but might also include quasichiral Higgs, lepton, or quark doublet pairs.

We have performed a systematic analysis of all allowed sets of matter additions with up to five fields and
present the results for the Madrid embedding in Table \ref{table:3-node madrid additions}.
An ``allowed" matter addition to an existing
quiver must cancel any T-charge or M-charge. We also require that a quiver with matter
additions be completely absent of vector pairs\footnote{ 
This is because vector pairs are allowed in string perturbation theory, but obtain string
scale masses at generic points in moduli space, in which case they  can be integrated out, i.e., they ``vector up''.
Often times
 MSSM quivers $Q_1$ and   $Q_2$ with matter additions $A_1$ and $A_2$ will be equivalent due to change in family structure from vectoring up. A vector pair can also be
added arbitrarily to a quiver, since they have no T-charge or M-charge, and
thus aren't particularly important for the purposes of this work. }, by which we mean vector with respect to all symmetries in the
theory, including anomalous $U(1)$'s. As mentioned above, this assumption  eliminates three-node quivers with no additions. 
Finally,
for a three-node quiver in the Madrid embedding there is an equivalent quiver\footnote{In a type II string compactification, this happens
by switching which brane in an orientifold invariant pair is designated to be the image brane.} obtained
by replacing $b \leftrightarrow \ov b$. We remove this overcounting by fixing $H_u$ to transform as $(b,c)$. Similar statements apply to the other embeddings.

A major difficulty with the three-node quivers is that the lepton doublets $L$ and down-type Higgs doublet $H_d$ are indistinguishable in both their gauge and anomalous $U(1)$ quantum numbers, because once we exclude vector pairs they have only one allowed quiver assignment. This implies that  lepton number (and R-parity) violating couplings such as $QL\dr$, $LL\Er$, and $LH_u$ are allowed at the same level (perturbatively or by non-perturbative D-instanton effects) as  the corresponding lepton-number conserving couplings  $QH_d\dr$, $LH_d \Er$, and $H_d H_u$. We ignore this difficulty in this section, as we are more concerned with the gauge quantum numbers
of the matter additions. However, we will discuss it in more detail in section  \ref{sec:4-node}, where it will be seen that many of the four-node quivers do allow the desired distinction between the lepton and Higgs fields.

We see from the results of the table that a large number of quivers indeed involve MSSM singlets and triplets\footnote{There is no quiver with only an added triplet,
since this would require the presence of a vector pair $H_u L$ or $H_uH_d$.} of $SU(2)$ with $Y=0$. As emphasized previously, these are always charged under the anomalous
$U(1)$'s, which strongly affects the structure of couplings. 
Some of the singlets are candidates for NMSSM-type Higgs singlets or for right-handed neutrinos, depending on the $L$ and $H_d$ assignments, as will be discussed in section \ref{sec:4-node}.
Quasichiral quark doublet pairs and down-type quark isosinglet pairs are quite common, but
there are no up-type isosinglet pairs. Quasichiral $SU(2)$-doublet pairs, which may be interpreted as additional Higgs pairs, lepton pairs, or both,
are also common, though perhaps not as much as one
might expect. This is because in the three-node Madrid embedding the only way to add quasichiral Higgs pairs without any vectoring up is to add more copies of the MSSM $H_u$ and $H_d$ fields.
Two of the quivers involve additions which have the MSSM quantum numbers of a   \textbf{5}+\textbf{5}$^\ast$+\textbf{1} of $SU(5)$. They therefore would not
modify the MSSM-type running of the gauge coupling constants at one loop\footnote{However, as is well known, intersecting brane constructions generically do not lead to simple gauge unification conditions at the string or GUT scale~\cite{Blumenhagen:2005mu,Blumenhagen:2006ci,Cvetic:2011vz}.}.
There are also five additions which correspond to a fourth generation $q_L$, $\ur$, $\dr$, $L$, $\Er$. There are no unusually charged states that could lead to doubly or fractionally charged color singlets.

\begin{table}[htb]
	\centering
\scalebox{.8}{
	\begin{tabular}{|c|c|c|c|c|c|}  \hline
		Multiplicity & \multicolumn{5}{c|}{Matter Additions} \\ \hline
$4$ & $\Ysymm_{b}$, $(\textbf{1},\textbf{3})_{0}$ & $\Ysymm_{b}$, $(\textbf{1},\textbf{3})_{0}$ & $\Yasymm_{b}$, $(\textbf{1},\textbf{1})_{0}$ & $(a,\ov{b})$, $(\textbf{3},\textbf{2})_{\frac{1}{6}}$ & $(\ov{a},\ov{b})$, $(\ov{\textbf{3}},\textbf{2})_{-\frac{1}{6}}$ \\ \hline
$4$ & $\Ysymm_{b}$, $(\textbf{1},\textbf{3})_{0}$ & $\Yasymm_{b}$, $(\textbf{1},\textbf{1})_{0}$ & & & \\ \hline
$4$ & $\ov \Ysymm_{b}$, $(\textbf{1},\textbf{3})_{0}$ & $\Yasymm_{b}$, $(\textbf{1},\textbf{1})_{0}$ & & & \\ \hline
$4$ & $\Ysymm_{b}$, $(\textbf{1},\textbf{3})_{0}$ & $\Yasymm_{b}$, $(\textbf{1},\textbf{1})_{0}$ & $\Yasymm_{b}$, $(\textbf{1},\textbf{1})_{0}$ & $(b,\ov{c})$, $(\textbf{1},\textbf{2})_{-\frac{1}{2}}$ & $(b,c)$, $(\textbf{1},\textbf{2})_{\frac{1}{2}}$ \\ \hline
$4$ & $\ov \Ysymm_{b}$, $(\textbf{1},\textbf{3})_{0}$ & $\Yasymm_{b}$, $(\textbf{1},\textbf{1})_{0}$ & $\Yasymm_{b}$, $(\textbf{1},\textbf{1})_{0}$ & $(b,\ov{c})$, $(\textbf{1},\textbf{2})_{-\frac{1}{2}}$ & $(b,c)$, $(\textbf{1},\textbf{2})_{\frac{1}{2}}$ \\ \hline
$4$ & $\Ysymm_{b}$, $(\textbf{1},\textbf{3})_{0}$ & $\ov \Yasymm_{b}$, $(\textbf{1},\textbf{1})_{0}$ & $\ov \Yasymm_{b}$, $(\textbf{1},\textbf{1})_{0}$ & $(a,\ov{b})$, $(\textbf{3},\textbf{2})_{\frac{1}{6}}$ & $(\ov{a},\ov{b})$, $(\ov{\textbf{3}},\textbf{2})_{-\frac{1}{6}}$ \\ \hline
$4$ & $\ov \Yasymm_{b}$, $(\textbf{1},\textbf{1})_{0}$ & $\ov \Yasymm_{b}$, $(\textbf{1},\textbf{1})_{0}$ & & & \\ \hline
$4$ & $\ov \Yasymm_{b}$, $(\textbf{1},\textbf{1})_{0}$ & $(b,\ov{c})$, $(\textbf{1},\textbf{2})_{-\frac{1}{2}}$ & $(b,c)$, $(\textbf{1},\textbf{2})_{\frac{1}{2}}$ & & \\ \hline
$4$ & $(b,\ov{c})$, $(\textbf{1},\textbf{2})_{-\frac{1}{2}}$ & $(b,\ov{c})$, $(\textbf{1},\textbf{2})_{-\frac{1}{2}}$ & $(b,c)$, $(\textbf{1},\textbf{2})_{\frac{1}{2}}$ & $(b,c)$, $(\textbf{1},\textbf{2})_{\frac{1}{2}}$ & \\ \hline
$4$ & $(a,\ov{b})$, $(\textbf{3},\textbf{2})_{\frac{1}{6}}$ & $\Yasymm_{a}$, $(\ov{\textbf{3}},\textbf{1})_{\frac{1}{3}}$ & $(b,\ov{c})$, $(\textbf{1},\textbf{2})_{-\frac{1}{2}}$ & $(\ov{a},\ov{c})$, $(\ov{\textbf{3}},\textbf{1})_{-\frac{2}{3}}$ & $\Ysymm_{c}$, $(\textbf{1},\textbf{1})_{1}$ \\ \hline
$4$ & $\Ysymm_{b}$, $(\textbf{1},\textbf{3})_{0}$ & $\Yasymm_{b}$, $(\textbf{1},\textbf{1})_{0}$ & $\Yasymm_{b}$, $(\textbf{1},\textbf{1})_{0}$ & $\Yasymm_{b}$, $(\textbf{1},\textbf{1})_{0}$ & $\Yasymm_{b}$, $(\textbf{1},\textbf{1})_{0}$ \\ \hline
$4$ & $\ov \Ysymm_{b}$, $(\textbf{1},\textbf{3})_{0}$ & $\Yasymm_{b}$, $(\textbf{1},\textbf{1})_{0}$ & $\Yasymm_{b}$, $(\textbf{1},\textbf{1})_{0}$ & $\Yasymm_{b}$, $(\textbf{1},\textbf{1})_{0}$ & $\Yasymm_{b}$, $(\textbf{1},\textbf{1})_{0}$ \\ \hline
$4$ & $\ov \Ysymm_{b}$, $(\textbf{1},\textbf{3})_{0}$ & $\ov \Yasymm_{b}$, $(\textbf{1},\textbf{1})_{0}$ & $\ov \Yasymm_{b}$, $(\textbf{1},\textbf{1})_{0}$ & & \\ \hline
$4$ & $\ov \Ysymm_{b}$, $(\textbf{1},\textbf{3})_{0}$ & $\ov \Yasymm_{b}$, $(\textbf{1},\textbf{1})_{0}$ & $(b,\ov{c})$, $(\textbf{1},\textbf{2})_{-\frac{1}{2}}$ & $(b,c)$, $(\textbf{1},\textbf{2})_{\frac{1}{2}}$ & \\ \hline
$4$ & $\ov \Ysymm_{b}$, $(\textbf{1},\textbf{3})_{0}$ & $(b,\ov{c})$, $(\textbf{1},\textbf{2})_{-\frac{1}{2}}$ & $(b,\ov{c})$, $(\textbf{1},\textbf{2})_{-\frac{1}{2}}$ & $(b,c)$, $(\textbf{1},\textbf{2})_{\frac{1}{2}}$ & $(b,c)$, $(\textbf{1},\textbf{2})_{\frac{1}{2}}$ \\ \hline
$4$ & $\Yasymm_{b}$, $(\textbf{1},\textbf{1})_{0}$ & & & & \\ \hline
$4$ & $\Yasymm_{b}$, $(\textbf{1},\textbf{1})_{0}$ & $\Yasymm_{b}$, $(\textbf{1},\textbf{1})_{0}$ & $(b,\ov{c})$, $(\textbf{1},\textbf{2})_{-\frac{1}{2}}$ & $(b,c)$, $(\textbf{1},\textbf{2})_{\frac{1}{2}}$ & \\ \hline
$4$ & $\ov \Ysymm_{b}$, $(\textbf{1},\textbf{3})_{0}$ & $\ov \Ysymm_{b}$, $(\textbf{1},\textbf{3})_{0}$ & $\ov \Yasymm_{b}$, $(\textbf{1},\textbf{1})_{0}$ & $\ov \Yasymm_{b}$, $(\textbf{1},\textbf{1})_{0}$ & \\ \hline
$4$ & $\ov \Ysymm_{b}$, $(\textbf{1},\textbf{3})_{0}$ & $\ov \Ysymm_{b}$, $(\textbf{1},\textbf{3})_{0}$ & $\ov \Yasymm_{b}$, $(\textbf{1},\textbf{1})_{0}$ & $(b,\ov{c})$, $(\textbf{1},\textbf{2})_{-\frac{1}{2}}$ & $(b,c)$, $(\textbf{1},\textbf{2})_{\frac{1}{2}}$ \\ \hline
$4$ & $\Yasymm_{b}$, $(\textbf{1},\textbf{1})_{0}$ & $\Yasymm_{b}$, $(\textbf{1},\textbf{1})_{0}$ & $\Yasymm_{b}$, $(\textbf{1},\textbf{1})_{0}$ & $\Yasymm_{b}$, $(\textbf{1},\textbf{1})_{0}$ & \\ \hline
$4$ & $\ov \Ysymm_{b}$, $(\textbf{1},\textbf{3})_{0}$ & $\ov \Ysymm_{b}$, $(\textbf{1},\textbf{3})_{0}$ & $\ov \Ysymm_{b}$, $(\textbf{1},\textbf{3})_{0}$ & $\ov \Yasymm_{b}$, $(\textbf{1},\textbf{1})_{0}$ & $\ov \Yasymm_{b}$, $(\textbf{1},\textbf{1})_{0}$ \\ \hline
$4$ & $\ov \Ysymm_{b}$, $(\textbf{1},\textbf{3})_{0}$ & $\ov \Ysymm_{b}$, $(\textbf{1},\textbf{3})_{0}$ & $\Yasymm_{b}$, $(\textbf{1},\textbf{1})_{0}$ & & \\ \hline
$1$ & $\Yasymm_{a}$, $(\ov{\textbf{3}},\textbf{1})_{\frac{1}{3}}$ & $\Ysymm_{b}$, $(\textbf{1},\textbf{3})_{0}$ & $\Yasymm_{b}$, $(\textbf{1},\textbf{1})_{0}$ & $(a,\ov{c})$, $(\textbf{3},\textbf{1})_{-\frac{1}{3}}$ & \\ \hline
$1$ & $\ov \Yasymm_{a}$, $(\textbf{3},\textbf{1})_{-\frac{1}{3}}$ & $\Ysymm_{b}$, $(\textbf{1},\textbf{3})_{0}$ & $\Yasymm_{b}$, $(\textbf{1},\textbf{1})_{0}$ & $(\ov{a},c)$, $(\ov{\textbf{3}},\textbf{1})_{\frac{1}{3}}$ & \\ \hline
$1$ & $\Yasymm_{a}$, $(\ov{\textbf{3}},\textbf{1})_{\frac{1}{3}}$ & $\ov \Ysymm_{b}$, $(\textbf{1},\textbf{3})_{0}$ & $\Yasymm_{b}$, $(\textbf{1},\textbf{1})_{0}$ & $(a,\ov{c})$, $(\textbf{3},\textbf{1})_{-\frac{1}{3}}$ & \\ \hline
$1$ & $\ov \Yasymm_{a}$, $(\textbf{3},\textbf{1})_{-\frac{1}{3}}$ & $\ov \Ysymm_{b}$, $(\textbf{1},\textbf{3})_{0}$ & $\Yasymm_{b}$, $(\textbf{1},\textbf{1})_{0}$ & $(\ov{a},c)$, $(\ov{\textbf{3}},\textbf{1})_{\frac{1}{3}}$ & \\ \hline
$1$ & $\Yasymm_{a}$, $(\ov{\textbf{3}},\textbf{1})_{\frac{1}{3}}$ & $\ov \Yasymm_{b}$, $(\textbf{1},\textbf{1})_{0}$ & $\ov \Yasymm_{b}$, $(\textbf{1},\textbf{1})_{0}$ & $(a,\ov{c})$, $(\textbf{3},\textbf{1})_{-\frac{1}{3}}$ & \\ \hline
$1$ & $\ov \Yasymm_{a}$, $(\textbf{3},\textbf{1})_{-\frac{1}{3}}$ & $\ov \Yasymm_{b}$, $(\textbf{1},\textbf{1})_{0}$ & $\ov \Yasymm_{b}$, $(\textbf{1},\textbf{1})_{0}$ & $(\ov{a},c)$, $(\ov{\textbf{3}},\textbf{1})_{\frac{1}{3}}$ & \\ \hline
$1$ & $\Yasymm_{a}$, $(\ov{\textbf{3}},\textbf{1})_{\frac{1}{3}}$ & $\ov \Yasymm_{b}$, $(\textbf{1},\textbf{1})_{0}$ & $(b,\ov{c})$, $(\textbf{1},\textbf{2})_{-\frac{1}{2}}$ & $(b,c)$, $(\textbf{1},\textbf{2})_{\frac{1}{2}}$ & $(a,\ov{c})$, $(\textbf{3},\textbf{1})_{-\frac{1}{3}}$ \\ \hline
$1$ & $\ov \Yasymm_{a}$, $(\textbf{3},\textbf{1})_{-\frac{1}{3}}$ & $\ov \Yasymm_{b}$, $(\textbf{1},\textbf{1})_{0}$ & $(b,\ov{c})$, $(\textbf{1},\textbf{2})_{-\frac{1}{2}}$ & $(b,c)$, $(\textbf{1},\textbf{2})_{\frac{1}{2}}$ & $(\ov{a},c)$, $(\ov{\textbf{3}},\textbf{1})_{\frac{1}{3}}$ \\ \hline
$1$ & $(a,\ov{b})$, $(\textbf{3},\textbf{2})_{\frac{1}{6}}$ & $(b,\ov{c})$, $(\textbf{1},\textbf{2})_{-\frac{1}{2}}$ & $(\ov{a},c)$, $(\ov{\textbf{3}},\textbf{1})_{\frac{1}{3}}$ & $(\ov{a},\ov{c})$, $(\ov{\textbf{3}},\textbf{1})_{-\frac{2}{3}}$ & $\Ysymm_{c}$, $(\textbf{1},\textbf{1})_{1}$ \\ \hline
$1$ & $\Yasymm_{a}$, $(\ov{\textbf{3}},\textbf{1})_{\frac{1}{3}}$ & $\ov \Ysymm_{b}$, $(\textbf{1},\textbf{3})_{0}$ & $\ov \Yasymm_{b}$, $(\textbf{1},\textbf{1})_{0}$ & $\ov \Yasymm_{b}$, $(\textbf{1},\textbf{1})_{0}$ & $(a,\ov{c})$, $(\textbf{3},\textbf{1})_{-\frac{1}{3}}$ \\ \hline
$1$ & $\ov \Yasymm_{a}$, $(\textbf{3},\textbf{1})_{-\frac{1}{3}}$ & $\ov \Ysymm_{b}$, $(\textbf{1},\textbf{3})_{0}$ & $\ov \Yasymm_{b}$, $(\textbf{1},\textbf{1})_{0}$ & $\ov \Yasymm_{b}$, $(\textbf{1},\textbf{1})_{0}$ & $(\ov{a},c)$, $(\ov{\textbf{3}},\textbf{1})_{\frac{1}{3}}$ \\ \hline
$1$ & $\Yasymm_{a}$, $(\ov{\textbf{3}},\textbf{1})_{\frac{1}{3}}$ & $\Yasymm_{a}$, $(\ov{\textbf{3}},\textbf{1})_{\frac{1}{3}}$ & $\Yasymm_{b}$, $(\textbf{1},\textbf{1})_{0}$ & $(a,\ov{c})$, $(\textbf{3},\textbf{1})_{-\frac{1}{3}}$ & $(a,\ov{c})$, $(\textbf{3},\textbf{1})_{-\frac{1}{3}}$ \\ \hline
$1$ & $\ov \Yasymm_{a}$, $(\textbf{3},\textbf{1})_{-\frac{1}{3}}$ & $\ov \Yasymm_{a}$, $(\textbf{3},\textbf{1})_{-\frac{1}{3}}$ & $\Yasymm_{b}$, $(\textbf{1},\textbf{1})_{0}$ & $(\ov{a},c)$, $(\ov{\textbf{3}},\textbf{1})_{\frac{1}{3}}$ & $(\ov{a},c)$, $(\ov{\textbf{3}},\textbf{1})_{\frac{1}{3}}$ \\ \hline
$1$ & $\Yasymm_{a}$, $(\ov{\textbf{3}},\textbf{1})_{\frac{1}{3}}$ & $\Yasymm_{b}$, $(\textbf{1},\textbf{1})_{0}$ & $(a,\ov{c})$, $(\textbf{3},\textbf{1})_{-\frac{1}{3}}$ & & \\ \hline
$1$ & $\ov \Yasymm_{a}$, $(\textbf{3},\textbf{1})_{-\frac{1}{3}}$ & $\Yasymm_{b}$, $(\textbf{1},\textbf{1})_{0}$ & $(\ov{a},c)$, $(\ov{\textbf{3}},\textbf{1})_{\frac{1}{3}}$ & & \\ \hline
$1$ & $\Yasymm_{a}$, $(\ov{\textbf{3}},\textbf{1})_{\frac{1}{3}}$ & $\ov \Ysymm_{b}$, $(\textbf{1},\textbf{3})_{0}$ & $\ov \Ysymm_{b}$, $(\textbf{1},\textbf{3})_{0}$ & $\Yasymm_{b}$, $(\textbf{1},\textbf{1})_{0}$ & $(a,\ov{c})$, $(\textbf{3},\textbf{1})_{-\frac{1}{3}}$ \\ \hline
$1$ & $\ov \Yasymm_{a}$, $(\textbf{3},\textbf{1})_{-\frac{1}{3}}$ & $\ov \Ysymm_{b}$, $(\textbf{1},\textbf{3})_{0}$ & $\ov \Ysymm_{b}$, $(\textbf{1},\textbf{3})_{0}$ & $\Yasymm_{b}$, $(\textbf{1},\textbf{1})_{0}$ & $(\ov{a},c)$, $(\ov{\textbf{3}},\textbf{1})_{\frac{1}{3}}$ \\ \hline
	\end{tabular}}
	\caption{Matter additions and their multiplicity for the 105 three-node quivers with the Madrid embedding, up to five additions, and no vector pairs. Listed for
	each set of additions is its multiplicity and the field transformation behavior with respect to both the quiver and the MSSM.}
	\label{table:3-node madrid additions}
\end{table}

Of the $105$ MSSM quivers whose matter additions are presented in Table \ref{table:3-node madrid additions}, only
one allows for an additional (non-anomalous) massless $U(1)'$ gauge symmetry. This quiver is presented in Table \ref{table:3-node madrid u1}. In this
quiver, any linear combination of the form
\begin{equation}
	U(1)_{\text{massless}} = a \, U(1)_a + b \, U(1)_b + (3a - 2b) \, U(1)_c
	\label{eqn:3-node u1'}
\end{equation}
is left massless\footnote{Technically, this $U(1)$ only satisfies constraints necessary (but not sufficient) for a massless gauge boson. In a globally defined compactification, the $U(1)$ may still become massive, though this possibility cannot be addressed at the quiver level. This is true
of all ``massless" $U(1)$'s in this paper.} by the Green-Schwarz mechanism, where hypercharge is the linear combination with
$a = \frac{1}{6}$ and $b = 0$. This quiver exhibits
a number of features that we will see are characteristic of $U(1)'$ symmetries in many quivers with
a higher number of nodes, such as family non-universality of the $U(1)'$ and an interesting structure
of exotics. It has some nice phenomenological properties, such as a perturbatively realized
top-quark mass for two generations and superpotential couplings which allow for a perturbative dynamical $\mu$-term
\begin{equation}
	W \supset (S_1 + S_2)\,H_u\,H_d,
\end{equation}
where $S_1$ and $S_2$ are the two singlets added to the quiver for consistency, realized as antisymmetrics of $SU(2)$. 
As with the other 3-node quivers, however, the   leptons $L$ are indistinguishable from $H_d$.
\begin{table}[htb]
	\centering
	\begin{tabular}{|c|c|} \hline
		$q_L$ & $(a, \ov b)_{1,-1}$\qquad  $(a, \ov b)_{1,-1}$\qquad  $(a, b)_{1,1}$\\ \hline
		$\ur$ & $(\ov a, \ov c)_{-4, 2}$\qquad  $(\ov a,\ov c)_{-4, 2}$\qquad  $(\ov a,\ov c)_{-4, 2}$ \\ \hline
		$\dr$ & $(\Yasymm_a)_{2, 0}$\qquad  $(\ov a,c)_{2, -2}$\qquad  $(\ov a,c)_{2, -2}$ \\ \hline
		$L$ & $(b,\ov c)_{-3, 3}$\qquad  $(b,\ov c)_{-3, 3}$\qquad  $(b,\ov c)_{-3, 3}$ \\ \hline
		$\Er$ & $(\Ysymm_c)_{6,-4}$\qquad $(\Ysymm_c)_{6,-4}$\qquad $(\Ysymm_c)_{6,-4}$\\ \hline
		$H_u$ & $(b,c)_{3,-1}$\\ \hline
		$H_d$ & $(b,\ov c)_{-3, 3}$ \\ \hline
		Add & $(\ov \Ysymm_b)_{0,-2}$ \qquad $(\ov \Yasymm_b)_{0,-2}$ \qquad $(\ov \Yasymm_b)_{0,-2}$\\ \hline
	\end{tabular}
	\caption{The only three-node quiver with a $U(1)'$ symmetry with up to five matter additions that contains
	the MSSM spectrum and has no vector pairs. ``Add'' refers to the additional matter that must be added to satisfy the stringy constraints. Subscripts denote the values of $a$ and $b$ in \eqref{eqn:3-node u1'} for that field.}
	\label{table:3-node madrid u1}
\end{table}

\subsection{The non-Madrid Embedding \label{sec:3-node non-madrid}} 
The other three-node hypercharge embedding, which we call the ``non-Madrid" embedding,
is given by
\begin{equation}
	\label{eqn:3-node non-madrid}
	U(1)_Y = -\frac{1}{3}\, U(1)_a - \frac{1}{2}\, U(1)_b.
\end{equation}
As mentioned in the introduction, this could actually emerge from a non-abelian   Georgi-Glashow  $SU(5)$
subgroup of $U(5)$ by brane splitting, with the $\ur$ realized as $\Yasymm_a$ in the $\textbf{10}$. 
Every possible field which might
appear in a three-stack quiver is displayed in Table \ref{table:3-node arb}. Of those, the ones
which correspond to MSSM fields are listed in Table \ref{table:3-node non-madrid spectrum}.

Enumerating every possible three-node quiver with the non-Madrid embedding and the MSSM spectrum,
the overshoot in the tadpole takes the form
\begin{equation}
	\label{eqn:3-node non-madrid T}
	T_a = 0 \,\,\,\,\,\,\,\, T_b= 0 \,\,\,\,\,\,\,\, T_c \in \{-15,-11,-9,-7,-5,-3,-1,1,3,5,7,9,11,13,15,19\}.
\end{equation}
Those quivers with $T_c\ne 0\,\,\text{mod} \,\,3$ violate the string consistency constraints
\eqref{eqn:chiral tadpole constraint}. The only possible single-particle addition
is an MSSM singlet.
The $41$ quivers from a systematic analysis with up to five matter additions are presented in Table \ref{table:3-node non-madrid additions}. We fix $H_u$ to transform
as $(\ov b, c)$ in order to account for the quiver symmetry $c \leftrightarrow \ov c$. There are many MSSM-singlets,  many quasichiral Higgs/lepton doublet pairs, and many quasichiral
down-type quark isosinglet pairs.  Two of the quivers involve additional matter with the MSSM quantum numbers of  \textbf{5}+\textbf{5}$^\ast$+\textbf{1}
of $SU(5)$, and there are five examples of a fourth family.
None have a massless linear combination of anomalous $U(1)$'s that can be interpreted as a massless $U(1)'$ symmetry.
Similar to the three-node Madrid embedding, there is no distinction between the quantum numbers of the lepton and down-Higgs doublets.

\begin{table}[htb]
	\centering
	\begin{tabular}{|c|c|c|c|c|c|c|c|}\hline
		Field & Transformation & $T_a$ & $T_b$ & $T_c$ & $M_a$ & $M_b$ & $M_c$ \\ \hline \hline
		$q_L$ & $(a, \ov b)$ & $2$ & $-3$ & $0$ & $-1$ & $1$ & $0$ \\ \hline 
		$\ur$ & $\Yasymm_a$ & $-1$ & $0$ & $0$ & $1$ & $0$ & $0$ \\ \hline
		$\dr$ & $(\ov a, c)$ & $-1$ & $0$ & $3$ & $0$ & $0$ & $-1$ \\ \hline
		$\dr$ & $(\ov a, \ov c)$ &$-1$ & $0$ & $-3$ & $0$ & $0$ & $-1$ \\ \hline
		$H_u$ & $(\ov b, c)$ & $0$ & $-1$ & $2$ & $0$ & $0$ & $-1$ \\ \hline
		$H_u$ & $(\ov b, \ov c)$ & $0$ & $-1$ & $2$ & $0$ & $0$ & $-1$ \\ \hline
		$L$ & $(b, \ov c)$ & $0$ & $1$ & $-2$ & $0$ & $0$ & $1$ \\ \hline
		$L$ & $(b, c)$ & $0$ & $1$ & $2$ & $0$ & $0$ & $1$ \\ \hline
		$\Er$ & $\ov \Yasymm_b$ & $0$ & $2$ & $0$ & $0$ & $-1$ & $0$ \\ \hline
	\end{tabular}
	\caption{All possible MSSM field transformations for the hypercharge embedding 
		$U(1)_Y = -\frac{1}{3}\, U(1)_a - \frac{1}{2}\, U(1)_b$, along with their contributions to the conditions
		necessary for tadpole cancellation and a massless $U(1)_Y$. }
	\label{table:3-node non-madrid spectrum}
\end{table}

\begin{table}[htb]
	\centering
	\scalebox{.9}{
	\begin{tabular}{|c|c|c|c|c|c|}\hline
		Multiplicity & \multicolumn{5}{c|}{Matter Additions} \\ \hline
$4$ & $\Ysymm_{c}$, $(\textbf{1},\textbf{1})_{0}$ & & & & \\ \hline
$4$ & $\Ysymm_{c}$, $(\textbf{1},\textbf{1})_{0}$ & $\Ysymm_{c}$, $(\textbf{1},\textbf{1})_{0}$ & $\Ysymm_{c}$, $(\textbf{1},\textbf{1})_{0}$ & $\Ysymm_{c}$, $(\textbf{1},\textbf{1})_{0}$ & \\ \hline
$4$ & $\ov \Ysymm_{c}$, $(\textbf{1},\textbf{1})_{0}$ & $\ov \Ysymm_{c}$, $(\textbf{1},\textbf{1})_{0}$ & & & \\ \hline
$4$ & $\ov \Ysymm_{c}$, $(\textbf{1},\textbf{1})_{0}$ & $\ov \Ysymm_{c}$, $(\textbf{1},\textbf{1})_{0}$ & $\ov \Ysymm_{c}$, $(\textbf{1},\textbf{1})_{0}$ & $\ov \Ysymm_{c}$, $(\textbf{1},\textbf{1})_{0}$ & $\ov \Ysymm_{c}$, $(\textbf{1},\textbf{1})_{0}$ \\ \hline
$4$ & $(a,\ov{b})$, $(\textbf{3},\textbf{2})_{\frac{1}{6}}$ & $\Yasymm_{a}$, $(\ov{\textbf{3}},\textbf{1})_{-\frac{2}{3}}$ & $(b,c)$, $(\textbf{1},\textbf{2})_{-\frac{1}{2}}$ & $(\ov{a},c)$, $(\ov{\textbf{3}},\textbf{1})_{\frac{1}{3}}$ & $\ov \Yasymm_{b}$, $(\textbf{1},\textbf{1})_{1}$ \\ \hline
$4$ & $(\ov{b},c)$, $(\textbf{1},\textbf{2})_{\frac{1}{2}}$ & $(\ov{b},c)$, $(\textbf{1},\textbf{2})_{\frac{1}{2}}$ & $(b,c)$, $(\textbf{1},\textbf{2})_{-\frac{1}{2}}$ & $(b,c)$, $(\textbf{1},\textbf{2})_{-\frac{1}{2}}$ & \\ \hline
$4$ & $(\ov{b},c)$, $(\textbf{1},\textbf{2})_{\frac{1}{2}}$ & $(b,c)$, $(\textbf{1},\textbf{2})_{-\frac{1}{2}}$ & $\Ysymm_{c}$, $(\textbf{1},\textbf{1})_{0}$ & $\Ysymm_{c}$, $(\textbf{1},\textbf{1})_{0}$ & \\ \hline
$4$ & $(\ov{b},c)$, $(\textbf{1},\textbf{2})_{\frac{1}{2}}$ & $(b,c)$, $(\textbf{1},\textbf{2})_{-\frac{1}{2}}$ & $\ov \Ysymm_{c}$, $(\textbf{1},\textbf{1})_{0}$ & & \\ \hline
$1$ & $\Ysymm_{c}$, $(\textbf{1},\textbf{1})_{0}$ & $(\ov{a},c)$, $(\ov{\textbf{3}},\textbf{1})_{\frac{1}{3}}$ & $(\ov{a},c)$, $(\ov{\textbf{3}},\textbf{1})_{\frac{1}{3}}$ & $(a,c)$, $(\textbf{3},\textbf{1})_{-\frac{1}{3}}$ & $(a,c)$, $(\textbf{3},\textbf{1})_{-\frac{1}{3}}$ \\ \hline
$1$ & $\Ysymm_{c}$, $(\textbf{1},\textbf{1})_{0}$ & $(a,\ov{c})$, $(\textbf{3},\textbf{1})_{-\frac{1}{3}}$ & $(a,\ov{c})$, $(\textbf{3},\textbf{1})_{-\frac{1}{3}}$ & $(\ov{a},\ov{c})$, $(\ov{\textbf{3}},\textbf{1})_{\frac{1}{3}}$ & $(\ov{a},\ov{c})$, $(\ov{\textbf{3}},\textbf{1})_{\frac{1}{3}}$ \\ \hline
$1$ & $\Ysymm_{c}$, $(\textbf{1},\textbf{1})_{0}$ & $(\ov{a},c)$, $(\ov{\textbf{3}},\textbf{1})_{\frac{1}{3}}$ & $(a,c)$, $(\textbf{3},\textbf{1})_{-\frac{1}{3}}$ & & \\ \hline
$1$ & $\Ysymm_{c}$, $(\textbf{1},\textbf{1})_{0}$ & $(a,\ov{c})$, $(\textbf{3},\textbf{1})_{-\frac{1}{3}}$ & $(\ov{a},\ov{c})$, $(\ov{\textbf{3}},\textbf{1})_{\frac{1}{3}}$ & & \\ \hline
$1$ & $\ov \Ysymm_{c}$, $(\textbf{1},\textbf{1})_{0}$ & $\ov \Ysymm_{c}$, $(\textbf{1},\textbf{1})_{0}$ & $(\ov{a},c)$, $(\ov{\textbf{3}},\textbf{1})_{\frac{1}{3}}$ & $(a,c)$, $(\textbf{3},\textbf{1})_{-\frac{1}{3}}$ & \\ \hline
$1$ & $\ov \Ysymm_{c}$, $(\textbf{1},\textbf{1})_{0}$ & $\ov \Ysymm_{c}$, $(\textbf{1},\textbf{1})_{0}$ & $(a,\ov{c})$, $(\textbf{3},\textbf{1})_{-\frac{1}{3}}$ & $(\ov{a},\ov{c})$, $(\ov{\textbf{3}},\textbf{1})_{\frac{1}{3}}$ & \\ \hline
$1$ & $(a,\ov{b})$, $(\textbf{3},\textbf{2})_{\frac{1}{6}}$ & $\Yasymm_{a}$, $(\ov{\textbf{3}},\textbf{1})_{-\frac{2}{3}}$ & $(b,c)$, $(\textbf{1},\textbf{2})_{-\frac{1}{2}}$ & $(\ov{a},\ov{c})$, $(\ov{\textbf{3}},\textbf{1})_{\frac{1}{3}}$ & $\ov \Yasymm_{b}$, $(\textbf{1},\textbf{1})_{1}$ \\ \hline
$1$ & $(\ov{b},c)$, $(\textbf{1},\textbf{2})_{\frac{1}{2}}$ & $(b,c)$, $(\textbf{1},\textbf{2})_{-\frac{1}{2}}$ & $\ov \Ysymm_{c}$, $(\textbf{1},\textbf{1})_{0}$ & $(\ov{a},c)$, $(\ov{\textbf{3}},\textbf{1})_{\frac{1}{3}}$ & $(a,c)$, $(\textbf{3},\textbf{1})_{-\frac{1}{3}}$ \\ \hline
$1$ & $(\ov{b},c)$, $(\textbf{1},\textbf{2})_{\frac{1}{2}}$ & $(b,c)$, $(\textbf{1},\textbf{2})_{-\frac{1}{2}}$ & $\ov \Ysymm_{c}$, $(\textbf{1},\textbf{1})_{0}$ & $(a,\ov{c})$, $(\textbf{3},\textbf{1})_{-\frac{1}{3}}$ & $(\ov{a},\ov{c})$, $(\ov{\textbf{3}},\textbf{1})_{\frac{1}{3}}$ \\ \hline
	\end{tabular}}
	\caption{Matter additions and their multiplicity for the 41 three-node quivers with the non-Madrid embedding, up to five additions, and no vector pairs.}
	\label{table:3-node non-madrid additions}
\end{table}

\section{Statistics and Trends in Four-Node Quivers}\label{sec:4-node}
There are a number of interesting physical issues that arise in the systematic study of four-node quivers with matter additions. We divide our analysis into two sections, with one studying the physics of the full set of quivers, and the other restricted to those quivers with a $U(1)'$ symmetry in the low energy theory. In each section, we will discuss them in the order of
least to most phenomenological input. 

Let us first summarize our assumptions. We study quivers with a number
of different hypercharge embeddings\footnote{This is the subset of hypercharge embeddings
which can accommodate consistent quivers exhibiting the MSSM plus three singlets.}, given by \cite{Anastasopoulos:2006da}
\begin{align}
\label{eqn:4-node hypercharge embeddings}
U(1)_Y &= -\frac{1}{3} U(1)_a - \frac{1}{2}U(1)_b \qquad\qquad \qquad \,\,\,\,\,\,\,\,\,\,
U(1)_Y = -\frac{1}{3} U(1)_a - \frac{1}{2}U(1)_b + \frac{1}{2}U(1)_d\notag \\
U(1)_Y &= -\frac{1}{3} U(1)_a - \frac{1}{2}U(1)_b + U(1)_d \qquad\qquad
U(1)_Y = \frac{1}{6} U(1)_a + \frac{1}{2}U(1)_c  \qquad \qquad \qquad\notag \\
U(1)_Y &= \frac{1}{6} U(1)_a + \frac{1}{2}U(1)_c + \frac{1}{2}U(1)_d \qquad \qquad
U(1)_Y = \frac{1}{6} U(1)_a + \frac{1}{2}U(1)_c + \frac{3}{2}U(1)_d
\end{align}
where the generic $U(3)_a\times U(2)_b \times U(1)_c \times U(1)_d$ gauge symmetry 
is broken to the standard model (perhaps with additional $U(1)'$ factors) when anomalous $U(1)$ bosons receive a Stuckelberg mass. We construct all quivers with up to five additional matter fields beyond the exact MSSM and ensure that there are no vector pairs, where vector here means with respect to all symmetries in the theory, including the anomalous $U(1)$'s. We place no restrictions on the structure of matter additions except that they be in the class of fields which can be realized in the quiver (bifundamental, symmetric, and antisymmetric representations) and that the MSSM quiver plus additions satisfies the constraints \eqref{eqn:chiral tadpole constraint}, \eqref{eqn:chiral masslessness constraint} and \eqref{eqn:chiral masslessness constraint N1}. The possible fields for each embedding are listed in tables \ref{4node1}-\ref{4node6} in appendix \ref{mattertables}.

\subsection{Trends in Generic Quivers and Preferred Matter Additions}
\label{sec:4node generic}

There are $146$ three-node quivers and $89818$ four-node quivers which satisfy our assumptions. Given the hypercharge embeddings that we study, these quivers have no phenomenological input
other than that they contain the spectrum of the exact MSSM. Since the matter beyond the MSSM is entirely determined by the stringy constraints on chiral matter, it is interesting to see if any particular type of matter field is ``preferred". We present the results of this analysis
in Table \ref{table:particle addition table}, where matter additions are classified with respect to their representation
under the standard model gauge group. 
\begin{table}[htb]
\centering
\scalebox{.95}{
\begin{tabular}{|c|c|c|c|c|}\hline
SM Rep & Total Multiplicity & Int. El. & $4^\text{th}$ Gen. Removed & Shifted $4^\text{th}$ Gen. Also Removed \\ \hline \hline
$(\textbf{1},\textbf{1})_{0}$ & $174276$ & $173578$ & $173578$ & $173578$ \\ \hline
$(\textbf{1},\textbf{3})_{0}$ & $48291$ & $48083$ & $48083$ & $48083$ \\ \hline
$(\textbf{1},\textbf{2})_{-\frac{1}{2}}$ & $39600$ & $39560$ & $38814$ & $38814$ \\ \hline
$(\textbf{1},\textbf{2})_{\frac{1}{2}}$ & $38854$ & $38814$ & $38814$ & $38814$ \\ \hline
$(\ov{\textbf{3}},\textbf{1})_{\frac{1}{3}}$ & $25029$ & $25007$ & $24261$ & $24241$ \\ \hline
$(\textbf{3},\textbf{1})_{-\frac{1}{3}}$ & $24299$ & $24277$ & $24277$ & $24241$ \\ \hline
$(\textbf{1},\textbf{1})_{1}$ & $15232$ & $15228$ & $14482$ & $14482$ \\ \hline
$(\textbf{1},\textbf{1})_{-1}$ & $14486$ & $14482$ & $14482$ & $14482$ \\ \hline
$(\ov{\textbf{3}},\textbf{1})_{-\frac{2}{3}}$ & $3501$ & $3501$ & $2755$ & $2755$ \\ \hline
$(\textbf{3},\textbf{1})_{\frac{2}{3}}$ & $2755$ & $2755$ & $2755$ & $2755$ \\ \hline
$(\textbf{3},\textbf{2})_{\frac{1}{6}}$ & $1784$ & $1784$ & $1038$ & $1038$ \\ \hline
$(\ov{\textbf{3}},\textbf{2})_{-\frac{1}{6}}$ & $1038$ & $1038$ & $1038$ & $1038$ \\ \hline
$(\textbf{1},\textbf{2})_{0}$ & $852$ & $0$ & $0$ & $0$ \\ \hline
$(\textbf{1},\textbf{2})_{\frac{3}{2}}$ & $220$ & $220$ & $220$ & $184$ \\ \hline
$(\textbf{1},\textbf{2})_{-\frac{3}{2}}$ & $204$ & $204$ & $204$ & $184$ \\ \hline
$(\textbf{1},\textbf{1})_{\frac{1}{2}}$ & $152$ & $0$ & $0$ & $0$ \\ \hline
$(\textbf{1},\textbf{1})_{-\frac{1}{2}}$ & $152$ & $0$ & $0$ & $0$ \\ \hline
$(\textbf{3},\textbf{1})_{\frac{1}{6}}$ & $124$ & $0$ & $0$ & $0$ \\ \hline
$(\ov{\textbf{3}},\textbf{1})_{-\frac{1}{6}}$ & $124$ & $0$ & $0$ & $0$ \\ \hline
$(\textbf{3},\textbf{1})_{-\frac{4}{3}}$ & $36$ & $36$ & $36$ & $0$ \\ \hline
$(\textbf{1},\textbf{3})_{-1}$ & $36$ & $36$ & $36$ & $0$ \\ \hline
$(\ov{\textbf{3}},\textbf{2})_{\frac{5}{6}}$ & $36$ & $36$ & $36$ & $0$ \\ \hline
$(\ov{\textbf{3}},\textbf{1})_{\frac{4}{3}}$ & $20$ & $20$ & $20$ & $0$ \\ \hline
$(\textbf{1},\textbf{3})_{1}$ & $20$ & $20$ & $20$ & $0$ \\ \hline
$(\textbf{3},\textbf{2})_{-\frac{5}{6}}$ & $20$ & $20$ & $20$ & $0$ \\ \hline
\end{tabular}}
\caption{Displayed are the standard model representation of matter additions, together with their total multiplicity
across all three-node and four-node quivers in the hypercharge embeddings we have studied. The third column excludes quivers
involving states that would lead to fractionally-charged color singlets. The fourth further excludes those where the matter additions correspond to a fourth generation, while the last also excludes  a shifted fourth generation. The remaining additions correspond to MSSM singlets, $SU(2)$ triplets with $Y=0$, and  quasichiral pairs.}\label{table:particle addition table}
\end{table}

MSSM singlets are by far the most preferred type of matter addition, with $SU(2)$ triplets with $Y=0$ and 
quasichiral pairs (such as Higgs/lepton doublets and isosinglet down-type quarks) 
 being the next most common, corroborating the intuition gained in the three-node case. 
 We emphasize that all of these vector pairs are quasichiral with respect to the anomalous $U(1)$'s. This is because our analysis does not allow for the appearance of pairs which are vector with respect to all symmetries in the theory for the reasons stated above.

Some of the four-node quivers involve fields with the ``wrong'' relation between their $SU(2)$ and $Y$ assignments, viz,
\begin{equation}
(\textbf{1},\textbf{2})_{0} \qquad (\textbf{1},\textbf{1})_{\frac{1}{2}} \qquad (\textbf{1},\textbf{1})_{-\frac{1}{2}} \qquad (\textbf{3},\textbf{1})_{\frac{1}{6}} \qquad (\ov{\textbf{3}},\textbf{1})_{-\frac{1}{6}} . 
\end{equation}
These have fractional electric charge, or, for the color triplets, would bind to form fractionally-charged color singlets.
Such fractional charges are a typical feature of string constructions, and are very strongly constrained by experiment unless they are extremely heavy or confined. (For a recent discussion, see,~\cite{Langacker:2011db}.) We therefore 
eliminate these quivers from further consideration.

Since we consider up to five matter additions beyond the MSSM, one possible set is a fourth generation $q_L$, $\ur$, $\dr$, $L$, $\Er$. This is not the only possible chiral anomaly-free set of five fields, though, as the combination
\begin{equation}
(\textbf{3},\textbf{2})_{-\frac{5}{6}} \qquad
		(\ov{\textbf{3}},\textbf{1})_{\frac{1}{3}}\qquad
		(\ov{\textbf{3}},\textbf{1})_{\frac{4}{3}}\qquad
		(\textbf{1},\textbf{2})_{-\frac{3}{2}}\qquad
		(\textbf{1},\textbf{3})_{1}
\end{equation}
also turns out to be anomaly free. We call this combination (or its charge conjugate) a ``shifted fourth family". 
It consists of a quark doublet with electric charges $-1/3$ and $-4/3$ and a lepton doublet with charges $-1,-2$.
The right-handed leptons\footnote{Similar to the $\Nr$ for an ordinary family, a singlet $(1,1)_0$ could be added 
phenomenologically to partner with the neutral component, but this would require the study of a sixth addition.} occur in an $SU(2)$ triplet with charges $0,-1,-2$. 
As far as we know this curious shifted family has not been commented on\footnote{A vector pair $(3,2)_{-5/6}+
(\ov{3},2)_{5/6}$ has been suggested in connection with the LEP forward-backward
asymmetry into $b \bar b$~\cite{Choudhury:2001hs}, but no such pairs emerge in our analysis.}.
Relatively few of the quivers involve a fourth family, e.g., only 595 of the quivers with
the four-node Madrid embedding $U(1)_Y = \frac{1}{6} U(1)_a + \frac{1}{2} U(1)_c + \frac{1}{2} U(1)_d$. Furthermore,
as described in the introduction, ordinary or shifted chiral families lead to Landau poles at low energy, and 
also require delicate cancellations for precision electroweak physics.
For these reasons,  we will instead focus on the remaining quivers after these are removed.

The last column in Table \ref{table:particle addition table} lists the number of quivers with each type of matter addition, 
after removing those with fractional charges or an ordinary or shifted fourth family.
There is a clear symmetry in the multiplicity of matter additions which indicates that standard model vector pairs are  common. For example, there are $38814$ occurrences of both $(\textbf{1},\textbf{2})_{-\frac{1}{2}}$ and $(\textbf{1},\textbf{2})_{\frac{1}{2}}$ that are not a part of a fourth family. This pair, or any other standard model vector pair, must be quasichiral with respect to the anomalous $U(1)$'s in the quiver given the assumptions
of our analysis, i.e., no true vector pairs are allowed. 
The fact that a pair $X$ and $\ov{X}$ are quasichiral implies that the MSSM gauge invariant term $X\ov{X}$ has net anomalous $U(1)$ charge, forbidding
it in the superpotential. This term could be generated by non-perturbative effects, such as D-instantons, in which case the  mass of the pair cannot be determined without
a detailed global construction. However,  it is also possible that
$XX$ couples to some singlet $S$ so that $SXX$ is allowed (perturbatively). In that case the mass would be generated by the VEV of $S$, which (if nonzero) would likely be
at the electroweak-TeV scale.

The most common quasichiral pair, as shown in Table \ref{table:particle addition table}, is a  $(\textbf{1},\textbf{2})_{-\frac{1}{2}}+(\textbf{1},\textbf{2})_{\frac{1}{2}}$ pair, which could be Higgs-like or lepton-like, or both, depending on their allowed couplings.
The next most common  are down-type quark isosinglets, $(\ov{\textbf{3}},\textbf{1})_{\frac{1}{3}} +(\textbf{3},\textbf{1})_{-\frac{1}{3}}$, similar to those which occur in $E_6$ models~\cite{Langacker:1980js,Hewett:1988xc}.
Together, these two types of pairs have the MSSM quantum numbers of
$\textbf{5}+\textbf{5}^*$ of $SU(5)$. There are $2832$ quivers where the additional matter
has the quantum numbers of $\textbf{5}+\textbf{5}^*$ or $\textbf{5}+\textbf{5}^*+\textbf{1}$.
Also occurring, but less frequently, are isoscalar up-type pairs $(\ov{\textbf{3}},\textbf{1})_{-\frac{2}{3}} + (\textbf{3},\textbf{1})_{\frac{2}{3}}$, and isodoublet quark pairs, $(\ov{\textbf{3}},\textbf{2})_{-\frac{1}{6}} + (\textbf{3},\textbf{2})_{\frac{1}{6}}$. Such quasichiral quarks can be produced prolifically by QCD
processes at the LHC, with a variety of interesting decay signatures~\cite{Kang:2007ib,Nath:2010zj,Atre:2011ae,Athron:2011wu,Alves:2011wf,Gopalakrishna:2011ef}.
Typically, the heavier quasichiral quarks and their scalar partners undergo cascade decays into the lighter ones,
with the lightest decaying by mixing, leptoquark or diquark interactions, or higher-dimensional operators. The decays
may be rapid, delayed, or  stable (on collider time scales). Finally, there are a significant number of quivers with isosinglet
charged lepton or Higgs pairs, $(\textbf{1},\textbf{1})_{1}+(\textbf{1},\textbf{1})_{-1}$, and a few with isodoublets with electric charges $(\pm1,\pm2)$, $(\textbf{1},\textbf{2})_{\frac{3}{2}}+(\textbf{1},\textbf{2})_{-\frac{3}{2}}$.

In the MSSM the lepton doublets $L$ and the down-type Higgs doublet $H_d$ have the same gauge quantum numbers,
transforming as  $(\textbf{1},\textbf{2})_{-\frac{1}{2}}$. If there is no distinction between them in an $\cN=1$ gauge theory, and therefore also in the quivers we study, then the lepton and R-parity violating couplings $q_L L \dr$, $L L \Er$,  $H_u L$, and $S_\mu H_u L$ (where $S_\mu $ is a possible MSSM-singlet addition)
are
indistinguishable from the ordinary  couplings $q_L H_d \dr$,  $L H_d \Er$,   $\mu$-term $H_u H_d$, or dynamical $\mu$-term  $S_\mu H_u H_d$, and therefore are expected to be present at comparable  scales. For example, if one is allowed perturbatively, then so is the other, although the coefficients could differ, with similar statements for non-perturbative terms generated by D-instantons. 
These do not necessarily lead to proton decay (depending on possible baryon number violation), but are nevertheless strongly constrained by limits on rare processes and other experiments~\cite{Chemtob:2004xr,Barbier:2004ez}.
We therefore impose the further phenomenological restriction of only considering those quivers in which there
is a clear distinction between the lepton and down-Higgs doublets.
(This is a necessary but not sufficient condition for lepton number and R-parity conservation.)
This can occur when they are quiver distinct, i.e., they are realized at the intersection of different stacks of D-branes, and therefore have different anomalous
$U(1)$ charges.
Specifically, the quivers that we study must have between 4 and 9 fields transforming as $(\textbf{1},\textbf{2})_{-\frac{1}{2}}$.
We call a given field of this type an ``$H_d$ candidate'' if it is quiver distinct from at least three other $(\textbf{1},\textbf{2})_{-\frac{1}{2}}$ fields, which we identify as lepton doublets $L$.

\begin{table}[htb]	
	\centering
	\begin{tabular}{|r||ccc|ccc|}\hline
		& \multicolumn{6}{c|}{Multiplicity of Quivers} \\[2pt] \hline
		Hypercharge & Total & Int. El. & $H_d$ Candidate & No 4th Gen & $S_\mu H_uH_d$ & $\Nr H_u L$\\ \hline \hline
		$(-\frac{1}{3},-\frac{1}{2},0)$ & $41$& $41$ & $0$ & $0$ & $0$ & $0$ \\[2pt] \hline
		$(\frac{1}{6},0,\frac{1}{2})$ & $105$& $105$ & $0$ & $0$ & $0$ & $0$ \\[2pt] \hline
		$(-\frac{1}{3},-\frac{1}{2},0,0)$ & $6974$& $6974$ & $4954$ & $4938$ & $1824$ & $2066$  \\[2pt] \hline
		$(-\frac{1}{3},-\frac{1}{2},0,\frac{1}{2})$ & $70$& $0$ & $0$ & $0$ & $0$ & $0$ \\[2pt] \hline
		$(-\frac{1}{3},-\frac{1}{2},0,1)$ & $4176$& $4176$ & $1842$ & $1792$ & $0$ & $80$ \\[2pt] \hline
		$(\frac{1}{6},0,\frac{1}{2},0)$ & $480$& $16$ & $0$ & $0$ & $0$ & $0$ \\[2pt] \hline
		$(\frac{1}{6},0,\frac{1}{2},\frac{1}{2})$ & $77853$& $77853$ & $54119$ & $53654$ & $16754$ & $15524$ \\[2pt] \hline
		$(\frac{1}{6},0,\frac{1}{2},\frac{3}{2})$ & $265$& $265$ & $0$ & $0$ & $0$ & $0$ \\[2pt] \hline
	\end{tabular}
	\caption{Counts of quiver with certain properties. The first column displays the coefficients of the $U(1)$ factors for the hypercharge. The second  gives the total number of quivers satisfying string constraints on chiral matter and anomaly cancellation, given the assumptions of our analysis. The third represents those quivers in the second column that do not give rise to fractional electric charge.  The fourth represents those quivers in the third column that have a candidate for a distinct $H_d$. The remaining
	columns represent those quivers in column four where the matter additions do not correspond to a fourth family, those  which have a singlet $S_\mu$ with a  perturbative  $S_\mu H_uH_d$ term, and those  which have at least one $\Nr$-candidate with a
	Dirac coupling $\Nr H_u L$ term, respectively. }
	\label{table:quiver count w hypercharge rows}
\end{table}

We also consider possible couplings of the MSSM singlets, $(\textbf{1},\textbf{1})_{0}$. A singlet $S_\mu$ is ``NMSSM-like'' if the anomalous $U(1)$ charges allow the perturbative coupling $S_\mu H_uH_d$  (independent of other discrete, global, or gauge symmetries that may affect $S_\mu$).
 A  dynamical $\mu$ term is generated if $S_\mu$ acquires a vacuum expectation value, as in the NMSSM
 and related models~\cite{Ellis:1988er,Barger:2006dh,Ellwanger:2009dp,Maniatis:2009re,Cvetic:2010dz}. 
Of course, an effective $\mu$ parameter can instead be generated by non-perturbative D-instantons~\cite{Blumenhagen:2006xt,Ibanez:2006da}. On the other hand, 
a singlet $\Nr$ is  ``$\Nr$-like''
if the coupling $\Nr L H_u$ is perturbatively allowed. This can lead to the Dirac neutrino mass term that is needed in
a conventional seesaw model, perhaps with the Majorana mass generated by
D-instantons~\cite{Blumenhagen:2006xt,Ibanez:2006da}. (Other possibilities for small neutrino masses in this context include
small non-perturbative Dirac masses~\cite{Cvetic:2008hi}, or a non-perturbative  Weinberg operator~\cite{Cvetic:2010mm},
either being generated by D-instantons.) In a given quiver there may be multiple singlets, multiple up-type  $H_u$ fields/candidates, multiple $H_d$ candidates, and/or more than four leptons $L$. The determination of whether
a given singlet is NMSSM-like,  $\Nr$-like, or neither must be done separately for each identification of the $H_u$, $H_d$, and $L$ fields, and the possibilities iterated for each quiver. It is possible that a given quiver identification contains both NMSSM and $\Nr$-like singlets, but a given singlet cannot be both simultaneously if the $H_d$ and $L$ are distinct.

The quiver counts for each hypercharge embedding  are presented in Table \ref{table:quiver count w hypercharge rows},
where the various columns list the total number of quivers, those with fractional charges removed, and those which also have a
unique $H_d$ candidate. The last three columns impose the further  respective restrictions of no fourth family, the
existence of an NMSSM-type singlet, and the existence of at least one $\Nr$-type singlet
(for at least one identification of the matter fields in the quiver).
The four-node Madrid  and non-Madrid embeddings, with 
$U(1)_Y = \frac{1}{6} U(1)_a + \frac{1}{2} U(1)_c + \frac{1}{2} U(1)_d$ and 
$U(1)_Y = -\frac{1}{3}U(1)_a - \frac{1}{2}U(1)_b$ respectively,  give rise to almost ninety percent of the quivers, with the
four-node Madrid embedding alone yielding over eighty percent. This is because
they allow for the MSSM fields to be realized in more ways than the other  embeddings. Only two of
the hypercharge embeddings give rise to quivers with fractional electric charge.

Only three of the  embeddings allow for a distinct $H_d$ candidate\footnote{It was already emphasized in section \ref{sec:three-node} that the
three-node quivers do not distinguish $H_d$ from $L$.}, because there must be at least two possible realizations of $(\textbf{1},\textbf{2})_{-\frac{1}{2}}$ distinct from the vector
partner of the $H_u$. For example, in a four-node quiver in the Madrid embedding with $H_u$ transforming
as $(b,c)$, $H_d$ or $L$ could
be realized as $(b, \ov c)$, $(b, \ov d)$ or $(\ov b, \ov d)$. More hypercharge embeddings could give rise to an $H_d$ candidate
in an analysis where
vector pairs are allowed, but at the expense of introducing
a perturbative $\mu$-term, which is string scale at generic points in moduli space.

It is clear from Table
\ref{table:quiver count w hypercharge rows} that most quivers do not have a fourth family. Combinatorically, this
is simply because there are a very large number of ways to construct five-particle additions (though many do
not give consistent quivers) and a relatively small number of those possibilities correspond to a fourth
family.

Of the 60915 quivers with integral electric charge and a distinct $H_d$ candidate, 51743  contain
at least one MSSM singlet. In 25399 there is either a singlet $S_\mu$ which can couple as $S_\mu H_u H_d$, at least one singlet $\Nr$ which can couple as  $\Nr H_u L$, or both. There are 2322
quivers that have both $S_\mu$ and $\Nr$-type singlets.

\subsection{Quivers with $U(1)'$}
\label{sec:4node u1prime}

We now restrict our attention to the subset of quivers  in section \ref{sec:4node generic}  with a massless non-anomalous $U(1)'$ gauge symmetry
in the low energy theory. This can occur if there is another linear combination,
\begin{equation}
\label{eqn:u1prime linear comb}
U(1)' = \sum_x q_x^{} U(1)_x,
\end{equation}
which satisfies the constraints \eqref{eqn:chiral masslessness constraint}
and \eqref{eqn:chiral masslessness constraint N1}, and is therefore
 left massless by the generalized Green-Schwarz mechanism, just as in the case
of the hypercharge boson. As stated in the introduction, these sometimes can emerge from non-abelian factors
in larger $U(n)$ stacks (similar to  the non-Madrid three-stack embedding of hypercharge), while in other cases they 
are not associated with underlying non-abelian symmetries (analogous to the
Madrid three-stack hypercharge embedding). We again emphasize that these $U(1)'$s may be present at the TeV scale
and observable at the LHC, acquiring mass when singlet fields $S$ (such as NMSSM-like singlets discussed previously) acquire a TeV scale VEV.  The $Z'$ does not obtain a string scale Stuckelberg mass and may therefore   be observable even for large string scales  $M_s$ close to the Planck scale. This is in contrast to the scenario~\cite{Ghilencea:2002da,Berenstein:2006pk,Berenstein:2008xg,Armillis:2007tb,Kumar:2007zza,Dudas:2009uq,Anchordoqui:2011ag,Anchordoqui:2011eg}
in which anomalous bosons are only observable for  $M_s$ at the  TeV scale. For recent globally consistent type IIa models with an extra non-anomalous $U(1)'$ symmetry, see \cite{Maxin:2011ne}, for example.

The number of quivers for each hypercharge embedding are listed in Table \ref{u1primequivers}. 
Comparing with Table \ref{table:quiver count w hypercharge rows} we see that only a small fraction of the total
involve a $U(1)'$, and most of those are in the four-node Madrid embedding. The table also lists the
 number with an $H_d$ candidate, and the numbers with NMSSM-like  or $\Nr$-like singlets\footnote{The possibilities  for neutrino mass in  $U(1)'$ models are surveyed in~\cite{Kang:2004ix}, but a detailed quiver analysis  is beyond the scope of this paper.}. The three families of quarks and leptons are frequently quiver distinct. This can lead to family non-universal
 $U(1)'$ charges for two or more families of $q_L$, $L$, $\ur$, $\dr$, or $\Er$. In fact, it can be seen in  Table \ref{u1primequivers}
 that only about 30\% of the $U(1)'$ quivers are completely universal. Family non-universality leads to a violation of the
 GIM mechanism and therefore to flavor changing neutral current (FCNC) couplings of the $Z'$ when fermion
 family  mixing is turned on~\cite{Langacker:2000ju}, and also to new constraints on the possible Yukawa matrices.
 Non-universality between the third family and the first two is most likely (because of strong experimental constraints from the neutral kaons  and from $\mu$ interactions/decays), and has been suggested as an explanation of possible anomalies in the neutral $B$ system~\cite{Langacker:2008yv,Barger:2009eq,Barger:2009qs,Everett:2009cn,Deshpande:2010hy,delAguila:2010mx}, where even a heavy $Z'$ exchange may be important because it is a tree-level exchange competing with
 SM or MSSM loop effects.

\begin{table}[htb]	
	\centering
	\begin{tabular}{|r||cc|ccc|}\hline
		& \multicolumn{5}{c|}{Multiplicity of Quivers} \\ \hline
		Hypercharge & $U(1)'$ & $H_d$ Candidate & Fam. Univ & $S_\mu H_uH_d$ & $LH_u\Nr$\\ \hline \hline
		$(-\frac{1}{3},-\frac{1}{2},0)$ & $0$ & $0$ & $0$ & $0$  & $0$\\[2pt] \hline
		$(\frac{1}{6},0,\frac{1}{2})$ & $1$ & $0$ & $0$& $0$& $0$\\[2pt] \hline
		$(-\frac{1}{3},-\frac{1}{2},0,0)$ & $198$ & $146$ & $56$ & $70$ & $94$\\[2pt] \hline
		$(-\frac{1}{3},-\frac{1}{2},0,\frac{1}{2})$ & $0$ & $0$ & $0$ & $0$& $0$\\[2pt] \hline
		$(-\frac{1}{3},-\frac{1}{2},0,1)$  & $78$ & $16$ & $10$ & $0$& $5$\\[2pt] \hline
		$(\frac{1}{6},0,\frac{1}{2},0)$ & $0$ & $0$ & $0$& $0$ & $0$\\[2pt] \hline
		$(\frac{1}{6},0,\frac{1}{2},\frac{1}{2})$  & $1803$ & $1466$ & $629$ & $610$ & $600$ \\[2pt] \hline
		$(\frac{1}{6},0,\frac{1}{2},\frac{3}{2})$  & $82$ & $0$ & $0$ & $0$& $0$\\[2pt] \hline
	\end{tabular}
	\caption{Counts of $U(1)'$ quivers with certain properties. In all counts, we have already filtered out any quivers with fractional electric charge. The second column gives counts of $U(1)'$ quivers, and the third gives counts of those with at least one $H_d$ candidate. The remaining columns are  the numbers which in addition are  family universal with respect to $U(1)'$, have a perturbative $S_\mu H_uH_d$ term, or have at least one $\Nr$ with an $LH_u\Nr$ term, respectively.}
	\label{u1primequivers}
\end{table}

The multiplicities of matter additions in the $U(1)'$ models are listed in Table \ref{table:Z' particle addition table}. One sees a pattern
similar to the general case. There are large numbers of MSSM singlets and isotriplets with $Y=0$, a few ordinary and
shifted fourth families, and many quasichiral quark and lepton pairs, including some lepton
doublets with charges $(\pm 1, \pm 2)$. In this case, the pairs may be chiral under either the non-anomalous $U(1)'$
(as is familiar in $E_6$ models),
 the anomalous ones, or both.

\begin{table}[htb]
\centering
\begin{tabular}{|c|c|c|c|}\hline
SM Rep & Total Multiplicity & $4^\text{th}$ Gen. Removed & Shifted $4^\text{th}$ Gen. Also Removed \\ \hline \hline
$(\textbf{1},\textbf{1})_{0}$ & $4556$ & $4556$ & $4556$ \\ \hline
$(\textbf{1},\textbf{3})_{0}$ & $1290$ & $1290$ & $1290$ \\ \hline
$(\textbf{1},\textbf{2})_{-\frac{1}{2}}$ & $631$ & $619$ & $619$ \\ \hline
$(\textbf{1},\textbf{2})_{\frac{1}{2}}$ & $619$ & $619$ & $619$ \\ \hline
$(\ov{\textbf{3}},\textbf{1})_{\frac{1}{3}}$ & $478$ & $466$ & $458$ \\ \hline
$(\textbf{3},\textbf{1})_{-\frac{1}{3}}$ & $458$ & $458$ & $458$ \\ \hline
$(\textbf{1},\textbf{1})_{1}$ & $262$ & $250$ & $250$ \\ \hline
$(\textbf{1},\textbf{1})_{-1}$ & $250$ & $250$ & $250$ \\ \hline
$(\textbf{1},\textbf{2})_{-\frac{3}{2}}$ & $101$ & $101$ & $93$ \\ \hline
$(\textbf{1},\textbf{2})_{\frac{3}{2}}$ & $93$ & $93$ & $93$ \\ \hline
$(\textbf{3},\textbf{2})_{\frac{1}{6}}$ & $46$ & $34$ & $34$ \\ \hline
$(\ov{\textbf{3}},\textbf{2})_{-\frac{1}{6}}$ & $34$ & $34$ & $34$ \\ \hline
$(\ov{\textbf{3}},\textbf{1})_{-\frac{2}{3}}$ & $30$ & $18$ & $18$ \\  \hline
$(\textbf{3},\textbf{1})_{\frac{2}{3}}$ & $18$ & $18$ & $18$ \\ \hline
$(\textbf{1},\textbf{3})_{1}$ & $8$ & $8$ & $0$ \\ \hline
$(\textbf{3},\textbf{2})_{-\frac{5}{6}}$ & $8$ & $8$ & $0$ \\ \hline
$(\ov{\textbf{3}},\textbf{1})_{\frac{4}{3}}$ & $8$ & $8$ & $0$ \\ \hline
\end{tabular}
\caption{Displayed are the standard model representation of matter additions for $U(1)'$ quivers, together with their total multiplicity
across all three-node and four-node quivers in the hypercharge embeddings we have studied. In the second column we have already filtered
out quivers with fractional electric charge. The third excludes quivers where the matter additions correspond to a fourth generation. 
The fourth column additionally excludes quivers where the matter additions correspond to a shifted fourth generation, 
from which it can be seen that the majority of additions correspond to quasichiral pairs.}
\label{table:Z' particle addition table}
\end{table}

\section{Conclusions}
In this paper we studied constraints on chiral matter which arise in certain types of string compactifications, with
a focus on the implications for matter beyond the MSSM, which might be seen at the LHC.
These constraints include the standard constraints for cancellation of non-abelian anomalies, which are also present in ordinary field theories or in field theories augmented by string-motivated
concepts such as ``anomalous'' $U(1)$ factors and Chern-Simons terms. However, some
are genuinely stringy conditions associated with $U(2)$ and $U(1)$ tadpole cancellation (which ensures the cancellation of D-brane charge in the compact extra dimensions via Gauss' law), while others correspond to conditions necessary for a massless hypercharge boson.

We studied the impact of these stringy conditions as found in type IIa intersecting brane and
related constructions. The local constructions\footnote{The local constructions we study satisfy
the stringy conditions, which are necessary but not sufficient for global consistency.} we study  are described by quivers, which are graphs in which the nodes represent
the gauge factors (e.g., $U(N)$ groups  from  stacks of D6 branes), and the directed edges represent chiral matter,
which live at the brane intersections. The allowed representations are bifundamentals,  symmetric or antisymmetric products,
and adjoints (which are not restricted to the intersections). The $U(1)$ factors from the trace generators are generically ``anomalous'',
with their bosons acquiring string scale masses by the Stuckelberg mechanism and  the anomalies cancelled by
Chern-Simons terms. The ``anomalous'' $U(1)$'s survive as global symmetries on the low energy theory, restricting the 
allowed superpotential couplings at the perturbative level. (These otherwise forbidden couplings can sometimes be restored by
exponentially-suppressed non-perturbative D-instanton effects.) However, one or more linear combinations 
of the anomalous $U(1)$'s (such as weak hypercharge) may be non-anomalous with a massless boson
(before turning on the Higgs mechanism). We have studied the 
three-node ($U(3)\times U(2)\times U(1)$) and four-node ($U(3)\times U(2)\times U(1)\times U(1) $) models with the conditions: 
(a) There is a realistic embedding of weak hypercharge, with the corresponding $U(1)_Y$ satisfying the necessary
masslessness conditions. (b) The matter spectrum includes that of the MSSM (without right-handed neutrinos), with up to five additional matter fields. (c) There are no vector pairs of fields, because pairs that are vector under both the low energy gauge symmetries and the 
``anomalous'' $U(1)$'s are typically expected to acquire string-scale masses. Moreover, vector pairs automatically satisfy the stringy conditions. (d) The quiver must satisfy the stringy tadpole conditions.
The results are:
\begin{itemize}
\item There are no consistent three-node quivers and very few four-node ones that do not have at least one matter addition beyond the MSSM
fields. This motivates, within our framework and assumptions, the possibility that there may be additional matter accessible 
at the LHC.
\item We present three illustrative examples which demonstrate the power of 
	the constraints. One is a quiver with the spectrum of the MSSM which violates
	a stringy constraint, despite being free of
	any (currently known) field theoretic pathology. In two other examples,
	we present both an MSSM singlet and a standard model vector pair which
	(when added to a consistent quiver) force the hypercharge boson to obtain
	a Stuckelberg mass, despite the fact that these additions are rather mundane
	in field theory.
\item The violation of string consistency conditions by quivers with the exact MSSM spectrum
often takes a suggestive form. This is particularly true in the case of three-node quivers, 
where an inconsistent quiver with the MSSM spectrum violates only one of the six
string consistency conditions\footnote{There are six conditions in the case of three-node quivers
and eight in the case of four-node quivers.}. Given the specific structure of the violation, some
matter additions are much more likely than others to render the quiver consistent. In this sense,
these matter fields are ``preferred" by the string conditions.
\item The most common matter additions are MSSM singlets, isotriplets with $Y=0$, or
quasichiral pairs, i.e., nonchiral under the MSSM, but chiral under additional anomalous or non-anomalous $U(1)$ factors.
The latter include  lepton/Higgs doublets,  down-type isosinglet quarks, isodoublet or up-type isosinglet quarks, nonabelian singlets with charge $\pm 1$, and lepton/Higgs doublets with charges $(\pm 1,\pm 2)$. There are also smaller numbers of other additions that are excluded or
strongly constrained by experimental or theoretical considerations, including  chiral ordinary or charge-shifted fourth families,
or fields which lead to fractionally-charged color singlet states.
\item Many of the four-node quivers allow a distinction between the down-type Higgs doublet and the lepton doublets, due to their different anomalous
$U(1)$ charges (i.e., they are quiver distinct). This is a necessary condition for lepton number and R-parity conservation.
\item A small fraction of the four-node quivers (and one three-node one) allow an additional $U(1)'$ gauge symmetry in the low-energy theory.
In the majority of the cases at least one type of fermion field has family non-universal couplings, due to their different embeddings
at the quiver level. This implies flavor changing neutral currents when fermion mixing is turned on, as is possibly relevant
to neutral $B$ system.
\item Many of the quivers have MSSM-singlets $S_\mu$ with perturbative
couplings $S_\mu H_u H_d$ that can lead to an NMSSM-type dynamical $\mu$ term. ($\mu$ terms may instead be generated by D-instantons
in other cases.)
\item Many of the quivers have at least one MSSM-singlet $\Nr$  that can have a perturbative Dirac neutrino mass term $L H_d \Nr$,
which is needed for a conventional neutrino seesaw. (Other possibilities for neutrino mass include a non-perturbative Weinberg operator or a  small Dirac mass.)
\end{itemize}
This work may be viewed as a small step in the exploration of the subset of the string landscape that
is consistent with the standard model or MSSM but may include additional TeV scale physics observable at the LHC. We have used string consistency conditions as a guidepost for new
physics and note that our analyses allow
for the likely scenario that the  string scale $M_s$ is comparable to the Planck scale.

\acknowledgments
                                                                                                                                                                                                                                                                                                                                                                                                                                                                                                                                                                                                                                                                                                                                                                                                                                                                                                                                                                           
J.H. is grateful to I\~ naki Garc\' ia-Etxebarria, Joe Marsano, Godfrey Miller, Robert Richter and Wati Taylor for interesting
conversations about the content of this work. 
The work of J.H. is supported by an NSF Graduate Fellowship through the String Vacuum Project and a DOE Graduate Fellowship in High Energy Theory.
The work of P.L. is supported by the IBM Einstein Fellowship and by the NSF grant PHY-0503584.
The work of M.C. is supported by
the NSF grant PHY05-51164, DOE under grant
DE-FG05-95ER40893-A020, NSF RTG grant DMS-0636606, the Fay R. and Eugene L.
Langberg Chair, and the Slovenian Research Agency (ARRS).
This material is based upon work supported in part by the National Science Foundation under Grant No. 1066293 and the hospitality of the Aspen Center for Physics. 

\appendix

\section{Stringy Constraints on Chiral Matter\label{sec:stringy constraints}}
Here we briefly describe the origin of the constraints on chiral matter that are necessary for tadpole cancellation and a massless
hypercharge boson. They include the field-theoretic constraints for the absence of non-abelian anomalies,
though there are some (stringy) constraints that are not present in field theory.

Though the constraints necessary for tadpole cancellation and a massless hypercharge boson have often been discussed in the context of type 
IIa orientifold compactifications
with intersecting D6-branes, they are also applicable in other regions of
the landscape. It is straightforward to show that the same constraints hold in the
T-dual type IIb formulation, and it has been shown \cite{Bianchi:2000de,Dijkstra:2004ym,Anastasopoulos:2010hu}  that they
 also apply in the context of the non-geometric rational conformal field theory phase of type IIa. One might expect, via duality,
that the constraints apply in even broader patches of the landscape. 

We will see that some of the stringy constraints
are related to the ability to distinguish between $\textbf{2}$ and $\ov{\textbf{2}}$,
due to charge under the (trace) $U(1)$ of $U(2)$. One might therefore expect similar constraints
to apply in the heterotic string with holomorphic vector bundles with abelian factors in their structure group \cite{Blumenhagen:2005ga}. Such compactifications are S-dual to Type I
compactifications with D9-branes and D6-branes, which in turn are mirror symmetric to the
type IIa intersecting D6-brane models that we typically have in mind \cite{Blumenhagen:2005zg}.
It would also be interesting to see if these constraints appear directly in M-theory.
Let us briefly sketch their string theoretic origin
 in the type IIa corner of the landscape. 

In type IIa string theory, gauge theories live on D6-branes wrapping three-cycles\footnote{D6-branes and O6-planes in the compactifications we consider fill spacetime and wrap three-cycles in the
extra dimensions, which make up a Calabi-Yau manifold. We denote cycles wrapped by D6-branes with lower case latin subscripts, and denote the cycle wrapped by the O6-planes as $\pi_{O6}$. The topological intersection number of two three-cycles $\pi_1$ and $\pi_2$ is denoted by $\pi_1\circ \pi_2$, which satisfies $\pi_1\circ \pi_2 = -\pi_2\circ \pi_1$.} $\pi$
in the internal dimensions (a Calabi-Yau manifold). In the presence of O6-planes, every
D-brane has an ``image" brane under the orientifold involution. Chiral matter appears
at the intersection of two D6-branes, where the representation and multiplicity
are given in Table
\ref{table:intersections}. D-branes are sources for Ramond-Ramond flux, which must
be cancelled in the compact internal dimensions since (heuristically) the flux
lines have nowhere to go. By using Gauss' Law in the internal dimensions, one can
ensure the cancellation of Ramond-Ramond charge.
\begin{table}
\centering \vspace{3mm}
\label{table:spectrum}
\begin{tabular}{|c|c|}
\hline
Representation  & Multiplicity \\
\hline $\Yasymm_a$
 & ${1\over 2}\left(\pi_a\circ \pi'_a+ \pi_a\circ \pi_{{\rm O}6}
\right)$  \\
 $\Ysymm_a$
      & ${1\over 2}\left(\pi_a\circ \pi'_a-\pi_a\circ \pi_{{\rm O}6}
\right)$   \\
      $(\fund_a,\antifund_b)$
       & $\pi_a\circ \pi_{b}$   \\
        $(\fund_a,\fund_b)$
	 & $\pi_a\circ \pi'_{b}$
	 \\
	 \hline
	 \end{tabular}
	 \caption{Representations and multiplicities for chiral matter at the intersection of two D6-branes.} 
	 \label{table:intersections}
\end{table}
This gives the tadpole cancellation conditions
\begin{equation}
\label{eqn:tadpole}
\sum_b N_b \, (\pi_b + \pi_{b'}) = 4 \, \pi_{O6},
\end{equation}
which constrain the three-cycles wrapped by the D-branes. The integer $N_b$ is the number of D6-branes on $\pi_b$, and the image brane associated
to a brane on $\pi_b$ wraps the cycle $\pi_{b'}$.  These conditions must be satisfied
in a consistent type IIa string compactification. In the absence of orientifolds,
one can intersect a D-brane on a cycle $\pi_a$ with \eqref{eqn:tadpole} to obtain
\begin{equation}
	\label{eqn:tadpole no orientifolds}
	0 = \pi_a \circ \sum_b N_b \,\,\, \pi_b  = \sum_b N_b (\#(a,\ov b) - \#(\ov a,b)),
\end{equation}
which requires that $\# a = \#\ov a$. Typically the gauge symmetry on $\pi_a$ is 
$U(N_a)$, so that the condition \eqref{eqn:tadpole no orientifolds} for $N_a>2$ are
the conditions necessary for the absence of $SU(N_a)^3$ triangle anomalies. On the
other hand, for $N_a = 2$ and $N_a=1$, there are no such anomalies, yet these are
still stringy constraints necessary for tadpole cancellation.
In the presence of orientifold planes, the full condition is
\begin{align}
	\label{eqn:chiral tadpole constraint app}
	N_a \ge 2&: \qquad \# a - \# \ov a + (N_a+4)\,\, (\# \, \Ysymm_a - \#\, \ov \Ysymm_a) + (N_a-4) \,\, (\# \, \Yasymm_a - \# \, \ov \Yasymm_a) = 0 \notag \\ \notag \\
	N_a = 1&: \qquad \# a - \# \ov a + (N_a+4)\,\, (\# \, \Ysymm_a - \# \, \ov \Ysymm_a) = 0 \,\,\, \text{mod} \,\,\, 3,
\end{align}
where the mod 3 condition for the $N_a=1$ case comes from the fact that there is no antisymmetric representation of $U(1)$.
This particular constraint is not affected by the structure of the quark sector, as every bifundamental 
quark or antiquark 
contributes $\pm 3$ to the $N_a=1$ constraint. It is these $N_a=2$ and $N_a=1$ conditions that MSSM quivers discussed in section
\ref{sec:3-node madrid} and \ref{sec:3-node non-madrid} violate.
For more details of this derivation, we refer the reader to \cite{Cvetic:2009yh}.

The quivers studied in this paper generically have anomalous $U(1)$ factors, which
could have abelian and/or mixed anomalies. These anomalies are cancelled
\cite{Aldazabal:2000dg} by the generalized Green-Schwarz mechanism due to the presence
of Chern-Simons terms. One such Chern-Simons term is of the form $\int B \wedge Tr(F)$,
where $B$ is a $2$-form and $F$ is a gauge field strength. In addition to playing a role in anomaly cancellation, this term gives a Stuckelberg
mass to the  $U(1)$ gauge boson. However, sometimes these terms are absent for a linear
combination $U(1)_G = \sum_x q_x\, U(1)_x$. This occurs if
\begin{equation}
	\sum_x q_x \, N_x \, (\pi_x - \pi_{x^{'}}) = 0,
\end{equation}
in which case the $U(1)$ gauge boson is massless (except perhaps for a mass obtained via
the ordinary Higgs mechanism).
As with the condition on D6-brane three-cycles necessary for tadpole cancellation, this condition places constraints on
chiral matter:
\begin{equation}
\label{eqn:chiral masslessness constraint app}
-q_aN_a\,\,(\#\Ysymm_a - \#\ov\Ysymm_a + \#\Yasymm_a - \#\ov\Yasymm_a) + \sum_{x\ne a} q_x N_x \,\, (\#(a,\ov x) - \#(a,x)) = 0
\end{equation}
for $N_a\ge 2$, and
\begin{equation}
\label{eqn:chiral masslessness constraint N1 app}
-q_a \,\,\frac{\#(a) - \#(\ov a) + 8(\#\Ysymm_a)-\#\ov\Ysymm_a)}{3} + \sum_{x\ne a} q_x N_x \,\, (\#(a,\ov x) - \#(a,x)) = 0
\end{equation}
for $N_a=1$. These constraints are necessary for $U(1)_G$ to be left massless, but are not
sufficient. In MSSM quivers, one such linear combination must
be identifiable as hypercharge, and any additional linear combination that satisfies the constraints is a $U(1)'$
symmetry.

Let us describe an interesting case regarding Chern-Simons couplings and
a non-anomalous $U(1)$. In addition to terms of the form $\int B\wedge F$, cancellation of
abelian and mixed gauge anomalies requires the presence of terms $\int \phi\, F\wedge F$, and cancellation of mixed gravitational anomalies requires the  terms 
$\int \phi \, R \wedge R$. If a theory is such
that there are no abelian or mixed anomalies for some anomalous $U(1)$, then there
cannot be both $\phi$-type and $B$-type terms, as they would give rise to an anomaly
rather than cancelling one. However, in the scenario where a $B$-type term is
present but $\phi$-type terms are not,
the $U(1)$ boson can receive a Stuckelberg mass, despite being anomaly free. This unusual feature,
where a non-anomalous $U(1)$ nevertheless becomes massive, has been realized in explicit string compactifications \cite{Ibanez:2001nd}.

Let us end with a brief comment about the possibility of global $SU(2)$
anomalies \cite{Witten:1982fp}. Global consistency conditions of string theory
ensure the absence of such anomalies, which is closely related to the fact that
D-branes are classified by K-theory \cite{Uranga:2000xp}. Since we do not have
a global string embedding, it is interesting to address whether or not one
could add matter to an MSSM quiver in a way consistent with the stringy constraints
while introducing a global $SU(2)$ anomaly in the process. In the three-node Madrid embedding,
	it is easy to see from \eqref{eqn:3-node madrid T} that one can only add an
	even number of doublets, and thus there is no global $SU(2)$ anomaly due to matter
	additions. Generalizing this result, it is easy to argue that $T_b$ is even for
	any MSSM quiver of the type described in this paper, and therefore one will
	never introduce a global $SU(2)$ anomaly by adding matter to an MSSM quiver in a way
	consistent with stringy constraints.

\section{Matter Addition Tables}\label{mattertables}
In this appendix we list all fields, together with their standard model representation, that can be
realized in a 3-node or 4-node quiver with the hypercharge embeddings studied in this paper. The possibilities for the three-node Madrid and non-Madrid
embedding are presented in Table \ref{table:3-node arb} and the four-node possibilities are listed
in Tables \ref{4node1}-\ref{4node6}. Each table also lists the T-charge and M-charge of
each field.

\begin{table}
	\hspace{-.6cm}
	\begin{minipage}[t]{.5\linewidth}
	\scalebox{.8}{ \begin{tabular}{|c|c|c|c|c|c|c|}\hline
		Transformation & $T_a$ & $T_b$ & $T_c$ & $M_a$ & $M_b$ & $M_c$ \\ \hline \hline
		$\Ysymm_a$\qquad $(\textbf{6},\textbf{1})_{\frac{1}{3}}$ &$7$ &$0$ &$0$ &$-\frac{1}{2}$ &$0$ &$0$\\ \hline
		$\ov \Ysymm_a$\qquad $(\ov{\textbf{6}},\textbf{1})_{-\frac{1}{3}}$ &$-7$ &$0$ &$0$ &$\frac{1}{2}$ &$0$ &$0$\\ \hline
		$\Yasymm_a$\qquad $(\ov{\textbf{3}},\textbf{1})_{\frac{1}{3}}$ &$-1$ &$0$ &$0$ &$-\frac{1}{2}$ &$0$ &$0$\\ \hline
		$\ov \Yasymm_a$\qquad $(\textbf{3},\textbf{1})_{-\frac{1}{3}}$ &$1$ &$0$ &$0$ &$\frac{1}{2}$ &$0$ &$0$\\ \hline
		$\Ysymm_b$\qquad $(\textbf{1},\textbf{3})_{0}$ &$0$ &$6$ &$0$ &$0$ &$0$ &$0$\\ \hline
		$\ov \Ysymm_b$\qquad $(\textbf{1},\textbf{3})_{0}$ &$0$ &$-6$ &$0$ &$0$ &$0$ &$0$\\ \hline
		$\Yasymm_b$\qquad $(\textbf{1},\textbf{1})_{0}$ &$0$ &$-2$ &$0$ &$0$ &$0$ &$0$\\ \hline
		$\ov \Yasymm_b$\qquad $(\textbf{1},\textbf{1})_{0}$ &$0$ &$2$ &$0$ &$0$ &$0$ &$0$\\ \hline
		$(\ov b,c)$\qquad $(\textbf{1},\textbf{2})_{\frac{1}{2}}$ &$0$ &$-1$ &$2$ &$0$ &$-\frac{1}{2}$ &$-\frac{1}{3}$\\ \hline
		$(b,\ov c)$\qquad $(\textbf{1},\textbf{2})_{-\frac{1}{2}}$ &$0$ &$1$ &$-2$ &$0$ &$\frac{1}{2}$ &$\frac{1}{3}$\\ \hline
		$(b,c)$\qquad $(\textbf{1},\textbf{2})_{\frac{1}{2}}$ &$0$ &$1$ &$2$ &$0$ &$-\frac{1}{2}$ &$-\frac{1}{3}$\\ \hline
		$(\ov b,\ov c)$\qquad $(\textbf{1},\textbf{2})_{-\frac{1}{2}}$ &$0$ &$-1$ &$-2$ &$0$ &$\frac{1}{2}$ &$\frac{1}{3}$\\ \hline
		$\Ysymm_c$\qquad $(\textbf{1},\textbf{1})_{1}$ &$0$ &$0$ &$5$ &$0$ &$0$ &$-\frac{4}{3}$\\ \hline
		$\ov \Ysymm_c$\qquad $(\textbf{1},\textbf{1})_{-1}$ &$0$ &$0$ &$-5$ &$0$ &$0$ &$\frac{4}{3}$\\ \hline
		$(a,\ov b)$\qquad $(\textbf{3},\textbf{2})_{\frac{1}{6}}$ &$2$ &$-3$ &$0$ &$0$ &$-\frac{1}{2}$ &$0$\\ \hline
		$(\ov a,b)$\qquad $(\ov{\textbf{3}},\textbf{2})_{-\frac{1}{6}}$ &$-2$ &$3$ &$0$ &$0$ &$\frac{1}{2}$ &$0$\\ \hline
		$(a,b)$\qquad $(\textbf{3},\textbf{2})_{\frac{1}{6}}$ &$2$ &$3$ &$0$ &$0$ &$-\frac{1}{2}$ &$0$\\ \hline
		$(\ov a,\ov b)$\qquad $(\ov{\textbf{3}},\textbf{2})_{-\frac{1}{6}}$ &$-2$ &$-3$ &$0$ &$0$ &$\frac{1}{2}$ &$0$\\ \hline
		$(a,\ov c)$\qquad $(\textbf{3},\textbf{1})_{-\frac{1}{3}}$ &$1$ &$0$ &$-3$ &$\frac{1}{2}$ &$0$ &$0$\\ \hline
		$(\ov a,c)$\qquad $(\ov{\textbf{3}},\textbf{1})_{\frac{1}{3}}$ &$-1$ &$0$ &$3$ &$-\frac{1}{2}$ &$0$ &$0$\\ \hline
		$(a,c)$\qquad $(\textbf{3},\textbf{1})_{\frac{2}{3}}$ &$1$ &$0$ &$3$ &$-\frac{1}{2}$ &$0$ &$-1$\\ \hline
		$(\ov a,\ov c)$\qquad $(\ov{\textbf{3}},\textbf{1})_{-\frac{2}{3}}$ &$-1$ &$0$ &$-3$ &$\frac{1}{2}$ &$0$ &$1$\\ \hline
	\end{tabular}}
	 \\ \vspace{.5cm} \centering \small Madrid embedding additions.
	\end{minipage}
	\hspace{.5cm}
	\begin{minipage}[t]{.5\linewidth}
	\scalebox{.8}{
	\begin{tabular}{|c|c|c|c|c|c|c|}\hline
		Transformation & $T_a$ & $T_b$ & $T_c$ & $M_a$ & $M_b$ & $M_c$ \\ \hline \hline
		$\Ysymm_a$\qquad $(\textbf{6},\textbf{1})_{-\frac{2}{3}}$ &$7$ &$0$ &$0$ &$1$ &$0$ &$0$\\ \hline
		$\ov \Ysymm_a$\qquad $(\ov{\textbf{6}},\textbf{1})_{\frac{2}{3}}$ &$-7$ &$0$ &$0$ &$-1$ &$0$ &$0$\\ \hline
		$\Yasymm_a$\qquad $(\ov{\textbf{3}},\textbf{1})_{-\frac{2}{3}}$ &$-1$ &$0$ &$0$ &$1$ &$0$ &$0$\\ \hline
		$\ov \Yasymm_a$\qquad $(\textbf{3},\textbf{1})_{\frac{2}{3}}$ &$1$ &$0$ &$0$ &$-1$ &$0$ &$0$\\ \hline
		$\Ysymm_b$\qquad $(\textbf{1},\textbf{3})_{-1}$ &$0$ &$6$ &$0$ &$0$ &$1$ &$0$\\ \hline
		$\ov \Ysymm_b$\qquad $(\textbf{1},\textbf{3})_{1}$ &$0$ &$-6$ &$0$ &$0$ &$-1$ &$0$\\ \hline
		$\Yasymm_b$\qquad $(\textbf{1},\textbf{1})_{-1}$ &$0$ &$-2$ &$0$ &$0$ &$1$ &$0$\\ \hline
		$\ov \Yasymm_b$\qquad $(\textbf{1},\textbf{1})_{1}$ &$0$ &$2$ &$0$ &$0$ &$-1$ &$0$\\ \hline
		$(\ov b,c)$\qquad $(\textbf{1},\textbf{2})_{\frac{1}{2}}$ &$0$ &$-1$ &$2$ &$0$ &$0$ &$-1$\\ \hline
		$(b,\ov c)$\qquad $(\textbf{1},\textbf{2})_{-\frac{1}{2}}$ &$0$ &$1$ &$-2$ &$0$ &$0$ &$1$\\ \hline
		$(b,c)$\qquad $(\textbf{1},\textbf{2})_{-\frac{1}{2}}$ &$0$ &$1$ &$2$ &$0$ &$0$ &$1$\\ \hline
		$(\ov b,\ov c)$\qquad $(\textbf{1},\textbf{2})_{\frac{1}{2}}$ &$0$ &$-1$ &$-2$ &$0$ &$0$ &$-1$\\ \hline
		$\Ysymm_c$\qquad $(\textbf{1},\textbf{1})_{0}$ &$0$ &$0$ &$5$ &$0$ &$0$ &$0$\\ \hline
		$\ov \Ysymm_c$\qquad $(\textbf{1},\textbf{1})_{0}$ &$0$ &$0$ &$-5$ &$0$ &$0$ &$0$\\ \hline
		$(a,\ov b)$\qquad $(\textbf{3},\textbf{2})_{\frac{1}{6}}$ &$2$ &$-3$ &$0$ &$-1$ &$1$ &$0$\\ \hline
		$(\ov a,b)$\qquad $(\ov{\textbf{3}},\textbf{2})_{-\frac{1}{6}}$ &$-2$ &$3$ &$0$ &$1$ &$-1$ &$0$\\ \hline
		$(a,b)$\qquad $(\textbf{3},\textbf{2})_{-\frac{5}{6}}$ &$2$ &$3$ &$0$ &$1$ &$1$ &$0$\\ \hline
		$(\ov a,\ov b)$\qquad $(\ov{\textbf{3}},\textbf{2})_{\frac{5}{6}}$ &$-2$ &$-3$ &$0$ &$-1$ &$-1$ &$0$\\ \hline
		$(a,\ov c)$\qquad $(\textbf{3},\textbf{1})_{-\frac{1}{3}}$ &$1$ &$0$ &$-3$ &$0$ &$0$ &$1$\\ \hline
		$(\ov a,c)$\qquad $(\ov{\textbf{3}},\textbf{1})_{\frac{1}{3}}$ &$-1$ &$0$ &$3$ &$0$ &$0$ &$-1$\\ \hline
		$(a,c)$\qquad $(\textbf{3},\textbf{1})_{-\frac{1}{3}}$ &$1$ &$0$ &$3$ &$0$ &$0$ &$1$\\ \hline
		$(\ov a,\ov c)$\qquad $(\ov{\textbf{3}},\textbf{1})_{\frac{1}{3}}$ &$-1$ &$0$ &$-3$ &$0$ &$0$ &$-1$\\ \hline
	\end{tabular}
	}
	 \\ \vspace{.5cm} \centering \small Non-Madrid embedding additions.
	\end{minipage}
\vspace{-.5cm}
	\caption{All possible fields which might arise in  quivers with  $U(3)$, $U(2)$, and  $U(1)$ nodes,
	for each three-node hypercharge embedding.
	Listed for each field are its transformation behavior in the quiver, under the MSSM gauge group, its T-charge and its M-charge. None of these three-node candidate fields can lead to color singlet particles with fractional electric charge.}	\label{table:3-node arb} 
\end{table}

\begin{table}
\centering
\begin{tabular}{|c|c|c|c|c|c|c|c|c|}\hline
Transformation & $T_a$ & $T_b$ & $T_c$ & $T_d$ & $M_a$ & $M_b$ & $M_c$ & $M_d$ \\ \hline \hline
$\Ysymm_a$\qquad $(\textbf{6},\textbf{1})_{-\frac{2}{3}}$ &$7$ &$0$ &$0$ &$0$ &$1$ &$0$ &$0$ &$0$\\ \hline
$\ov \Ysymm_a$\qquad $(\ov{\textbf{6}},\textbf{1})_{\frac{2}{3}}$ &$-7$ &$0$ &$0$ &$0$ &$-1$ &$0$ &$0$ &$0$\\ \hline
$\Yasymm_a$\qquad $(\ov{\textbf{3}},\textbf{1})_{-\frac{2}{3}}$ &$-1$ &$0$ &$0$ &$0$ &$1$ &$0$ &$0$ &$0$\\ \hline
$\ov \Yasymm_a$\qquad $(\textbf{3},\textbf{1})_{\frac{2}{3}}$ &$1$ &$0$ &$0$ &$0$ &$-1$ &$0$ &$0$ &$0$\\ \hline
$\Ysymm_b$\qquad $(\textbf{1},\textbf{3})_{-1}$ &$0$ &$6$ &$0$ &$0$ &$0$ &$1$ &$0$ &$0$\\ \hline
$\ov \Ysymm_b$\qquad $(\textbf{1},\textbf{3})_{1}$ &$0$ &$-6$ &$0$ &$0$ &$0$ &$-1$ &$0$ &$0$\\ \hline
$\Yasymm_b$\qquad $(\textbf{1},\textbf{1})_{-1}$ &$0$ &$-2$ &$0$ &$0$ &$0$ &$1$ &$0$ &$0$\\ \hline
$\ov \Yasymm_b$\qquad $(\textbf{1},\textbf{1})_{1}$ &$0$ &$2$ &$0$ &$0$ &$0$ &$-1$ &$0$ &$0$\\ \hline
$\Ysymm_d$\qquad $(\textbf{1},\textbf{1})_{0}$ &$0$ &$0$ &$0$ &$5$ &$0$ &$0$ &$0$ &$0$\\ \hline
$\ov \Ysymm_d$\qquad $(\textbf{1},\textbf{1})_{0}$ &$0$ &$0$ &$0$ &$-5$ &$0$ &$0$ &$0$ &$0$\\ \hline
$(c,\ov d)$\qquad $(\textbf{1},\textbf{1})_{0}$ &$0$ &$0$ &$1$ &$-1$ &$0$ &$0$ &$0$ &$0$\\ \hline
$(\ov c,d)$\qquad $(\textbf{1},\textbf{1})_{0}$ &$0$ &$0$ &$-1$ &$1$ &$0$ &$0$ &$0$ &$0$\\ \hline
$(c,d)$\qquad $(\textbf{1},\textbf{1})_{0}$ &$0$ &$0$ &$1$ &$1$ &$0$ &$0$ &$0$ &$0$\\ \hline
$(\ov c,\ov d)$\qquad $(\textbf{1},\textbf{1})_{0}$ &$0$ &$0$ &$-1$ &$-1$ &$0$ &$0$ &$0$ &$0$\\ \hline
$(\ov b,c)$\qquad $(\textbf{1},\textbf{2})_{\frac{1}{2}}$ &$0$ &$-1$ &$2$ &$0$ &$0$ &$0$ &$-1$ &$0$\\ \hline
$(b,\ov c)$\qquad $(\textbf{1},\textbf{2})_{-\frac{1}{2}}$ &$0$ &$1$ &$-2$ &$0$ &$0$ &$0$ &$1$ &$0$\\ \hline
$(b,c)$\qquad $(\textbf{1},\textbf{2})_{-\frac{1}{2}}$ &$0$ &$1$ &$2$ &$0$ &$0$ &$0$ &$1$ &$0$\\ \hline
$(\ov b,\ov c)$\qquad $(\textbf{1},\textbf{2})_{\frac{1}{2}}$ &$0$ &$-1$ &$-2$ &$0$ &$0$ &$0$ &$-1$ &$0$\\ \hline
$\Ysymm_c$\qquad $(\textbf{1},\textbf{1})_{0}$ &$0$ &$0$ &$5$ &$0$ &$0$ &$0$ &$0$ &$0$\\ \hline
$\ov \Ysymm_c$\qquad $(\textbf{1},\textbf{1})_{0}$ &$0$ &$0$ &$-5$ &$0$ &$0$ &$0$ &$0$ &$0$\\ \hline
$(a,\ov d)$\qquad $(\textbf{3},\textbf{1})_{-\frac{1}{3}}$ &$1$ &$0$ &$0$ &$-3$ &$0$ &$0$ &$0$ &$1$\\ \hline
$(\ov a,d)$\qquad $(\ov{\textbf{3}},\textbf{1})_{\frac{1}{3}}$ &$-1$ &$0$ &$0$ &$3$ &$0$ &$0$ &$0$ &$-1$\\ \hline
$(a,d)$\qquad $(\textbf{3},\textbf{1})_{-\frac{1}{3}}$ &$1$ &$0$ &$0$ &$3$ &$0$ &$0$ &$0$ &$1$\\ \hline
$(\ov a,\ov d)$\qquad $(\ov{\textbf{3}},\textbf{1})_{\frac{1}{3}}$ &$-1$ &$0$ &$0$ &$-3$ &$0$ &$0$ &$0$ &$-1$\\ \hline
$(a,\ov b)$\qquad $(\textbf{3},\textbf{2})_{\frac{1}{6}}$ &$2$ &$-3$ &$0$ &$0$ &$-1$ &$1$ &$0$ &$0$\\ \hline
$(\ov a,b)$\qquad $(\ov{\textbf{3}},\textbf{2})_{-\frac{1}{6}}$ &$-2$ &$3$ &$0$ &$0$ &$1$ &$-1$ &$0$ &$0$\\ \hline
$(a,b)$\qquad $(\textbf{3},\textbf{2})_{-\frac{5}{6}}$ &$2$ &$3$ &$0$ &$0$ &$1$ &$1$ &$0$ &$0$\\ \hline
$(\ov a,\ov b)$\qquad $(\ov{\textbf{3}},\textbf{2})_{\frac{5}{6}}$ &$-2$ &$-3$ &$0$ &$0$ &$-1$ &$-1$ &$0$ &$0$\\ \hline
$(a,\ov c)$\qquad $(\textbf{3},\textbf{1})_{-\frac{1}{3}}$ &$1$ &$0$ &$-3$ &$0$ &$0$ &$0$ &$1$ &$0$\\ \hline
$(\ov a,c)$\qquad $(\ov{\textbf{3}},\textbf{1})_{\frac{1}{3}}$ &$-1$ &$0$ &$3$ &$0$ &$0$ &$0$ &$-1$ &$0$\\ \hline
$(a,c)$\qquad $(\textbf{3},\textbf{1})_{-\frac{1}{3}}$ &$1$ &$0$ &$3$ &$0$ &$0$ &$0$ &$1$ &$0$\\ \hline
$(\ov a,\ov c)$\qquad $(\ov{\textbf{3}},\textbf{1})_{\frac{1}{3}}$ &$-1$ &$0$ &$-3$ &$0$ &$0$ &$0$ &$-1$ &$0$\\ \hline
$(b,\ov d)$\qquad $(\textbf{1},\textbf{2})_{-\frac{1}{2}}$ &$0$ &$1$ &$0$ &$-2$ &$0$ &$0$ &$0$ &$1$\\ \hline
$(\ov b,d)$\qquad $(\textbf{1},\textbf{2})_{\frac{1}{2}}$ &$0$ &$-1$ &$0$ &$2$ &$0$ &$0$ &$0$ &$-1$\\ \hline
$(b,d)$\qquad $(\textbf{1},\textbf{2})_{-\frac{1}{2}}$ &$0$ &$1$ &$0$ &$2$ &$0$ &$0$ &$0$ &$1$\\ \hline
$(\ov b,\ov d)$\qquad $(\textbf{1},\textbf{2})_{\frac{1}{2}}$ &$0$ &$-1$ &$0$ &$-2$ &$0$ &$0$ &$0$ &$-1$\\ \hline
\end{tabular}
\caption{Possible fields which might arise in the 4-node quiver with hypercharge embedding $U(1)_Y = -\frac{1}{3}U(1)_a - \frac{1}{2} U(1)_b$.}
\label{4node1}
\end{table}

\begin{table}
\centering
\begin{tabular}{|c|c|c|c|c|c|c|c|c|}\hline
Transformation & $T_a$ & $T_b$ & $T_c$ & $T_d$ & $M_a$ & $M_b$ & $M_c$ & $M_d$ \\ \hline \hline
$\Ysymm_a$\qquad $(\textbf{6},\textbf{1})_{-\frac{2}{3}}$ &$7$ &$0$ &$0$ &$0$ &$1$ &$0$ &$0$ &$0$\\ \hline
$\ov \Ysymm_a$\qquad $(\ov{\textbf{6}},\textbf{1})_{\frac{2}{3}}$ &$-7$ &$0$ &$0$ &$0$ &$-1$ &$0$ &$0$ &$0$\\ \hline
$\Yasymm_a$\qquad $(\ov{\textbf{3}},\textbf{1})_{-\frac{2}{3}}$ &$-1$ &$0$ &$0$ &$0$ &$1$ &$0$ &$0$ &$0$\\ \hline
$\ov \Yasymm_a$\qquad $(\textbf{3},\textbf{1})_{\frac{2}{3}}$ &$1$ &$0$ &$0$ &$0$ &$-1$ &$0$ &$0$ &$0$\\ \hline
$\Ysymm_b$\qquad $(\textbf{1},\textbf{3})_{-1}$ &$0$ &$6$ &$0$ &$0$ &$0$ &$1$ &$0$ &$0$\\ \hline
$\ov \Ysymm_b$\qquad $(\textbf{1},\textbf{3})_{1}$ &$0$ &$-6$ &$0$ &$0$ &$0$ &$-1$ &$0$ &$0$\\ \hline
$\Yasymm_b$\qquad $(\textbf{1},\textbf{1})_{-1}$ &$0$ &$-2$ &$0$ &$0$ &$0$ &$1$ &$0$ &$0$\\ \hline
$\ov \Yasymm_b$\qquad $(\textbf{1},\textbf{1})_{1}$ &$0$ &$2$ &$0$ &$0$ &$0$ &$-1$ &$0$ &$0$\\ \hline
$\Ysymm_d$\qquad $(\textbf{1},\textbf{1})_{1}$ &$0$ &$0$ &$0$ &$5$ &$0$ &$0$ &$0$ &$-\frac{4}{3}$\\ \hline
$\ov \Ysymm_d$\qquad $(\textbf{1},\textbf{1})_{-1}$ &$0$ &$0$ &$0$ &$-5$ &$0$ &$0$ &$0$ &$\frac{4}{3}$\\ \hline
$(c,\ov d)$\qquad $(\textbf{1},\textbf{1})_{-\frac{1}{2}}$ &$0$ &$0$ &$1$ &$-1$ &$0$ &$0$ &$\frac{1}{2}$ &$\frac{1}{6}$\\ \hline
$(\ov c,d)$\qquad $(\textbf{1},\textbf{1})_{\frac{1}{2}}$ &$0$ &$0$ &$-1$ &$1$ &$0$ &$0$ &$-\frac{1}{2}$ &$-\frac{1}{6}$\\ \hline
$(c,d)$\qquad $(\textbf{1},\textbf{1})_{\frac{1}{2}}$ &$0$ &$0$ &$1$ &$1$ &$0$ &$0$ &$-\frac{1}{2}$ &$-\frac{1}{6}$\\ \hline
$(\ov c,\ov d)$\qquad $(\textbf{1},\textbf{1})_{-\frac{1}{2}}$ &$0$ &$0$ &$-1$ &$-1$ &$0$ &$0$ &$\frac{1}{2}$ &$\frac{1}{6}$\\ \hline
$(\ov b,c)$\qquad $(\textbf{1},\textbf{2})_{\frac{1}{2}}$ &$0$ &$-1$ &$2$ &$0$ &$0$ &$0$ &$-1$ &$0$\\ \hline
$(b,\ov c)$\qquad $(\textbf{1},\textbf{2})_{-\frac{1}{2}}$ &$0$ &$1$ &$-2$ &$0$ &$0$ &$0$ &$1$ &$0$\\ \hline
$(b,c)$\qquad $(\textbf{1},\textbf{2})_{-\frac{1}{2}}$ &$0$ &$1$ &$2$ &$0$ &$0$ &$0$ &$1$ &$0$\\ \hline
$(\ov b,\ov c)$\qquad $(\textbf{1},\textbf{2})_{\frac{1}{2}}$ &$0$ &$-1$ &$-2$ &$0$ &$0$ &$0$ &$-1$ &$0$\\ \hline
$\Ysymm_c$\qquad $(\textbf{1},\textbf{1})_{0}$ &$0$ &$0$ &$5$ &$0$ &$0$ &$0$ &$0$ &$0$\\ \hline
$\ov \Ysymm_c$\qquad $(\textbf{1},\textbf{1})_{0}$ &$0$ &$0$ &$-5$ &$0$ &$0$ &$0$ &$0$ &$0$\\ \hline
$(a,\ov d)$\qquad $(\textbf{3},\textbf{1})_{-\frac{5}{6}}$ &$1$ &$0$ &$0$ &$-3$ &$\frac{1}{2}$ &$0$ &$0$ &$\frac{3}{2}$\\ \hline
$(\ov a,d)$\qquad $(\ov{\textbf{3}},\textbf{1})_{\frac{5}{6}}$ &$-1$ &$0$ &$0$ &$3$ &$-\frac{1}{2}$ &$0$ &$0$ &$-\frac{3}{2}$\\ \hline
$(a,d)$\qquad $(\textbf{3},\textbf{1})_{\frac{1}{6}}$ &$1$ &$0$ &$0$ &$3$ &$-\frac{1}{2}$ &$0$ &$0$ &$\frac{1}{2}$\\ \hline
$(\ov a,\ov d)$\qquad $(\ov{\textbf{3}},\textbf{1})_{-\frac{1}{6}}$ &$-1$ &$0$ &$0$ &$-3$ &$\frac{1}{2}$ &$0$ &$0$ &$-\frac{1}{2}$\\ \hline
$(a,\ov b)$\qquad $(\textbf{3},\textbf{2})_{\frac{1}{6}}$ &$2$ &$-3$ &$0$ &$0$ &$-1$ &$1$ &$0$ &$0$\\ \hline
$(\ov a,b)$\qquad $(\ov{\textbf{3}},\textbf{2})_{-\frac{1}{6}}$ &$-2$ &$3$ &$0$ &$0$ &$1$ &$-1$ &$0$ &$0$\\ \hline
$(a,b)$\qquad $(\textbf{3},\textbf{2})_{-\frac{5}{6}}$ &$2$ &$3$ &$0$ &$0$ &$1$ &$1$ &$0$ &$0$\\ \hline
$(\ov a,\ov b)$\qquad $(\ov{\textbf{3}},\textbf{2})_{\frac{5}{6}}$ &$-2$ &$-3$ &$0$ &$0$ &$-1$ &$-1$ &$0$ &$0$\\ \hline
$(a,\ov c)$\qquad $(\textbf{3},\textbf{1})_{-\frac{1}{3}}$ &$1$ &$0$ &$-3$ &$0$ &$0$ &$0$ &$1$ &$0$\\ \hline
$(\ov a,c)$\qquad $(\ov{\textbf{3}},\textbf{1})_{\frac{1}{3}}$ &$-1$ &$0$ &$3$ &$0$ &$0$ &$0$ &$-1$ &$0$\\ \hline
$(a,c)$\qquad $(\textbf{3},\textbf{1})_{-\frac{1}{3}}$ &$1$ &$0$ &$3$ &$0$ &$0$ &$0$ &$1$ &$0$\\ \hline
$(\ov a,\ov c)$\qquad $(\ov{\textbf{3}},\textbf{1})_{\frac{1}{3}}$ &$-1$ &$0$ &$-3$ &$0$ &$0$ &$0$ &$-1$ &$0$\\ \hline
$(b,\ov d)$\qquad $(\textbf{1},\textbf{2})_{-1}$ &$0$ &$1$ &$0$ &$-2$ &$0$ &$\frac{1}{2}$ &$0$ &$\frac{4}{3}$\\ \hline
$(\ov b,d)$\qquad $(\textbf{1},\textbf{2})_{1}$ &$0$ &$-1$ &$0$ &$2$ &$0$ &$-\frac{1}{2}$ &$0$ &$-\frac{4}{3}$\\ \hline
$(b,d)$\qquad $(\textbf{1},\textbf{2})_{0}$ &$0$ &$1$ &$0$ &$2$ &$0$ &$-\frac{1}{2}$ &$0$ &$\frac{2}{3}$\\ \hline
$(\ov b,\ov d)$\qquad $(\textbf{1},\textbf{2})_{0}$ &$0$ &$-1$ &$0$ &$-2$ &$0$ &$\frac{1}{2}$ &$0$ &$-\frac{2}{3}$\\ \hline
\end{tabular}
\caption{Possible fields which might arise in the 4-node quiver with hypercharge embedding $U(1)_Y = -\frac{1}{3}U(1)_a - \frac{1}{2} U(1)_b + \frac{1}{2} U(1)_d$.
The $(\textbf{1},\textbf{1})_{\frac{1}{2}}$, $(\textbf{1},\textbf{2})_{1,0}$, and $(\textbf{3},\textbf{1})_{\frac{1}{6},-\frac{5}{6}}$
states and their conjugates would  lead to fractional charge.}
\end{table}

\begin{table}
\centering
\begin{tabular}{|c|c|c|c|c|c|c|c|c|}\hline
Transformation & $T_a$ & $T_b$ & $T_c$ & $T_d$ & $M_a$ & $M_b$ & $M_c$ & $M_d$ \\ \hline \hline
$\Ysymm_a$\qquad $(\textbf{6},\textbf{1})_{-\frac{2}{3}}$ &$7$ &$0$ &$0$ &$0$ &$1$ &$0$ &$0$ &$0$\\ \hline
$\ov \Ysymm_a$\qquad $(\ov{\textbf{6}},\textbf{1})_{\frac{2}{3}}$ &$-7$ &$0$ &$0$ &$0$ &$-1$ &$0$ &$0$ &$0$\\ \hline
$\Yasymm_a$\qquad $(\ov{\textbf{3}},\textbf{1})_{-\frac{2}{3}}$ &$-1$ &$0$ &$0$ &$0$ &$1$ &$0$ &$0$ &$0$\\ \hline
$\ov \Yasymm_a$\qquad $(\textbf{3},\textbf{1})_{\frac{2}{3}}$ &$1$ &$0$ &$0$ &$0$ &$-1$ &$0$ &$0$ &$0$\\ \hline
$\Ysymm_b$\qquad $(\textbf{1},\textbf{3})_{-1}$ &$0$ &$6$ &$0$ &$0$ &$0$ &$1$ &$0$ &$0$\\ \hline
$\ov \Ysymm_b$\qquad $(\textbf{1},\textbf{3})_{1}$ &$0$ &$-6$ &$0$ &$0$ &$0$ &$-1$ &$0$ &$0$\\ \hline
$\Yasymm_b$\qquad $(\textbf{1},\textbf{1})_{-1}$ &$0$ &$-2$ &$0$ &$0$ &$0$ &$1$ &$0$ &$0$\\ \hline
$\ov \Yasymm_b$\qquad $(\textbf{1},\textbf{1})_{1}$ &$0$ &$2$ &$0$ &$0$ &$0$ &$-1$ &$0$ &$0$\\ \hline
$\Ysymm_d$\qquad $(\textbf{1},\textbf{1})_{1}$ &$0$ &$0$ &$0$ &$5$ &$0$ &$0$ &$0$ &$-\frac{4}{3}$\\ \hline
$\ov \Ysymm_d$\qquad $(\textbf{1},\textbf{1})_{-1}$ &$0$ &$0$ &$0$ &$-5$ &$0$ &$0$ &$0$ &$\frac{4}{3}$\\ \hline
$(c,\ov d)$\qquad $(\textbf{1},\textbf{1})_{-\frac{1}{2}}$ &$0$ &$0$ &$1$ &$-1$ &$0$ &$0$ &$\frac{1}{2}$ &$\frac{1}{6}$\\ \hline
$(\ov c,d)$\qquad $(\textbf{1},\textbf{1})_{\frac{1}{2}}$ &$0$ &$0$ &$-1$ &$1$ &$0$ &$0$ &$-\frac{1}{2}$ &$-\frac{1}{6}$\\ \hline
$(c,d)$\qquad $(\textbf{1},\textbf{1})_{\frac{1}{2}}$ &$0$ &$0$ &$1$ &$1$ &$0$ &$0$ &$-\frac{1}{2}$ &$-\frac{1}{6}$\\ \hline
$(\ov c,\ov d)$\qquad $(\textbf{1},\textbf{1})_{-\frac{1}{2}}$ &$0$ &$0$ &$-1$ &$-1$ &$0$ &$0$ &$\frac{1}{2}$ &$\frac{1}{6}$\\ \hline
$(\ov b,c)$\qquad $(\textbf{1},\textbf{2})_{\frac{1}{2}}$ &$0$ &$-1$ &$2$ &$0$ &$0$ &$0$ &$-1$ &$0$\\ \hline
$(b,\ov c)$\qquad $(\textbf{1},\textbf{2})_{-\frac{1}{2}}$ &$0$ &$1$ &$-2$ &$0$ &$0$ &$0$ &$1$ &$0$\\ \hline
$(b,c)$\qquad $(\textbf{1},\textbf{2})_{-\frac{1}{2}}$ &$0$ &$1$ &$2$ &$0$ &$0$ &$0$ &$1$ &$0$\\ \hline
$(\ov b,\ov c)$\qquad $(\textbf{1},\textbf{2})_{\frac{1}{2}}$ &$0$ &$-1$ &$-2$ &$0$ &$0$ &$0$ &$-1$ &$0$\\ \hline
$\Ysymm_c$\qquad $(\textbf{1},\textbf{1})_{0}$ &$0$ &$0$ &$5$ &$0$ &$0$ &$0$ &$0$ &$0$\\ \hline
$\ov \Ysymm_c$\qquad $(\textbf{1},\textbf{1})_{0}$ &$0$ &$0$ &$-5$ &$0$ &$0$ &$0$ &$0$ &$0$\\ \hline
$(a,\ov d)$\qquad $(\textbf{3},\textbf{1})_{-\frac{5}{6}}$ &$1$ &$0$ &$0$ &$-3$ &$\frac{1}{2}$ &$0$ &$0$ &$\frac{3}{2}$\\ \hline
$(\ov a,d)$\qquad $(\ov{\textbf{3}},\textbf{1})_{\frac{5}{6}}$ &$-1$ &$0$ &$0$ &$3$ &$-\frac{1}{2}$ &$0$ &$0$ &$-\frac{3}{2}$\\ \hline
$(a,d)$\qquad $(\textbf{3},\textbf{1})_{\frac{1}{6}}$ &$1$ &$0$ &$0$ &$3$ &$-\frac{1}{2}$ &$0$ &$0$ &$\frac{1}{2}$\\ \hline
$(\ov a,\ov d)$\qquad $(\ov{\textbf{3}},\textbf{1})_{-\frac{1}{6}}$ &$-1$ &$0$ &$0$ &$-3$ &$\frac{1}{2}$ &$0$ &$0$ &$-\frac{1}{2}$\\ \hline
$(a,\ov b)$\qquad $(\textbf{3},\textbf{2})_{\frac{1}{6}}$ &$2$ &$-3$ &$0$ &$0$ &$-1$ &$1$ &$0$ &$0$\\ \hline
$(\ov a,b)$\qquad $(\ov{\textbf{3}},\textbf{2})_{-\frac{1}{6}}$ &$-2$ &$3$ &$0$ &$0$ &$1$ &$-1$ &$0$ &$0$\\ \hline
$(a,b)$\qquad $(\textbf{3},\textbf{2})_{-\frac{5}{6}}$ &$2$ &$3$ &$0$ &$0$ &$1$ &$1$ &$0$ &$0$\\ \hline
$(\ov a,\ov b)$\qquad $(\ov{\textbf{3}},\textbf{2})_{\frac{5}{6}}$ &$-2$ &$-3$ &$0$ &$0$ &$-1$ &$-1$ &$0$ &$0$\\ \hline
$(a,\ov c)$\qquad $(\textbf{3},\textbf{1})_{-\frac{1}{3}}$ &$1$ &$0$ &$-3$ &$0$ &$0$ &$0$ &$1$ &$0$\\ \hline
$(\ov a,c)$\qquad $(\ov{\textbf{3}},\textbf{1})_{\frac{1}{3}}$ &$-1$ &$0$ &$3$ &$0$ &$0$ &$0$ &$-1$ &$0$\\ \hline
$(a,c)$\qquad $(\textbf{3},\textbf{1})_{-\frac{1}{3}}$ &$1$ &$0$ &$3$ &$0$ &$0$ &$0$ &$1$ &$0$\\ \hline
$(\ov a,\ov c)$\qquad $(\ov{\textbf{3}},\textbf{1})_{\frac{1}{3}}$ &$-1$ &$0$ &$-3$ &$0$ &$0$ &$0$ &$-1$ &$0$\\ \hline
$(b,\ov d)$\qquad $(\textbf{1},\textbf{2})_{-1}$ &$0$ &$1$ &$0$ &$-2$ &$0$ &$\frac{1}{2}$ &$0$ &$\frac{4}{3}$\\ \hline
$(\ov b,d)$\qquad $(\textbf{1},\textbf{2})_{1}$ &$0$ &$-1$ &$0$ &$2$ &$0$ &$-\frac{1}{2}$ &$0$ &$-\frac{4}{3}$\\ \hline
$(b,d)$\qquad $(\textbf{1},\textbf{2})_{0}$ &$0$ &$1$ &$0$ &$2$ &$0$ &$-\frac{1}{2}$ &$0$ &$\frac{2}{3}$\\ \hline
$(\ov b,\ov d)$\qquad $(\textbf{1},\textbf{2})_{0}$ &$0$ &$-1$ &$0$ &$-2$ &$0$ &$\frac{1}{2}$ &$0$ &$-\frac{2}{3}$\\ \hline
\end{tabular}
\caption{Possible fields which might arise in the 4-node quiver with hypercharge embedding $U(1)_Y = -\frac{1}{3}U(1)_a - \frac{1}{2} U(1)_b + U(1)_d$.
The $(\textbf{1},\textbf{1})_{\frac{1}{2}}$, $(\textbf{1},\textbf{2})_{1,0}$, and $(\textbf{3},\textbf{1})_{\frac{1}{6},-\frac{5}{6}}$
states and their conjugates would  lead to fractional charge.}
\end{table}

\begin{table}
\centering
\begin{tabular}{|c|c|c|c|c|c|c|c|c|}\hline
Transformation & $T_a$ & $T_b$ & $T_c$ & $T_d$ & $M_a$ & $M_b$ & $M_c$ & $M_d$ \\ \hline \hline$\Ysymm_a$\qquad $(\textbf{6},\textbf{1})_{\frac{1}{3}}$ &$7$ &$0$ &$0$ &$0$ &$-\frac{1}{2}$ &$0$ &$0$ &$0$\\ \hline
$\ov \Ysymm_a$\qquad $(\ov{\textbf{6}},\textbf{1})_{-\frac{1}{3}}$ &$-7$ &$0$ &$0$ &$0$ &$\frac{1}{2}$ &$0$ &$0$ &$0$\\ \hline
$\Yasymm_a$\qquad $(\ov{\textbf{3}},\textbf{1})_{\frac{1}{3}}$ &$-1$ &$0$ &$0$ &$0$ &$-\frac{1}{2}$ &$0$ &$0$ &$0$\\ \hline
$\ov \Yasymm_a$\qquad $(\textbf{3},\textbf{1})_{-\frac{1}{3}}$ &$1$ &$0$ &$0$ &$0$ &$\frac{1}{2}$ &$0$ &$0$ &$0$\\ \hline
$\Ysymm_b$\qquad $(\textbf{1},\textbf{3})_{0}$ &$0$ &$6$ &$0$ &$0$ &$0$ &$0$ &$0$ &$0$\\ \hline
$\ov \Ysymm_b$\qquad $(\textbf{1},\textbf{3})_{0}$ &$0$ &$-6$ &$0$ &$0$ &$0$ &$0$ &$0$ &$0$\\ \hline
$\Yasymm_b$\qquad $(\textbf{1},\textbf{1})_{0}$ &$0$ &$-2$ &$0$ &$0$ &$0$ &$0$ &$0$ &$0$\\ \hline
$\ov \Yasymm_b$\qquad $(\textbf{1},\textbf{1})_{0}$ &$0$ &$2$ &$0$ &$0$ &$0$ &$0$ &$0$ &$0$\\ \hline
$\Ysymm_d$\qquad $(\textbf{1},\textbf{1})_{0}$ &$0$ &$0$ &$0$ &$5$ &$0$ &$0$ &$0$ &$0$\\ \hline
$\ov \Ysymm_d$\qquad $(\textbf{1},\textbf{1})_{0}$ &$0$ &$0$ &$0$ &$-5$ &$0$ &$0$ &$0$ &$0$\\ \hline
$(c,\ov d)$\qquad $(\textbf{1},\textbf{1})_{\frac{1}{2}}$ &$0$ &$0$ &$1$ &$-1$ &$0$ &$0$ &$-\frac{1}{6}$ &$-\frac{1}{2}$\\ \hline
$(\ov c,d)$\qquad $(\textbf{1},\textbf{1})_{-\frac{1}{2}}$ &$0$ &$0$ &$-1$ &$1$ &$0$ &$0$ &$\frac{1}{6}$ &$\frac{1}{2}$\\ \hline
$(c,d)$\qquad $(\textbf{1},\textbf{1})_{\frac{1}{2}}$ &$0$ &$0$ &$1$ &$1$ &$0$ &$0$ &$-\frac{1}{6}$ &$-\frac{1}{2}$\\ \hline
$(\ov c,\ov d)$\qquad $(\textbf{1},\textbf{1})_{-\frac{1}{2}}$ &$0$ &$0$ &$-1$ &$-1$ &$0$ &$0$ &$\frac{1}{6}$ &$\frac{1}{2}$\\ \hline
$(\ov b,c)$\qquad $(\textbf{1},\textbf{2})_{\frac{1}{2}}$ &$0$ &$-1$ &$2$ &$0$ &$0$ &$-\frac{1}{2}$ &$-\frac{1}{3}$ &$0$\\ \hline
$(b,\ov c)$\qquad $(\textbf{1},\textbf{2})_{-\frac{1}{2}}$ &$0$ &$1$ &$-2$ &$0$ &$0$ &$\frac{1}{2}$ &$\frac{1}{3}$ &$0$\\ \hline
$(b,c)$\qquad $(\textbf{1},\textbf{2})_{\frac{1}{2}}$ &$0$ &$1$ &$2$ &$0$ &$0$ &$-\frac{1}{2}$ &$-\frac{1}{3}$ &$0$\\ \hline
$(\ov b,\ov c)$\qquad $(\textbf{1},\textbf{2})_{-\frac{1}{2}}$ &$0$ &$-1$ &$-2$ &$0$ &$0$ &$\frac{1}{2}$ &$\frac{1}{3}$ &$0$\\ \hline
$\Ysymm_c$\qquad $(\textbf{1},\textbf{1})_{1}$ &$0$ &$0$ &$5$ &$0$ &$0$ &$0$ &$-\frac{4}{3}$ &$0$\\ \hline
$\ov \Ysymm_c$\qquad $(\textbf{1},\textbf{1})_{-1}$ &$0$ &$0$ &$-5$ &$0$ &$0$ &$0$ &$\frac{4}{3}$ &$0$\\ \hline
$(a,\ov d)$\qquad $(\textbf{3},\textbf{1})_{\frac{1}{6}}$ &$1$ &$0$ &$0$ &$-3$ &$0$ &$0$ &$0$ &$-\frac{1}{2}$\\ \hline
$(\ov a,d)$\qquad $(\ov{\textbf{3}},\textbf{1})_{-\frac{1}{6}}$ &$-1$ &$0$ &$0$ &$3$ &$0$ &$0$ &$0$ &$\frac{1}{2}$\\ \hline
$(a,d)$\qquad $(\textbf{3},\textbf{1})_{\frac{1}{6}}$ &$1$ &$0$ &$0$ &$3$ &$0$ &$0$ &$0$ &$-\frac{1}{2}$\\ \hline
$(\ov a,\ov d)$\qquad $(\ov{\textbf{3}},\textbf{1})_{-\frac{1}{6}}$ &$-1$ &$0$ &$0$ &$-3$ &$0$ &$0$ &$0$ &$\frac{1}{2}$\\ \hline
$(a,\ov b)$\qquad $(\textbf{3},\textbf{2})_{\frac{1}{6}}$ &$2$ &$-3$ &$0$ &$0$ &$0$ &$-\frac{1}{2}$ &$0$ &$0$\\ \hline
$(\ov a,b)$\qquad $(\ov{\textbf{3}},\textbf{2})_{-\frac{1}{6}}$ &$-2$ &$3$ &$0$ &$0$ &$0$ &$\frac{1}{2}$ &$0$ &$0$\\ \hline
$(a,b)$\qquad $(\textbf{3},\textbf{2})_{\frac{1}{6}}$ &$2$ &$3$ &$0$ &$0$ &$0$ &$-\frac{1}{2}$ &$0$ &$0$\\ \hline
$(\ov a,\ov b)$\qquad $(\ov{\textbf{3}},\textbf{2})_{-\frac{1}{6}}$ &$-2$ &$-3$ &$0$ &$0$ &$0$ &$\frac{1}{2}$ &$0$ &$0$\\ \hline
$(a,\ov c)$\qquad $(\textbf{3},\textbf{1})_{-\frac{1}{3}}$ &$1$ &$0$ &$-3$ &$0$ &$\frac{1}{2}$ &$0$ &$0$ &$0$\\ \hline
$(\ov a,c)$\qquad $(\ov{\textbf{3}},\textbf{1})_{\frac{1}{3}}$ &$-1$ &$0$ &$3$ &$0$ &$-\frac{1}{2}$ &$0$ &$0$ &$0$\\ \hline
$(a,c)$\qquad $(\textbf{3},\textbf{1})_{\frac{2}{3}}$ &$1$ &$0$ &$3$ &$0$ &$-\frac{1}{2}$ &$0$ &$-1$ &$0$\\ \hline
$(\ov a,\ov c)$\qquad $(\ov{\textbf{3}},\textbf{1})_{-\frac{2}{3}}$ &$-1$ &$0$ &$-3$ &$0$ &$\frac{1}{2}$ &$0$ &$1$ &$0$\\ \hline
$(b,\ov d)$\qquad $(\textbf{1},\textbf{2})_{0}$ &$0$ &$1$ &$0$ &$-2$ &$0$ &$0$ &$0$ &$0$\\ \hline
$(\ov b,d)$\qquad $(\textbf{1},\textbf{2})_{0}$ &$0$ &$-1$ &$0$ &$2$ &$0$ &$0$ &$0$ &$0$\\ \hline
$(b,d)$\qquad $(\textbf{1},\textbf{2})_{0}$ &$0$ &$1$ &$0$ &$2$ &$0$ &$0$ &$0$ &$0$\\ \hline
$(\ov b,\ov d)$\qquad $(\textbf{1},\textbf{2})_{0}$ &$0$ &$-1$ &$0$ &$-2$ &$0$ &$0$ &$0$ &$0$\\ \hline
\end{tabular}
\caption{Possible fields which might arise in the 4-node quiver with hypercharge embedding $U(1)_Y = \frac{1}{6} U(1)_a + \frac{1}{2} U(1)_c$.
The $(\textbf{1},\textbf{1})_{\frac{1}{2}}$, $(\textbf{1},\textbf{2})_{0}$, and $(\textbf{3},\textbf{1})_{\frac{1}{6}}$
states and their conjugates would  lead to fractional charge.}
\end{table}

\begin{table}
\centering
\begin{tabular}{|c|c|c|c|c|c|c|c|c|}\hline
Transformation & $T_a$ & $T_b$ & $T_c$ & $T_d$ & $M_a$ & $M_b$ & $M_c$ & $M_d$ \\ \hline \hline
$\Ysymm_a$\qquad $(\textbf{6},\textbf{1})_{\frac{1}{3}}$ &$7$ &$0$ &$0$ &$0$ &$-\frac{1}{2}$ &$0$ &$0$ &$0$\\ \hline
$\ov \Ysymm_a$\qquad $(\ov{\textbf{6}},\textbf{1})_{-\frac{1}{3}}$ &$-7$ &$0$ &$0$ &$0$ &$\frac{1}{2}$ &$0$ &$0$ &$0$\\ \hline
$\Yasymm_a$\qquad $(\ov{\textbf{3}},\textbf{1})_{\frac{1}{3}}$ &$-1$ &$0$ &$0$ &$0$ &$-\frac{1}{2}$ &$0$ &$0$ &$0$\\ \hline
$\ov \Yasymm_a$\qquad $(\textbf{3},\textbf{1})_{-\frac{1}{3}}$ &$1$ &$0$ &$0$ &$0$ &$\frac{1}{2}$ &$0$ &$0$ &$0$\\ \hline
$\Ysymm_b$\qquad $(\textbf{1},\textbf{3})_{0}$ &$0$ &$6$ &$0$ &$0$ &$0$ &$0$ &$0$ &$0$\\ \hline
$\ov \Ysymm_b$\qquad $(\textbf{1},\textbf{3})_{0}$ &$0$ &$-6$ &$0$ &$0$ &$0$ &$0$ &$0$ &$0$\\ \hline
$\Yasymm_b$\qquad $(\textbf{1},\textbf{1})_{0}$ &$0$ &$-2$ &$0$ &$0$ &$0$ &$0$ &$0$ &$0$\\ \hline
$\ov \Yasymm_b$\qquad $(\textbf{1},\textbf{1})_{0}$ &$0$ &$2$ &$0$ &$0$ &$0$ &$0$ &$0$ &$0$\\ \hline
$\Ysymm_d$\qquad $(\textbf{1},\textbf{1})_{1}$ &$0$ &$0$ &$0$ &$5$ &$0$ &$0$ &$0$ &$-\frac{4}{3}$\\ \hline
$\ov \Ysymm_d$\qquad $(\textbf{1},\textbf{1})_{-1}$ &$0$ &$0$ &$0$ &$-5$ &$0$ &$0$ &$0$ &$\frac{4}{3}$\\ \hline
$(c,\ov d)$\qquad $(\textbf{1},\textbf{1})_{0}$ &$0$ &$0$ &$1$ &$-1$ &$0$ &$0$ &$\frac{1}{3}$ &$-\frac{1}{3}$\\ \hline
$(\ov c,d)$\qquad $(\textbf{1},\textbf{1})_{0}$ &$0$ &$0$ &$-1$ &$1$ &$0$ &$0$ &$-\frac{1}{3}$ &$\frac{1}{3}$\\ \hline
$(c,d)$\qquad $(\textbf{1},\textbf{1})_{1}$ &$0$ &$0$ &$1$ &$1$ &$0$ &$0$ &$-\frac{2}{3}$ &$-\frac{2}{3}$\\ \hline
$(\ov c,\ov d)$\qquad $(\textbf{1},\textbf{1})_{-1}$ &$0$ &$0$ &$-1$ &$-1$ &$0$ &$0$ &$\frac{2}{3}$ &$\frac{2}{3}$\\ \hline
$(\ov b,c)$\qquad $(\textbf{1},\textbf{2})_{\frac{1}{2}}$ &$0$ &$-1$ &$2$ &$0$ &$0$ &$-\frac{1}{2}$ &$-\frac{1}{3}$ &$0$\\ \hline
$(b,\ov c)$\qquad $(\textbf{1},\textbf{2})_{-\frac{1}{2}}$ &$0$ &$1$ &$-2$ &$0$ &$0$ &$\frac{1}{2}$ &$\frac{1}{3}$ &$0$\\ \hline
$(b,c)$\qquad $(\textbf{1},\textbf{2})_{\frac{1}{2}}$ &$0$ &$1$ &$2$ &$0$ &$0$ &$-\frac{1}{2}$ &$-\frac{1}{3}$ &$0$\\ \hline
$(\ov b,\ov c)$\qquad $(\textbf{1},\textbf{2})_{-\frac{1}{2}}$ &$0$ &$-1$ &$-2$ &$0$ &$0$ &$\frac{1}{2}$ &$\frac{1}{3}$ &$0$\\ \hline
$\Ysymm_c$\qquad $(\textbf{1},\textbf{1})_{1}$ &$0$ &$0$ &$5$ &$0$ &$0$ &$0$ &$-\frac{4}{3}$ &$0$\\ \hline
$\ov \Ysymm_c$\qquad $(\textbf{1},\textbf{1})_{-1}$ &$0$ &$0$ &$-5$ &$0$ &$0$ &$0$ &$\frac{4}{3}$ &$0$\\ \hline
$(a,\ov d)$\qquad $(\textbf{3},\textbf{1})_{-\frac{1}{3}}$ &$1$ &$0$ &$0$ &$-3$ &$\frac{1}{2}$ &$0$ &$0$ &$0$\\ \hline
$(\ov a,d)$\qquad $(\ov{\textbf{3}},\textbf{1})_{\frac{1}{3}}$ &$-1$ &$0$ &$0$ &$3$ &$-\frac{1}{2}$ &$0$ &$0$ &$0$\\ \hline
$(a,d)$\qquad $(\textbf{3},\textbf{1})_{\frac{2}{3}}$ &$1$ &$0$ &$0$ &$3$ &$-\frac{1}{2}$ &$0$ &$0$ &$-1$\\ \hline
$(\ov a,\ov d)$\qquad $(\ov{\textbf{3}},\textbf{1})_{-\frac{2}{3}}$ &$-1$ &$0$ &$0$ &$-3$ &$\frac{1}{2}$ &$0$ &$0$ &$1$\\ \hline
$(a,\ov b)$\qquad $(\textbf{3},\textbf{2})_{\frac{1}{6}}$ &$2$ &$-3$ &$0$ &$0$ &$0$ &$-\frac{1}{2}$ &$0$ &$0$\\ \hline
$(\ov a,b)$\qquad $(\ov{\textbf{3}},\textbf{2})_{-\frac{1}{6}}$ &$-2$ &$3$ &$0$ &$0$ &$0$ &$\frac{1}{2}$ &$0$ &$0$\\ \hline
$(a,b)$\qquad $(\textbf{3},\textbf{2})_{\frac{1}{6}}$ &$2$ &$3$ &$0$ &$0$ &$0$ &$-\frac{1}{2}$ &$0$ &$0$\\ \hline
$(\ov a,\ov b)$\qquad $(\ov{\textbf{3}},\textbf{2})_{-\frac{1}{6}}$ &$-2$ &$-3$ &$0$ &$0$ &$0$ &$\frac{1}{2}$ &$0$ &$0$\\ \hline
$(a,\ov c)$\qquad $(\textbf{3},\textbf{1})_{-\frac{1}{3}}$ &$1$ &$0$ &$-3$ &$0$ &$\frac{1}{2}$ &$0$ &$0$ &$0$\\ \hline
$(\ov a,c)$\qquad $(\ov{\textbf{3}},\textbf{1})_{\frac{1}{3}}$ &$-1$ &$0$ &$3$ &$0$ &$-\frac{1}{2}$ &$0$ &$0$ &$0$\\ \hline
$(a,c)$\qquad $(\textbf{3},\textbf{1})_{\frac{2}{3}}$ &$1$ &$0$ &$3$ &$0$ &$-\frac{1}{2}$ &$0$ &$-1$ &$0$\\ \hline
$(\ov a,\ov c)$\qquad $(\ov{\textbf{3}},\textbf{1})_{-\frac{2}{3}}$ &$-1$ &$0$ &$-3$ &$0$ &$\frac{1}{2}$ &$0$ &$1$ &$0$\\ \hline
$(b,\ov d)$\qquad $(\textbf{1},\textbf{2})_{-\frac{1}{2}}$ &$0$ &$1$ &$0$ &$-2$ &$0$ &$\frac{1}{2}$ &$0$ &$\frac{1}{3}$\\ \hline
$(\ov b,d)$\qquad $(\textbf{1},\textbf{2})_{\frac{1}{2}}$ &$0$ &$-1$ &$0$ &$2$ &$0$ &$-\frac{1}{2}$ &$0$ &$-\frac{1}{3}$\\ \hline
$(b,d)$\qquad $(\textbf{1},\textbf{2})_{\frac{1}{2}}$ &$0$ &$1$ &$0$ &$2$ &$0$ &$-\frac{1}{2}$ &$0$ &$-\frac{1}{3}$\\ \hline
$(\ov b,\ov d)$\qquad $(\textbf{1},\textbf{2})_{-\frac{1}{2}}$ &$0$ &$-1$ &$0$ &$-2$ &$0$ &$\frac{1}{2}$ &$0$ &$\frac{1}{3}$\\ \hline
\end{tabular}
\caption{Possible fields which might arise in the 4-node quiver with hypercharge embedding $U(1)_Y = \frac{1}{6} U(1)_a + \frac{1}{2} U(1)_c + \frac{1}{2} U(1)_d$.}
\end{table}

\begin{table}
\centering
\begin{tabular}{|c|c|c|c|c|c|c|c|c|}\hline
Transformation & $T_a$ & $T_b$ & $T_c$ & $T_d$ & $M_a$ & $M_b$ & $M_c$ & $M_d$ \\ \hline \hline
$\Ysymm_a$\qquad $(\textbf{6},\textbf{1})_{\frac{1}{3}}$ &$7$ &$0$ &$0$ &$0$ &$-\frac{1}{2}$ &$0$ &$0$ &$0$\\ \hline
$\ov \Ysymm_a$\qquad $(\ov{\textbf{6}},\textbf{1})_{-\frac{1}{3}}$ &$-7$ &$0$ &$0$ &$0$ &$\frac{1}{2}$ &$0$ &$0$ &$0$\\ \hline
$\Yasymm_a$\qquad $(\ov{\textbf{3}},\textbf{1})_{\frac{1}{3}}$ &$-1$ &$0$ &$0$ &$0$ &$-\frac{1}{2}$ &$0$ &$0$ &$0$\\ \hline
$\ov \Yasymm_a$\qquad $(\textbf{3},\textbf{1})_{-\frac{1}{3}}$ &$1$ &$0$ &$0$ &$0$ &$\frac{1}{2}$ &$0$ &$0$ &$0$\\ \hline
$\Ysymm_b$\qquad $(\textbf{1},\textbf{3})_{0}$ &$0$ &$6$ &$0$ &$0$ &$0$ &$0$ &$0$ &$0$\\ \hline
$\ov \Ysymm_b$\qquad $(\textbf{1},\textbf{3})_{0}$ &$0$ &$-6$ &$0$ &$0$ &$0$ &$0$ &$0$ &$0$\\ \hline
$\Yasymm_b$\qquad $(\textbf{1},\textbf{1})_{0}$ &$0$ &$-2$ &$0$ &$0$ &$0$ &$0$ &$0$ &$0$\\ \hline
$\ov \Yasymm_b$\qquad $(\textbf{1},\textbf{1})_{0}$ &$0$ &$2$ &$0$ &$0$ &$0$ &$0$ &$0$ &$0$\\ \hline
$\Ysymm_d$\qquad $(\textbf{1},\textbf{1})_{3}$ &$0$ &$0$ &$0$ &$5$ &$0$ &$0$ &$0$ &$-4$\\ \hline
$\ov \Ysymm_d$\qquad $(\textbf{1},\textbf{1})_{-3}$ &$0$ &$0$ &$0$ &$-5$ &$0$ &$0$ &$0$ &$4$\\ \hline
$(c,\ov d)$\qquad $(\textbf{1},\textbf{1})_{-1}$ &$0$ &$0$ &$1$ &$-1$ &$0$ &$0$ &$\frac{4}{3}$ &$0$\\ \hline
$(\ov c,d)$\qquad $(\textbf{1},\textbf{1})_{1}$ &$0$ &$0$ &$-1$ &$1$ &$0$ &$0$ &$-\frac{4}{3}$ &$0$\\ \hline
$(c,d)$\qquad $(\textbf{1},\textbf{1})_{2}$ &$0$ &$0$ &$1$ &$1$ &$0$ &$0$ &$-\frac{5}{3}$ &$-1$\\ \hline
$(\ov c,\ov d)$\qquad $(\textbf{1},\textbf{1})_{-2}$ &$0$ &$0$ &$-1$ &$-1$ &$0$ &$0$ &$\frac{5}{3}$ &$1$\\ \hline
$(\ov b,c)$\qquad $(\textbf{1},\textbf{2})_{\frac{1}{2}}$ &$0$ &$-1$ &$2$ &$0$ &$0$ &$-\frac{1}{2}$ &$-\frac{1}{3}$ &$0$\\ \hline
$(b,\ov c)$\qquad $(\textbf{1},\textbf{2})_{-\frac{1}{2}}$ &$0$ &$1$ &$-2$ &$0$ &$0$ &$\frac{1}{2}$ &$\frac{1}{3}$ &$0$\\ \hline
$(b,c)$\qquad $(\textbf{1},\textbf{2})_{\frac{1}{2}}$ &$0$ &$1$ &$2$ &$0$ &$0$ &$-\frac{1}{2}$ &$-\frac{1}{3}$ &$0$\\ \hline
$(\ov b,\ov c)$\qquad $(\textbf{1},\textbf{2})_{-\frac{1}{2}}$ &$0$ &$-1$ &$-2$ &$0$ &$0$ &$\frac{1}{2}$ &$\frac{1}{3}$ &$0$\\ \hline
$\Ysymm_c$\qquad $(\textbf{1},\textbf{1})_{1}$ &$0$ &$0$ &$5$ &$0$ &$0$ &$0$ &$-\frac{4}{3}$ &$0$\\ \hline
$\ov \Ysymm_c$\qquad $(\textbf{1},\textbf{1})_{-1}$ &$0$ &$0$ &$-5$ &$0$ &$0$ &$0$ &$\frac{4}{3}$ &$0$\\ \hline
$(a,\ov d)$\qquad $(\textbf{3},\textbf{1})_{-\frac{4}{3}}$ &$1$ &$0$ &$0$ &$-3$ &$\frac{3}{2}$ &$0$ &$0$ &$1$\\ \hline
$(\ov a,d)$\qquad $(\ov{\textbf{3}},\textbf{1})_{\frac{4}{3}}$ &$-1$ &$0$ &$0$ &$3$ &$-\frac{3}{2}$ &$0$ &$0$ &$-1$\\ \hline
$(a,d)$\qquad $(\textbf{3},\textbf{1})_{\frac{5}{3}}$ &$1$ &$0$ &$0$ &$3$ &$-\frac{3}{2}$ &$0$ &$0$ &$-2$\\ \hline
$(\ov a,\ov d)$\qquad $(\ov{\textbf{3}},\textbf{1})_{-\frac{5}{3}}$ &$-1$ &$0$ &$0$ &$-3$ &$\frac{3}{2}$ &$0$ &$0$ &$2$\\ \hline
$(a,\ov b)$\qquad $(\textbf{3},\textbf{2})_{\frac{1}{6}}$ &$2$ &$-3$ &$0$ &$0$ &$0$ &$-\frac{1}{2}$ &$0$ &$0$\\ \hline
$(\ov a,b)$\qquad $(\ov{\textbf{3}},\textbf{2})_{-\frac{1}{6}}$ &$-2$ &$3$ &$0$ &$0$ &$0$ &$\frac{1}{2}$ &$0$ &$0$\\ \hline
$(a,b)$\qquad $(\textbf{3},\textbf{2})_{\frac{1}{6}}$ &$2$ &$3$ &$0$ &$0$ &$0$ &$-\frac{1}{2}$ &$0$ &$0$\\ \hline
$(\ov a,\ov b)$\qquad $(\ov{\textbf{3}},\textbf{2})_{-\frac{1}{6}}$ &$-2$ &$-3$ &$0$ &$0$ &$0$ &$\frac{1}{2}$ &$0$ &$0$\\ \hline
$(a,\ov c)$\qquad $(\textbf{3},\textbf{1})_{-\frac{1}{3}}$ &$1$ &$0$ &$-3$ &$0$ &$\frac{1}{2}$ &$0$ &$0$ &$0$\\ \hline
$(\ov a,c)$\qquad $(\ov{\textbf{3}},\textbf{1})_{\frac{1}{3}}$ &$-1$ &$0$ &$3$ &$0$ &$-\frac{1}{2}$ &$0$ &$0$ &$0$\\ \hline
$(a,c)$\qquad $(\textbf{3},\textbf{1})_{\frac{2}{3}}$ &$1$ &$0$ &$3$ &$0$ &$-\frac{1}{2}$ &$0$ &$-1$ &$0$\\ \hline
$(\ov a,\ov c)$\qquad $(\ov{\textbf{3}},\textbf{1})_{-\frac{2}{3}}$ &$-1$ &$0$ &$-3$ &$0$ &$\frac{1}{2}$ &$0$ &$1$ &$0$\\ \hline
$(b,\ov d)$\qquad $(\textbf{1},\textbf{2})_{-\frac{3}{2}}$ &$0$ &$1$ &$0$ &$-2$ &$0$ &$\frac{3}{2}$ &$0$ &$1$\\ \hline
$(\ov b,d)$\qquad $(\textbf{1},\textbf{2})_{\frac{3}{2}}$ &$0$ &$-1$ &$0$ &$2$ &$0$ &$-\frac{3}{2}$ &$0$ &$-1$\\ \hline
$(b,d)$\qquad $(\textbf{1},\textbf{2})_{\frac{3}{2}}$ &$0$ &$1$ &$0$ &$2$ &$0$ &$-\frac{3}{2}$ &$0$ &$-1$\\ \hline
$(\ov b,\ov d)$\qquad $(\textbf{1},\textbf{2})_{-\frac{3}{2}}$ &$0$ &$-1$ &$0$ &$-2$ &$0$ &$\frac{3}{2}$ &$0$ &$1$\\ \hline
\end{tabular}
\caption{Possible fields which might arise in the 4-node quiver with hypercharge embedding $U(1)_Y = \frac{1}{6} U(1)_a + \frac{1}{2} U(1)_c + \frac{3}{2} U(1)_d$.}
\label{4node6}
\end{table}

\bibliographystyle{JHEP}
\bibliography{refs}

\providecommand{\href}[2]{#2}\begingroup\raggedright\begin{thebibliography}{10}

\bibitem{Blumenhagen:2000wh}
R.~Blumenhagen, L.~Goerlich, B.~Kors, and D.~Lust, {\it {Noncommutative
  compactifications of type I strings on tori with magnetic background flux}},
  {\em JHEP} {\bf 10} (2000) 006,
  [\href{http://xxx.lanl.gov/abs/hep-th/0007024}{{\tt hep-th/0007024}}].

\bibitem{Aldazabal:2000dg}
G.~Aldazabal, S.~Franco, L.~E. Ibanez, R.~Rabadan, and A.~M. Uranga, {\it {D =
  4 Chiral String Compactifications from Intersecting Branes}},  {\em J. Math.
  Phys.} {\bf 42} (2001) 3103--3126,
  [\href{http://xxx.lanl.gov/abs/hep-th/0011073}{{\tt hep-th/0011073}}].

\bibitem{Aldazabal:2000cn}
G.~Aldazabal, S.~Franco, L.~E. Ibanez, R.~Rabadan, and A.~M. Uranga, {\it
  {Intersecting brane worlds}},  {\em JHEP} {\bf 02} (2001) 047,
  [\href{http://xxx.lanl.gov/abs/hep-ph/0011132}{{\tt hep-ph/0011132}}].

\bibitem{Blumenhagen:2001te}
R.~Blumenhagen, B.~Kors, D.~Lust, and T.~Ott, {\it {The standard model from
  stable intersecting brane world orbifolds}},  {\em Nucl. Phys.} {\bf B616}
  (2001) 3--33, [\href{http://xxx.lanl.gov/abs/hep-th/0107138}{{\tt
  hep-th/0107138}}].

\bibitem{Cvetic:2001nr}
M.~Cveti{\v c}, G.~Shiu, and A.~M. Uranga, {\it {Chiral Four-Dimensional
  ${\mathcal{N}}\!=1$ Supersymmetric Type IIA Orientifolds from Intersecting
  D6-Branes}},  {\em Nucl. Phys.} {\bf B615} (2001) 3--32,
  [\href{http://xxx.lanl.gov/abs/hep-th/0107166}{{\tt hep-th/0107166}}].

\bibitem{Cvetic:2001tj}
M.~Cveti{\v c}, G.~Shiu, and A.~M. Uranga, {\it {Three-Family Supersymmetric
  Standard Like Models from Intersecting Brane Worlds}},  {\em Phys. Rev.
  Lett.} {\bf 87} (2001) 201801,
  [\href{http://xxx.lanl.gov/abs/hep-th/0107143}{{\tt hep-th/0107143}}].

\bibitem{Blumenhagen:2005mu}
R.~Blumenhagen, M.~Cveti{\v c}, P.~Langacker, and G.~Shiu, {\it {Toward
  Realistic Intersecting D-Brane Models}},  {\em Ann. Rev. Nucl. Part. Sci.}
  {\bf 55} (2005) 71--139, [\href{http://xxx.lanl.gov/abs/hep-th/0502005}{{\tt
  hep-th/0502005}}].

\bibitem{Blumenhagen:2006ci}
R.~Blumenhagen, B.~Kors, D.~L{\" u}st, and S.~Stieberger, {\it
  {Four-Dimensional String Compactifications with D-Branes, Orientifolds and
  Fluxes}},  {\em Phys. Rept.} {\bf 445} (2007) 1--193,
  [\href{http://xxx.lanl.gov/abs/hep-th/0610327}{{\tt hep-th/0610327}}].

\bibitem{Cvetic:2011vz}
M.~Cveti{\v c} and J.~Halverson, {\it {Tasi Lectures: Particle Physics from
  Perturbative and Non- Perturbative Effects in D-Braneworlds}},
  \href{http://xxx.lanl.gov/abs/1101.2907}{{\tt arXiv:1101.2907}}.

\bibitem{Antoniadis:2000ena}
I.~Antoniadis, E.~Kiritsis, and T.~N. Tomaras, {\it {A D-Brane Alternative to
  Unification}},  {\em Phys. Lett.} {\bf B486} (2000) 186--193,
  [\href{http://xxx.lanl.gov/abs/hep-ph/0004214}{{\tt hep-ph/0004214}}].

\bibitem{Aldazabal:2000sa}
G.~Aldazabal, L.~E. Ibanez, F.~Quevedo, and A.~M. Uranga, {\it {D-Branes at
  Singularities: a Bottom-Up Approach to the String Embedding of the Standard
  Model}},  {\em JHEP} {\bf 08} (2000) 002,
  [\href{http://xxx.lanl.gov/abs/hep-th/0005067}{{\tt hep-th/0005067}}].

\bibitem{Cvetic:2010dz}
M.~Cveti{\v c}, J.~Halverson, and P.~Langacker, {\it {Singlet Extensions of the
  MSSM in the Quiver Landscape}},  {\em JHEP} {\bf 09} (2010) 076,
  [\href{http://xxx.lanl.gov/abs/1006.3341}{{\tt arXiv:1006.3341}}].

\bibitem{Bianchi:2000de}
M.~Bianchi and J.~F. Morales, {\it {Anomalies and Tadpoles}},  {\em JHEP} {\bf
  03} (2000) 030, [\href{http://xxx.lanl.gov/abs/hep-th/0002149}{{\tt
  hep-th/0002149}}].

\bibitem{Dijkstra:2004ym}
T.~P.~T. Dijkstra, L.~R. Huiszoon, and A.~N. Schellekens, {\it {Chiral
  Supersymmetric Standard Model Spectra from Orientifolds of Gepner Models}},
  {\em Phys. Lett.} {\bf B609} (2005) 408--417,
  [\href{http://xxx.lanl.gov/abs/hep-th/0403196}{{\tt hep-th/0403196}}].

\bibitem{Anastasopoulos:2010hu}
P.~Anastasopoulos, G.~K. Leontaris, R.~Richter, and A.~N. Schellekens, {\it
  {$SU(5)$ D-Brane Realizations, Yukawa Couplings and Proton Stability}},  {\em
  JHEP} {\bf 12} (2010) 011, [\href{http://xxx.lanl.gov/abs/1010.5188}{{\tt
  arXiv:1010.5188}}].

\bibitem{Blumenhagen:2005zg}
R.~Blumenhagen, G.~Honecker, and T.~Weigand, {\it {Non-Abelian Brane Worlds:
  the Heterotic String Story}},  {\em JHEP} {\bf 10} (2005) 086,
  [\href{http://xxx.lanl.gov/abs/hep-th/0510049}{{\tt hep-th/0510049}}].

\bibitem{Anastasopoulos:2006da}
P.~Anastasopoulos, T.~P.~T. Dijkstra, E.~Kiritsis, and A.~N. Schellekens, {\it
  {Orientifolds, Hypercharge Embeddings and the Standard Model}},  {\em Nucl.
  Phys.} {\bf B759} (2006) 83--146,
  [\href{http://xxx.lanl.gov/abs/hep-th/0605226}{{\tt hep-th/0605226}}].

\bibitem{Blumenhagen:2006xt}
R.~Blumenhagen, M.~Cveti{\v c}, and T.~Weigand, {\it {Spacetime Instanton
  Corrections in 4D String Vacua - the Seesaw Mechanism for D-Brane Models}},
  {\em Nucl. Phys.} {\bf B771} (2007) 113--142,
  [\href{http://xxx.lanl.gov/abs/hep-th/0609191}{{\tt hep-th/0609191}}].

\bibitem{Ibanez:2006da}
L.~E. Ibanez and A.~M. Uranga, {\it {Neutrino Majorana Masses from String
  Theory Instanton Effects}},  {\em JHEP} {\bf 03} (2007) 052,
  [\href{http://xxx.lanl.gov/abs/hep-th/0609213}{{\tt hep-th/0609213}}].

\bibitem{Florea:2006si}
B.~Florea, S.~Kachru, J.~McGreevy, and N.~Saulina, {\it {Stringy Instantons and
  Quiver Gauge Theories}},  {\em JHEP} {\bf 05} (2007) 024,
  [\href{http://xxx.lanl.gov/abs/hep-th/0610003}{{\tt hep-th/0610003}}].

\bibitem{Ibanez:2008my}
L.~E. Ibanez and .~Richter, Robert, {\it {Stringy Instantons and Yukawa
  Couplings in Mssm-Like Orientifold Models}},  {\em JHEP} {\bf 03} (2009) 090,
  [\href{http://xxx.lanl.gov/abs/0811.1583}{{\tt arXiv:0811.1583}}].

\bibitem{Anastasopoulos:2009mr}
P.~Anastasopoulos, E.~Kiritsis, and A.~Lionetto, {\it {On Mass Hierarchies in
  Orientifold Vacua}},  {\em JHEP} {\bf 08} (2009) 026,
  [\href{http://xxx.lanl.gov/abs/0905.3044}{{\tt arXiv:0905.3044}}].

\bibitem{Cvetic:2009yh}
M.~Cveti{\v c}, J.~Halverson, and .~Richter, Robert, {\it {Realistic Yukawa
  Structures from Orientifold Compactifications}},  {\em JHEP} {\bf 12} (2009)
  063, [\href{http://xxx.lanl.gov/abs/0905.3379}{{\tt arXiv:0905.3379}}].

\bibitem{Cvetic:2009ez}
M.~Cveti{\v c}, J.~Halverson, and R.~Richter, {\it {Mass Hierarchies from MSSM
  Orientifold Compactifications}},  {\em JHEP} {\bf 07} (2010) 005,
  [\href{http://xxx.lanl.gov/abs/0909.4292}{{\tt arXiv:0909.4292}}].

\bibitem{Cvetic:2009ng}
M.~Cveti{\v c}, J.~Halverson, and R.~Richter, {\it {Mass Hierarchies Vs. Proton
  Decay in MSSM Orientifold Compactifications}},
  \href{http://xxx.lanl.gov/abs/0910.2239}{{\tt arXiv:0910.2239}}.

\bibitem{Cvetic:2008hi}
M.~Cveti{\v c} and P.~Langacker, {\it {D-Instanton Generated Dirac Neutrino
  Masses}},  {\em Phys. Rev.} {\bf D78} (2008) 066012,
  [\href{http://xxx.lanl.gov/abs/0803.2876}{{\tt arXiv:0803.2876}}].

\bibitem{Kiritsis:2009sf}
E.~Kiritsis, M.~Lennek, and B.~Schellekens, {\it {$SU(5)$ Orientifolds, Yukawa
  Couplings, Stringy Instantons and Proton Decay}},  {\em Nucl. Phys.} {\bf
  B829} (2010) 298--324, [\href{http://xxx.lanl.gov/abs/0909.0271}{{\tt
  arXiv:0909.0271}}].

\bibitem{Cvetic:2010mm}
M.~Cveti{\v c}, J.~Halverson, P.~Langacker, and R.~Richter, {\it {The Weinberg
  Operator and a Lower String Scale in Orientifold Compactifications}},  {\em
  JHEP} {\bf 10} (2010) 094, [\href{http://xxx.lanl.gov/abs/1001.3148}{{\tt
  arXiv:1001.3148}}].

\bibitem{Kribs:2007nz}
G.~D. Kribs, T.~Plehn, M.~Spannowsky, and T.~M.~P. Tait, {\it {Four Generations
  and Higgs Physics}},  {\em Phys. Rev.} {\bf D76} (2007) 075016,
  [\href{http://xxx.lanl.gov/abs/0706.3718}{{\tt arXiv:0706.3718}}].

\bibitem{Erler:2010sk}
J.~Erler and P.~Langacker, {\it {Precision Constraints on Extra Fermion
  Generations}},  {\em Phys. Rev. Lett.} {\bf 105} (2010) 031801,
  [\href{http://xxx.lanl.gov/abs/1003.3211}{{\tt arXiv:1003.3211}}].

\bibitem{Baak:2011ze}
M.~Baak {\em et.~al.}, {\it {Updated Status of the Global Electroweak Fit and
  Constraints on New Physics}},  \href{http://xxx.lanl.gov/abs/1107.0975}{{\tt
  arXiv:1107.0975}}.

\bibitem{Godbole:2009sy}
R.~M. Godbole, S.~K. Vempati, and A.~Wingerter, {\it {Four Generations: SUSY
  and SUSY Breaking}},  {\em JHEP} {\bf 03} (2010) 023,
  [\href{http://xxx.lanl.gov/abs/0911.1882}{{\tt arXiv:0911.1882}}].

\bibitem{Kang:2007ib}
J.~Kang, P.~Langacker, and B.~D. Nelson, {\it {Theory and Phenomenology of
  Exotic Isosinglet Quarks and Squarks}},  {\em Phys. Rev.} {\bf D77} (2008)
  035003, [\href{http://xxx.lanl.gov/abs/0708.2701}{{\tt arXiv:0708.2701}}].

\bibitem{Nath:2010zj}
P.~Nath {\em et.~al.}, {\it {The Hunt for New Physics at the Large Hadron
  Collider}},  {\em Nucl. Phys. Proc. Suppl.} {\bf 200-202} (2010) 185--417,
  [\href{http://xxx.lanl.gov/abs/1001.2693}{{\tt arXiv:1001.2693}}].

\bibitem{Atre:2011ae}
A.~Atre {\em et.~al.}, {\it {Model-Independent Searches for New Quarks at the
  Lhc}},  \href{http://xxx.lanl.gov/abs/1102.1987}{{\tt arXiv:1102.1987}}.

\bibitem{Athron:2011wu}
P.~Athron, S.~F. King, D.~J. Miller, S.~Moretti, and R.~Nevzorov, {\it {Lhc
  Signatures of the Constrained Exceptional Supersymmetric Standard Model}},
  \href{http://xxx.lanl.gov/abs/1102.4363}{{\tt arXiv:1102.4363}}.

\bibitem{Alves:2011wf}
D.~Alves {\em et.~al.}, {\it {Simplified Models for Lhc New Physics Searches}},
   \href{http://xxx.lanl.gov/abs/1105.2838}{{\tt arXiv:1105.2838}}.

\bibitem{Gopalakrishna:2011ef}
S.~Gopalakrishna, T.~Mandal, S.~Mitra, and R.~Tibrewala, {\it {Lhc Signatures
  of a Vector-Like B'}},  \href{http://xxx.lanl.gov/abs/1107.4306}{{\tt
  arXiv:1107.4306}}.

\bibitem{Langacker:2008yv}
P.~Langacker, {\it {The Physics of Heavy Z-Prime Gauge Bosons}},  {\em Rev.
  Mod. Phys.} {\bf 81} (2009) 1199--1228,
  [\href{http://xxx.lanl.gov/abs/0801.1345}{{\tt arXiv:0801.1345}}].

\bibitem{Barger:2009eq}
V.~Barger {\em et.~al.}, {\it {Family Non-Universal U(1)' Gauge Symmetries and
  B To S Transitions}},  {\em Phys. Rev.} {\bf D80} (2009) 055008,
  [\href{http://xxx.lanl.gov/abs/0902.4507}{{\tt arXiv:0902.4507}}].

\bibitem{Barger:2009qs}
V.~Barger {\em et.~al.}, {\it {B to S Transitions in Family-Dependent U(1)'
  Models}},  {\em JHEP} {\bf 12} (2009) 048,
  [\href{http://xxx.lanl.gov/abs/0906.3745}{{\tt arXiv:0906.3745}}].

\bibitem{Everett:2009cn}
L.~L. Everett, J.~Jiang, P.~G. Langacker, and T.~Liu, {\it {Phenomenological
  Implications of Supersymmetric Family Non-Universal U(1)-Prime Models}},
  {\em Phys. Rev.} {\bf D82} (2010) 094024,
  [\href{http://xxx.lanl.gov/abs/0911.5349}{{\tt arXiv:0911.5349}}].

\bibitem{Deshpande:2010hy}
N.~G. Deshpande, X.-G. He, and G.~Valencia, {\it {D0 Dimuon Asymmetry in $B_S$
  - $\Bar B_S$ Mixing and Constraints on New Physics}},  {\em Phys. Rev.} {\bf
  D82} (2010) 056013, [\href{http://xxx.lanl.gov/abs/1006.1682}{{\tt
  arXiv:1006.1682}}].

\bibitem{delAguila:2010mx}
F.~del Aguila, J.~de~Blas, and M.~Perez-Victoria, {\it {Electroweak Limits on
  General New Vector Bosons}},  {\em JHEP} {\bf 09} (2010) 033,
  [\href{http://xxx.lanl.gov/abs/1005.3998}{{\tt arXiv:1005.3998}}].

\bibitem{Antoniadis:2001np}
I.~Antoniadis, E.~Kiritsis, and T.~Tomaras, {\it {D-Brane Standard Model}},
  {\em Fortsch. Phys.} {\bf 49} (2001) 573--580,
  [\href{http://xxx.lanl.gov/abs/hep-th/0111269}{{\tt hep-th/0111269}}].

\bibitem{Kiritsis:2002aj}
E.~Kiritsis and P.~Anastasopoulos, {\it {The Anomalous Magnetic Moment of the
  Muon in the D-Brane Realization of the Standard Model}},  {\em JHEP} {\bf 05}
  (2002) 054, [\href{http://xxx.lanl.gov/abs/hep-ph/0201295}{{\tt
  hep-ph/0201295}}].

\bibitem{Ghilencea:2002da}
D.~M. Ghilencea, L.~E. Ibanez, N.~Irges, and F.~Quevedo, {\it {Tev-Scale Z'
  Bosons from D-Branes}},  {\em JHEP} {\bf 08} (2002) 016,
  [\href{http://xxx.lanl.gov/abs/hep-ph/0205083}{{\tt hep-ph/0205083}}].

\bibitem{Berenstein:2006pk}
D.~Berenstein and S.~Pinansky, {\it {The Minimal Quiver Standard Model}},  {\em
  Phys. Rev.} {\bf D75} (2007) 095009,
  [\href{http://xxx.lanl.gov/abs/hep-th/0610104}{{\tt hep-th/0610104}}].

\bibitem{Berenstein:2008xg}
D.~Berenstein, R.~Mart{\'\i ne}z, F.~Ochoa, and S.~Pinansky, {\it {Z-Prime
  Boson Detection in the Minimal Quiver Standard Model}},  {\em Phys. Rev.}
  {\bf D79} (2009) 095005, [\href{http://xxx.lanl.gov/abs/0807.1126}{{\tt
  arXiv:0807.1126}}].

\bibitem{Armillis:2007tb}
R.~Armillis, C.~Coriano, and M.~Guzzi, {\it {Trilinear Anomalous Gauge
  Interactions from Intersecting Branes and the Neutral Currents Sector}},
  {\em JHEP} {\bf 05} (2008) 015,
  [\href{http://xxx.lanl.gov/abs/0711.3424}{{\tt arXiv:0711.3424}}].

\bibitem{Kumar:2007zza}
J.~Kumar, A.~Rajaraman, and J.~D. Wells, {\it {Probing the Green-Schwarz
  Mechanism at the Large Hadron Collider}},  {\em Phys. Rev.} {\bf D77} (2008)
  066011, [\href{http://xxx.lanl.gov/abs/0707.3488}{{\tt arXiv:0707.3488}}].

\bibitem{Dudas:2009uq}
E.~Dudas, Y.~Mambrini, S.~Pokorski, and A.~Romagnoni, {\it {(In)Visible Z' and
  Dark Matter}},  {\em JHEP} {\bf 08} (2009) 014,
  [\href{http://xxx.lanl.gov/abs/0904.1745}{{\tt arXiv:0904.1745}}].

\bibitem{Anchordoqui:2011ag}
L.~A. Anchordoqui, H.~Goldberg, X.~Huang, D.~L{\" u}st, and T.~R. Taylor, {\it
  {Stringy Origin of Tevatron Wjj Anomaly}},
  \href{http://xxx.lanl.gov/abs/1104.2302}{{\tt arXiv:1104.2302}}.

\bibitem{Anchordoqui:2011eg}
L.~A. Anchordoqui {\em et.~al.}, {\it {Z'-Gauge Bosons as Harbingers of Low
  Mass Strings}},  \href{http://xxx.lanl.gov/abs/1107.4309}{{\tt
  arXiv:1107.4309}}.

\bibitem{Anastasopoulos:2006cz}
P.~Anastasopoulos, M.~Bianchi, E.~Dudas, and E.~Kiritsis, {\it {Anomalies,
  Anomalous U(1)'s and Generalized Chern-Simons Terms}},  {\em JHEP} {\bf 11}
  (2006) 057, [\href{http://xxx.lanl.gov/abs/hep-th/0605225}{{\tt
  hep-th/0605225}}].

\bibitem{Mambrini:2009ad}
Y.~Mambrini, {\it {A Clear Dark Matter Gamma Ray Line Generated by the Green-
  Schwarz Mechanism}},  {\em JCAP} {\bf 0912} (2009) 005,
  [\href{http://xxx.lanl.gov/abs/0907.2918}{{\tt arXiv:0907.2918}}].

\bibitem{Kumar:2009us}
V.~Kumar and W.~Taylor, {\it {String Universality in Six Dimensions}},
  \href{http://xxx.lanl.gov/abs/0906.0987}{{\tt arXiv:0906.0987}}.

\bibitem{Kumar:2009ac}
V.~Kumar, D.~R. Morrison, and W.~Taylor, {\it {Mapping 6D ${\mathcal{N}}\!=1$
  Supergravities to F-Theory}},  {\em JHEP} {\bf 02} (2010) 099,
  [\href{http://xxx.lanl.gov/abs/0911.3393}{{\tt arXiv:0911.3393}}].

\bibitem{Kumar:2009ae}
V.~Kumar and W.~Taylor, {\it {A Bound on 6D ${\mathcal{N}}\!=1$
  Supergravities}},  {\em JHEP} {\bf 12} (2009) 050,
  [\href{http://xxx.lanl.gov/abs/0910.1586}{{\tt arXiv:0910.1586}}].

\bibitem{Witten:1982fp}
E.~Witten, {\it {An $SU(2)$ Anomaly}},  {\em Phys. Lett.} {\bf B117} (1982)
  324--328.

\bibitem{Ibanez:2001nd}
L.~E. Ibanez, F.~Marchesano, and R.~Rabadan, {\it {Getting Just the Standard
  Model at Intersecting Branes}},  {\em JHEP} {\bf 11} (2001) 002,
  [\href{http://xxx.lanl.gov/abs/hep-th/0105155}{{\tt hep-th/0105155}}].

\bibitem{Jiang:2006hf}
J.~Jiang, T.~Li, and D.~V. Nanopoulos, {\it {Testable Flipped $SU(5)$ $\times$
  $U(1)_X$ Models}},  {\em Nucl. Phys.} {\bf B772} (2007) 49--66,
  [\href{http://xxx.lanl.gov/abs/hep-ph/0610054}{{\tt hep-ph/0610054}}].

\bibitem{Georgi:1974sy}
H.~Georgi and S.~L. Glashow, {\it {Unity of All Elementary Particle Forces}},
  {\em Phys. Rev. Lett.} {\bf 32} (1974) 438--441.

\bibitem{Langacker:2011db}
P.~Langacker and G.~Steigman, {\it {Requiem for an Fchamp?}},
  \href{http://xxx.lanl.gov/abs/1107.3131}{{\tt arXiv:1107.3131}}.

\bibitem{Choudhury:2001hs}
D.~Choudhury, T.~M.~P. Tait, and C.~E.~M. Wagner, {\it {Beautiful Mirrors and
  Precision Electroweak Data}},  {\em Phys. Rev.} {\bf D65} (2002) 053002,
  [\href{http://xxx.lanl.gov/abs/hep-ph/0109097}{{\tt hep-ph/0109097}}].

\bibitem{Langacker:1980js}
P.~Langacker, {\it {Grand Unified Theories and Proton Decay}},  {\em Phys.
  Rept.} {\bf 72} (1981) 185.

\bibitem{Hewett:1988xc}
J.~L. Hewett and T.~G. Rizzo, {\it {Low-Energy Phenomenology of Superstring
  Inspired $E_{6}$ Models}},  {\em Phys. Rept.} {\bf 183} (1989) 193.

\bibitem{Chemtob:2004xr}
M.~Chemtob, {\it {Phenomenological Constraints on Broken R Parity Symmetry in
  Supersymmetry Models}},  {\em Prog. Part. Nucl. Phys.} {\bf 54} (2005)
  71--191, [\href{http://xxx.lanl.gov/abs/hep-ph/0406029}{{\tt
  hep-ph/0406029}}].

\bibitem{Barbier:2004ez}
R.~Barbier {\em et.~al.}, {\it {R-Parity Violating Supersymmetry}},  {\em Phys.
  Rept.} {\bf 420} (2005) 1--202,
  [\href{http://xxx.lanl.gov/abs/hep-ph/0406039}{{\tt hep-ph/0406039}}].

\bibitem{Ellis:1988er}
J.~R. Ellis, J.~F. Gunion, H.~E. Haber, L.~Roszkowski, and F.~Zwirner, {\it
  {Higgs Bosons in a Nonminimal Supersymmetric Model}},  {\em Phys. Rev.} {\bf
  D39} (1989) 844.

\bibitem{Barger:2006dh}
V.~Barger, P.~Langacker, H.-S. Lee, and G.~Shaughnessy, {\it {Higgs Sector in
  Extensions of the MSSM}},  {\em Phys. Rev.} {\bf D73} (2006) 115010,
  [\href{http://xxx.lanl.gov/abs/hep-ph/0603247}{{\tt hep-ph/0603247}}].

\bibitem{Ellwanger:2009dp}
U.~Ellwanger, C.~Hugonie, and A.~M. Teixeira, {\it {The Next-To-Minimal
  Supersymmetric Standard Model}},  {\em Phys. Rept.} {\bf 496} (2010) 1--77,
  [\href{http://xxx.lanl.gov/abs/0910.1785}{{\tt arXiv:0910.1785}}].

\bibitem{Maniatis:2009re}
M.~Maniatis, {\it {The Next-To-Minimal Supersymmetric Extension of the Standard
  Model Reviewed}},  {\em Int. J. Mod. Phys.} {\bf A25} (2010) 3505--3602,
  [\href{http://xxx.lanl.gov/abs/0906.0777}{{\tt arXiv:0906.0777}}].

\bibitem{Maxin:2011ne}
J.~A. Maxin, V.~E. Mayes, and D.~V. Nanopoulos, {\it {Proton Stability and Dark
  Matter in a Realistic String MSSM}},
  \href{http://xxx.lanl.gov/abs/1108.0887}{{\tt arXiv:1108.0887}}.

\bibitem{Kang:2004ix}
J.-h. Kang, P.~Langacker, and T.-j. Li, {\it {Neutrino Masses in Supersymmetric
  $SU(3)$ (C) $\times$ $SU(2)$ (L) $\times$ U(1) (Y) $\times$ U(1) -Prime
  Models}},  {\em Phys. Rev.} {\bf D71} (2005) 015012,
  [\href{http://xxx.lanl.gov/abs/hep-ph/0411404}{{\tt hep-ph/0411404}}].

\bibitem{Langacker:2000ju}
P.~Langacker and M.~Plumacher, {\it {Flavor Changing Effects in Theories with a
  Heavy Z-Prime Boson with Family Nonuniversal Couplings}},  {\em Phys. Rev.}
  {\bf D62} (2000) 013006, [\href{http://xxx.lanl.gov/abs/hep-ph/0001204}{{\tt
  hep-ph/0001204}}].

\bibitem{Blumenhagen:2005ga}
R.~Blumenhagen, G.~Honecker, and T.~Weigand, {\it {Loop-Corrected
  Compactifications of the Heterotic String with Line Bundles}},  {\em JHEP}
  {\bf 06} (2005) 020, [\href{http://xxx.lanl.gov/abs/hep-th/0504232}{{\tt
  hep-th/0504232}}].

\bibitem{Uranga:2000xp}
A.~M. Uranga, {\it {D-Brane Probes, Rr Tadpole Cancellation and K-Theory
  Charge}},  {\em Nucl. Phys.} {\bf B598} (2001) 225--246,
  [\href{http://xxx.lanl.gov/abs/hep-th/0011048}{{\tt hep-th/0011048}}].

\end{thebibliography}\endgroup

\end{document}